\documentclass[a4paper,11pt]{article}
\pdfoutput=1 

\usepackage{jheppub} 
                     
\usepackage{longtable}         
\usepackage{multirow}
\usepackage{changes}

\usepackage{booktabs}
\usepackage{array}
\usepackage{caption,subcaption}
\usepackage{xcolor,colortbl}

\usepackage[scaled=.85]{beramono}
\usepackage[T1]{fontenc}
\usepackage{color}

\definecolor{identifiercolor}{rgb}{.4,.6,.56}
\definecolor{stringcolor}{gray}{0.5}
\definecolor{inactivecolor}{rgb}{0.15,0.15,0.5}

\usepackage{listings}
\usepackage[T1]{fontenc} 

\newcommand{\C}[1]{{\cal C}_{#1}}

\begin{document} 

\begin{flushright}
{\small
LPT-Orsay-19-04}
\end{flushright}
$\ $
\vspace{-2mm}

\title{\boldmath What $R_K$ and $Q_5$ can tell us about New Physics in $b\to s\ell\ell$ transitions?
}

\author{Marcel Alguer\'o$^{a,b}$, Bernat Capdevila$^{a,b}$, S\'ebastien Descotes-Genon$^{c}$, Pere Masjuan$^{a,b}$ and Joaquim Matias$^{a,b}$ 
\vspace{0.3cm}}

\affiliation{
$^{a}$Grup de F\'isica Te\`orica (Departament de F\'isica), Universitat Aut\`onoma de Barcelona, E-08193 Bellaterra (Barcelona), Spain. \\
$^b$ Institut de F\'isica d'Altes Energies (IFAE), The Barcelona Institute of Science and Technology, Campus UAB, E-08193 Bellaterra (Barcelona), Spain.\\
$^{c}$Laboratoire de Physique Th\'eorique, UMR 8627, 
CNRS, Univ. Paris-Sud, Universit\'e Paris-Saclay, 91405 Orsay Cedex, France.\\
}

\abstract{
The deviations with respect to the Standard Model that are currently observed in $b \to s \ell\ell$ transitions, or $B$ anomalies, can be interpreted in terms of different New Physics (NP) scenarios within a model-independent effective approach. We identify a set of internal tensions of the fit that require further attention 
and whose  theoretical or experimental nature could be determined with more data.
In this landscape of NP, we discuss possible ways to discriminate among favoured NP hypotheses in the short term thanks to current and forthcoming observables. 
While an update of $R_K$ should help to disentangle the type of NP we may be observing (Lepton-Flavour Universality Violating and/or Lepton Flavour Universal),
 additional observables, in particular $Q_5$, turn out to be central to determine which NP hypothesis should be preferred. We also analyse the preferences shown by the current global fit concerning various NP hypotheses, using two different tools: the behaviour of the pulls of individual observables under NP scenarios and the directions favoured by approximate quadratic parametrisations of the observables in terms of Wilson coefficients.
  }

\maketitle

\section{Introduction and motivation}\label{sec:intro}

The LHCb~\cite{Aaij:2019wad} and Belle~\cite{Abdesselam:2019wac} collaborations have recently updated the measurements of the Lepton-Flavour Universality Violating (LFUV) ratios 
$R_K$ and $R_{K*}$:
\begin{equation}
R_{K(^*)}=\frac{{\cal B}(B\to K(^*)\mu^+\mu^-)}{{\cal B}(B\to K(^*)e^+e^-)}
\end{equation}
for various bins in the dilepton invariant mass. The LHCb collaboration observed hints of deviations from the SM between 2 and 3$\sigma$ in these observables, whereas the Belle measurements for $R_{K^*}$, affected by large uncertainties are compatible with the LHCb measurements as well as with SM. These deviations as well as others measured in $b\to s\mu\mu$ observables have triggered the combined analysis of LFUV and lepton-flavour dependent (LFD) observables performed in Ref.~\cite{Alguero:2019ptt}. The global analysis presented in the aforementiond reference shows that, using a model-independent approach, the Standard Model (SM) hypothesis is disfavoured compared to various hypotheses of New Physics (NP) contributions in $b\to s\ell\ell$ decays, with pulls w.r.t. the SM ranging from 5.3$\sigma$ to 5.9$\sigma$. Similar results were obtained by other groups using different treatments of hadronic uncertainties and sets of observables~\cite{Descotes-Genon:2013wba,Descotes-Genon:2015uva,Capdevila:2017bsm,Aebischer:2019mlg,Ciuchini:2019usw,Alok:2019ufo,Arbey:2019duh,Kumar:2019qbv}. These model-independent analyses constrain NP scenarios expressed
as contributions to the short-distance Wilson coefficients $\C{i\ell}$ in the effective Hamiltonian approach for $b\to s\ell\ell$ transitions. 

The first point to 
address is obviously whether NP has been discovered, but once this is established, it will prove important to determine
the specific pattern of NP discovered. Indeed, even if the amount of data obtained up to now for $b\to s\mu\mu$ makes sophisticated global fits to several Wilson coefficients possible~\cite{Descotes-Genon:2013wba,Descotes-Genon:2015uva,Capdevila:2017bsm,Alguero:2019ptt,Aebischer:2019mlg,Ciuchini:2019usw,Alok:2019ufo,Arbey:2019duh,Kumar:2019qbv}, the outcome is still not conclusive enough  to draw definite conclusions about the actual pattern of NP. 
Disentangling the realized pattern is an essential guide to build NP models in agreement with these observations. It is therefore usual to limit NP contributions to a few Wilson coefficients, that from now on we will refer as hypotheses, and to build NP models in agreement with these assumptions of the global fits.

Most scenarios discussed in the literature assumed that there is NP in muons only, i.e. the LFUV-NP contributions come from allowing the presence of NP in the muon channel and not in the electron one (or it is considered small). Three particularly interesting one-dimensional scenarios have emerged, namely NP in $\C{9\mu}$,  in $\C{9\mu}=-\C{9'\mu}$ and in  $\C{9\mu}=-\C{10\mu}$, with a larger significance for the first two scenarios and a smaller one for the latter in Ref.~\cite{Alguero:2019ptt}. On the contrary, we also found  that a fit restricted to a subset of mainly LFUV observables exhibits a marginal preference for the $\C{9\mu}^{\rm NP}=-\C{10\mu}^{\rm NP}$ scenario compared to the other scenarios. Moreover, we found several interesting two-dimensional scenarios, all involving $\C{9\mu}^{\rm NP}$, noticing that the inclusion of small but non-vanishing right-handed currents helps in solving some of the observed deviations with respect to SM expectations. 

Recently, several works  allowed for NP also in the electron channel, but no particular structure was envisaged from these fits by simply taking some of the electronic Wilson coefficients different from zero~\cite{Ciuchini:2017mik,Hurth:2017hxg,Ciuchini:2019usw,Kumar:2019qbv,Arbey:2019duh}.
However, in a recent article~\cite{Alguero:2018nvb}, we allowed the possibility of a specific structure, namely, that the $b\to s\ell\ell$ transitions get a common Lepton Flavour Universal (LFU) NP contribution for all
charged leptons (electrons, muons and tau leptons).  This permitted us  to identify new favoured NP hypotheses.
This idea was implemented by allowing two NP contributions inside the semileptonic Wilson coefficients:
\begin{equation}
\C{i\ell}^{\rm NP}=\C{i\ell}^{\rm V}+ \C{i}^{\rm U}
\end{equation}
with $\ell=e,\mu,\tau$ and where $\C{i\ell}^{\rm V}$ stands for LFUV-NP and  $\C{i}^{\rm U}$ for LFU-NP contributions. We distinguished the two contributions by imposing that $\C{ie}^{\rm V}=0$.  It is important at this point to emphasize the difference between simply allowing the presence of NP also in electrons or allowing the existence of two different kinds of NP contributions (LFU and LFUV). The case of simply allowing NP in the electron channel has been discussed quite extensively in Refs.~\cite{Hurth:2017hxg,Arbey:2019duh} (see also Refs.~\cite{Ciuchini:2017mik,Ciuchini:2019usw} for a smaller subset of scenarios with and without including low-recoil observables).  However, our approach of distinguishing LFU- and LFUV-NP structures provides new ideas to model building and extends the possible
 interpretations of the current fits. Performing the fits with this new setting~\cite{Alguero:2019ptt}, we obtained our previous results in Ref.~\cite{Capdevila:2017bsm} but also new scenarios different from 
 Refs.~\cite{Hurth:2017hxg,Ciuchini:2017mik,Ciuchini:2019usw,Arbey:2019duh}. This can be seen by translating LFU and LFUV contributions into NP contributions to muons and electrons (leaving $\tau$ aside at this stage)
\begin{eqnarray}
\C{9\mu}^{\rm NP}=\C{9\mu}^{\rm V}+\C{9}^{\rm U}, 
\quad 
\C{10\mu}^{\rm NP}=\C{10\mu}^{\rm V}+\C{10}^{\rm U},
\quad
\C{9e}^{\rm NP}=\C{9}^{\rm U}, \quad 
\C{10e}^{\rm NP}=\C{10}^{\rm U}\,.
\end{eqnarray}
This seemingly innocuous redefinition yields interesting consequences, as discussed in Ref.~\cite{Alguero:2018nvb}. 
It opens interesting perspectives to explain with different mechanisms the anomalies coming purely from the muon sector (like $\langle P_5^\prime \rangle_{[4,6]}$) and
the ones describing the violation of lepton flavour universality (like $\langle R_K\rangle_{[1.1,6]}$). These results were later updated in Ref.~\cite{Alguero:2019ptt}.
Let us notice in particular that this approach is different from all the analyses including NP in electrons \cite{Hurth:2017hxg,Ciuchini:2017mik,Ciuchini:2019usw,Kumar:2019qbv,Arbey:2019duh} where the muonic NP contribution is not correlated in any way with the electronic one.

When translated from one language to the other, the most interesting one- or two-dimensional scenarios in Refs.~\cite{Alguero:2018nvb,Alguero:2019ptt}
 become, for the purely LFUV cases:
\begin{eqnarray}
[{\rm Hyp.~I}]\qquad\qquad\qquad\qquad\qquad\qquad \{\C{9\mu}^{\rm V}\} &\to&
\{\C{9\mu}^{\rm NP}\}
\cr 
[{\rm Hyp.~II}]\qquad\qquad\qquad\qquad \{\C{9\mu}^{\rm V}=-\C{10\mu}^{\rm V} \} &\to&
\{\C{9\mu}^{\rm NP}=-\C{10\mu}^{\rm NP}\}
\cr
[{\rm Hyp.~III}]\qquad\qquad\qquad\qquad  \{\C{9\mu}^{\rm V}=-\C{9^\prime\mu}^{\rm V}\} &\to& \{\C{9\mu}^{\rm NP}=-\C{9^\prime\mu}^{\rm NP}\}
\cr
[{\rm Hyp.~IV}]\qquad\qquad\ \qquad\quad\qquad\   \{\C{9\mu}^{\rm V}, \C{10\mu}^{\rm V}\} &\to& \{\C{9\mu}^{\rm NP},\C{10\mu}^{\rm NP}\}
\cr
[{\rm Hyp.~V}] \qquad\qquad\qquad\quad\quad\qquad   \{\C{9\mu}^{\rm V}, \C{9'\mu}^{\rm V}\} &\to& \{\C{9\mu}^{\rm NP},\C{9'\mu}^{\rm NP}\}
\cr
[{\rm Hyp.~VI}]\qquad\qquad\qquad\qquad\quad  \{\C{9\mu}^{\rm V}, \C{10'\mu}^{\rm V}\} &\to& \{\C{9\mu}^{\rm NP},\C{10'\mu}^{\rm NP}\}
\cr
[{\rm Hyp.~VII}] \qquad  \{\C{9\mu}^{\rm V}=-\C{9'\mu}^{\rm V}, \C{10\mu}^{\rm V}=\C{10'\mu}^{\rm V}\} &\to& \{\C{9\mu}^{\rm NP}=-\C{9'\mu},\C{10\mu}^{\rm NP}=\C{10'\mu}\}
\cr
[{\rm Hyp.~VIII}] \qquad\qquad\quad  \{\C{9\mu}^{\rm V}, \C{9'\mu}^{\rm V}=-\C{10'\mu}^{\rm V}\} &\to& \{\C{9\mu}^{\rm NP},\C{9'\mu}^{\rm NP}=-\C{10'\mu}\}
\end{eqnarray}
and the scenarios allowing both LFUV and LFU contributions
\begin{eqnarray}
[{\rm Hyp. IX}]\quad \{\C{9\mu}^{\rm V}=-\C{10\mu}^{\rm V}, \C{9}^{\rm U}=\C{10}^{\rm U} \} &\to&
\{\C{9\mu}^{\rm NP}=-\C{10\mu}^{\rm NP}+ { 2 \C{9e}^{\rm NP}}, \C{9e}^{\rm NP}=\C{10e}^{\rm NP}\}
\cr
[{\rm Hyp. X}]\qquad \qquad\qquad\qquad \{\C{9\mu}^{\rm V}, \C{9}^{\rm U} \} &\to&
\{\C{9\mu}^{\rm NP},  \C{9e}^{\rm NP} \} 
\cr
[{\rm Hyp. XI}]\qquad\qquad \{\C{9\mu}^{\rm V}=-\C{10\mu}^{\rm V}, \C{9}^{\rm U} \} &\to& 
\{\C{9\mu}^{\rm NP}=-\C{10\mu}^{\rm NP}+\C{9e}^{\rm NP},\C{10\mu}^{\rm NP},\C{9e}^{\rm NP}\}
\cr
[{\rm Hyp. XII}]\quad\qquad\  \{\C{9\mu}^{\rm V}=-\C{10\mu}^{\rm V}, \C{10}^{\rm U} \} &\to& 
\{\C{9\mu}^{\rm NP},\C{10\mu}^{\rm NP}=-\C{9\mu}^{\rm NP}+\C{10e}^{\rm NP},\C{10e}^{\rm NP}\}
\cr
[{\rm Hyp. XIII}]\qquad \qquad\qquad\quad \{\C{9\mu}^{\rm V}, \C{10}^{\rm U} \} &\to&
\{\C{9\mu}^{\rm NP},  \C{10\mu}^{\rm NP}=\C{10e}^{\rm NP} \} 
\cr
[{\rm Hyp. XIV}]\qquad \qquad\qquad\quad \{\C{9\mu}^{\rm V}, \C{10'}^{\rm U} \} &\to&
\{\C{9\mu}^{\rm NP},  \C{10'\mu}^{\rm NP}=\C{10'e}^{\rm NP} \}
\end{eqnarray} 
The hypotheses VII and VIII correspond to the hypotheses 1 and 5 in Refs.~\cite{Capdevila:2017bsm,Alguero:2019ptt}.
The hypotheses IX to XIV correspond to the scenarios 6 to 11 in Ref.~\cite{Alguero:2018nvb,Alguero:2019ptt} (scenarios 5 and 13 are also interesting in terms of their ability to explain the deviations observed, but they require three or four free parameters and will not be considered in the following).

In this situation, it becomes clear that new data will be instrumental to disentangle the different hypotheses.
The goal of the present article is to scrutinize the results of the fit from a different perspective to prepare the next step, i.e. to discriminate the most relevant NP scenario among the ones already favoured, complementing our previous works~\cite{Capdevila:2017bsm, Capdevila:2017ert, Alguero:2018nvb, Alguero:2019ptt}. Currently, the most significant patterns 
identified exhibit a pull w.r.t the SM very close to each other (within a range of half a $\sigma$). We explore strategies to disentangle different scenarios and to identify the impact of a more precise measurement of $R_K$. We then combine information on $R_K$ and $Q_5$  in order to illustrate that 
$R_K$ by itself will not be sufficient to disentangle clearly one or a small subset of scenarios, but that a combination of $R_K$ and $Q_5$ can be useful, depending on the (future) measured value~\footnote{Up to now only the Belle experiment has been able to perform a measurement of $Q_{5}$, leading to $\langle Q_5^{\rm Belle}\rangle_{[1,6]}=+0.656\pm 0.485 \pm 0.103$~\cite{Wehle:2016yoi}.}.

In section~\ref{section tensions} we discuss the inner tensions of the fit in order to point those observables where further experimental or theoretical work would be required. In section~\ref{sec:disentangling} we explore how a forthcoming precise measurement of $R_K$ can disentangle or disfavour scenarios assuming that the statistical error is reduced and the central value stays within 2$\sigma$ of its present value. We analyse it considering two different fits, either with all observables or only the LFUV subset. We also discuss the impact of 
 a measurement of $Q_5$ in relation with its possible measurement by Belle II and LHCb. 
 We then discuss the structure of the current fits, looking more closely at the deviations of some observables in section~\ref{sec:pulls}, focusing on the change in their pulls depending on the NP scenario considered. 
In section~\ref{sec:param} we discuss the structure of the observables in terms of their Wilson coefficients
 to determine their sensitivities and the directions preferred by each of the anomalies, before drawing our conclusions.

\section{Inner tensions of the global fit}\label{section tensions}

In Refs.~\cite{Capdevila:2017bsm,Alguero:2019ptt}, we saw that different NP scenarios involving $\C{9\mu}^{\rm NP}$ led to a much better description of the data than the SM, with fits reaching p-values around 60-80\% (the SM being around 9\%) and providing pulls with respect to the SM above 5$\sigma$. The overall agreement is thus already very good within these NP scenarios and from a purely statistical point of view, it should be expected that these fits exhibit slight tensions. It is however interesting to look at these remaining tensions in more detail in order to determine where statistical fluctuations may be reduced with more data or where improved measurements might help to lift the degeneracy among NP scenarios. We focus on three main tensions that  we consider particularly relevant in the current global fit.

\subsection{$R_{K^*}$ in the first bin}\label{sec:firstbin}

\begin{figure}[t]
\includegraphics[width=7cm,height=5.1cm]{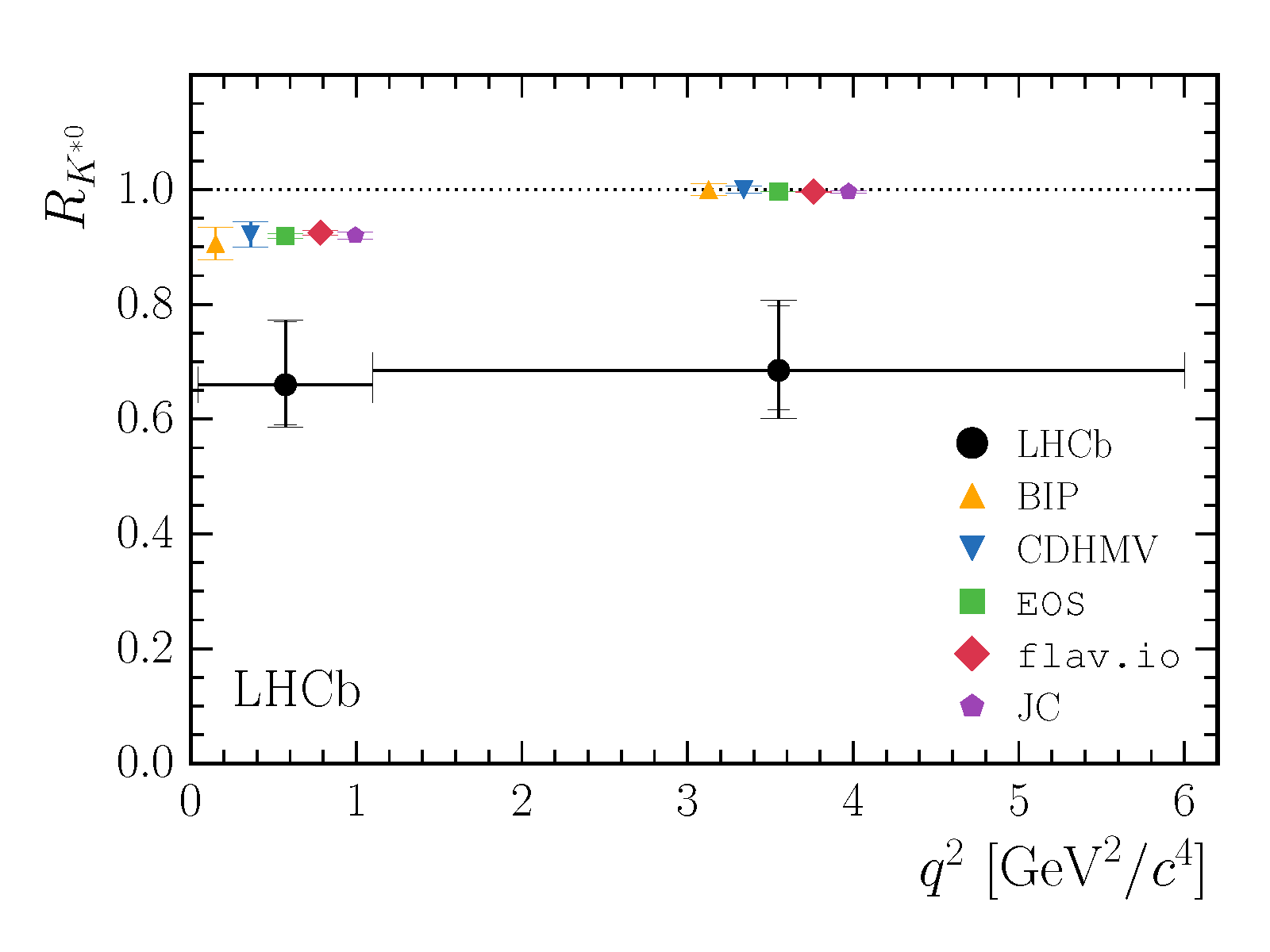}
\includegraphics[width=7.8cm]{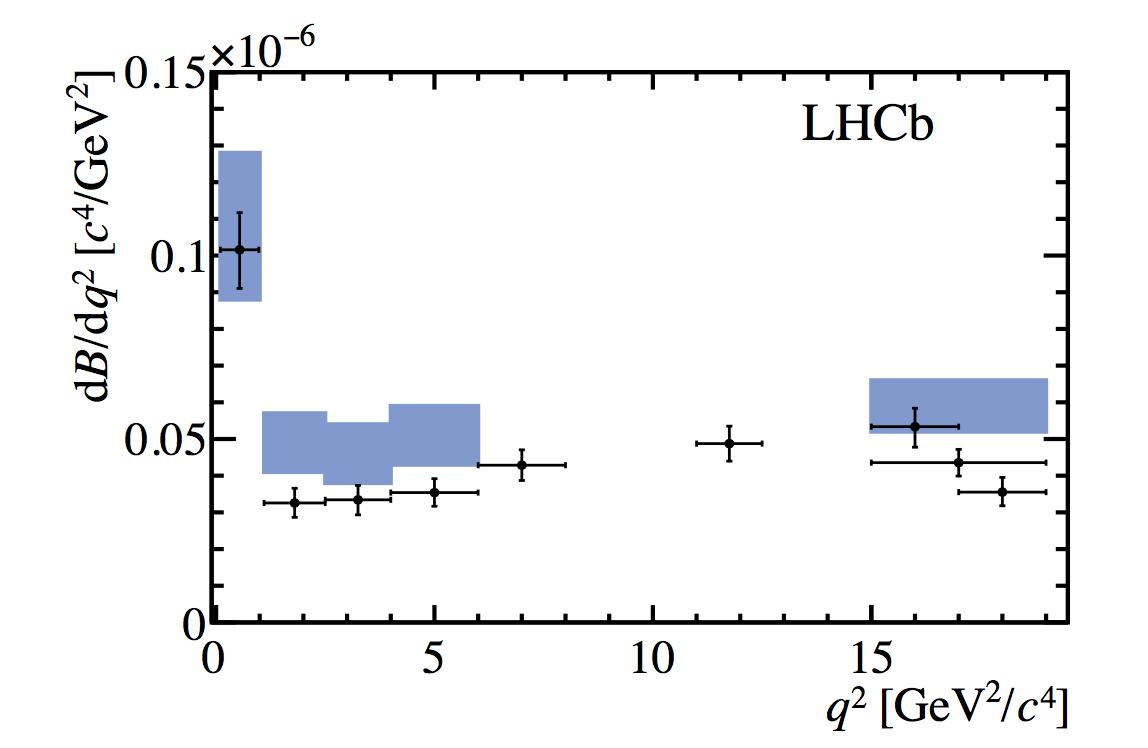}
\caption{$R_{K^*}$ (left panel) and ${\cal B}(B\to K^*\mu^{+}\mu^{-})$ (right panel) measured by LHCb. Figures extracted from Refs.~\cite{Aaij:2017vbb} and \cite{Aaij:2016flj} respectively.}\label{fig:RKstar}
\end{figure}

A first tension related to $R_{K^*}$ occurs in the global fit and it proves interesting to consider both $R_{K^*}$ and ${\cal B}(B \to K^*\mu^+\mu^-)$ as measured by LHCb in order to understand its nature (see Fig.~\ref{fig:RKstar}). 

Let us first consider the second bin (from 1 to 6 GeV$^2$) for $R_{K^*}$. Even though the deficit could be consistent with an excess in the electron channel with respect to the muon one, the study of the corresponding bins  of ${\cal B}(B \to K^*\mu^+\mu^-)$ points towards a deficit of muons. 
The mechanism that explains the deviation with respect to the SM in the long second bin of $R_{K^*}$ is consistent with all the deviations that have been observed in other channels and different invariant di-lepton mass square regions. 

The situation is different for the first bin of $R_{K^*}$, where ${\cal B}(B \to K^*\mu^+\mu^-)$ is clearly compatible with the SM (see Fig.~\ref{fig:RKstar}). An excess in the electron channel would then be needed in order to explain the observed deficit in $\langle R_{K^*}\rangle_{[1.1,6]}$. This difference of mechanism between the first and the second bins of $R_{K^*}$ can be understood in two ways: i) a specific
NP effect~\cite{Datta:2017ezo,Altmannshofer:2017bsz} localised at very low $q^2$ and able to compete with the dominant Wilson coefficient $\C{7}$ (well determined to be in agreement with the SM expectations from ${\cal B}(B \to X_s \gamma)$)~\cite{Capdevila:2017bsm,Alguero:2019ptt,Misiak:2006zs,Misiak:2006ab,DescotesGenon:2011yn,Asner:2010qj}; ii) some experimental issue in measuring di-electron pairs at very small invariant mass, close  to the photon pole. It would be very interesting that  LHCb keep on their efforts to understand the systematics in this bin. Interestingly the recent Belle measurement~\cite{Abdesselam:2019wac} indicates also a low central value in the same bin, even though the large uncertainty affecting the measurement prevents us from drawing any definite conclusion and makes it compatible also with the SM.

Another approach to  slightly reduce the tension between data and SM in the first bin of $R_{K^*}$ through a NP explanation consists in including NP contributions to the $b\to see$ channel, in particular, considering right-handed currents affecting electrons, as discussed in Ref.~\cite{Kumar:2019qbv}. 
 In the scenarios S8-S11 (using the notation of Ref.~\cite{Kumar:2019qbv}) the prediction of $\langle R_{K^*}\rangle_{[0.045,1.1]}$ is found to be within $\sim 1\sigma$ range of the current measurement. This could open a new window to explore the existence of right-handed currents and to explain some of the tensions found, even though more data is required in order to be conclusive.

\subsection{$B_s \to \phi \mu^{+}\mu^{-}$ versus $B \to K^*\mu^+\mu^-$}

Another tension in the fit concerns the branching ratio for $B_s \to\phi\mu^+\mu^-$, in particular when compared with the related decay $B \to K^*\mu^+\mu^-$. 

The prediction for the branching ratio ${\cal B}(B \to K^*\mu^+\mu^-)$ involves hadronic form factors to be determined using different theoretical approaches, depending on the di-lepton invariant mass region analysed: at large recoil, one can use light-cone sum-rules based on light-meson distribution amplitudes~\cite{Straub:2015ica}, while lattice form factors are available at low recoil. Due to the difficulty to assess precisely the uncertainties attached to light-cone sum rules, we perform our computation using  more conservative results from light-cone sum rules based on $B$-meson distribution amplitudes~\cite{Khodjamirian:2010vf} with conservative error estimates, exploiting QCD factorisation to restore correlations that were not available in Ref.~\cite{Khodjamirian:2010vf}. We checked that 
our results are compatible with those obtained in Ref.~\cite{Straub:2015ica} and that the two approaches yield very similar results for the fits~\cite{Capdevila:2017bsm,Descotes-Genon:2015uva,Altmannshofer:2017yso,Altmannshofer:2017fio}.

A recent update of these form factors is available in Ref.~\cite{Gubernari:2018wyi} using the same approach as Ref.~\cite{Khodjamirian:2010vf}, adding corrections from higher twists and providing correlations. We will update our results accordingly in a coming publication, but we do not expect very significant changes for the present article, based on our previous studies~\cite{Capdevila:2017ert}.  For instance, we checked that even if a large reduction of 50\% is achieved on the error of the form factors, the resulting uncertainty of key optimized observables like $P_5^\prime$ is minor 
(it would imply a reduction from 10\% to 8\% for the anomalous bins of $P_5^\prime$). On the contrary, a large impact is observed in unprotected observables like branching ratios or $S_i$ observables. As our fit is driven by the optimized observables, we expect only minor changes in the outcome of the fits. 

\begin{figure}[t]\begin{center}
{\includegraphics[width=9cm]{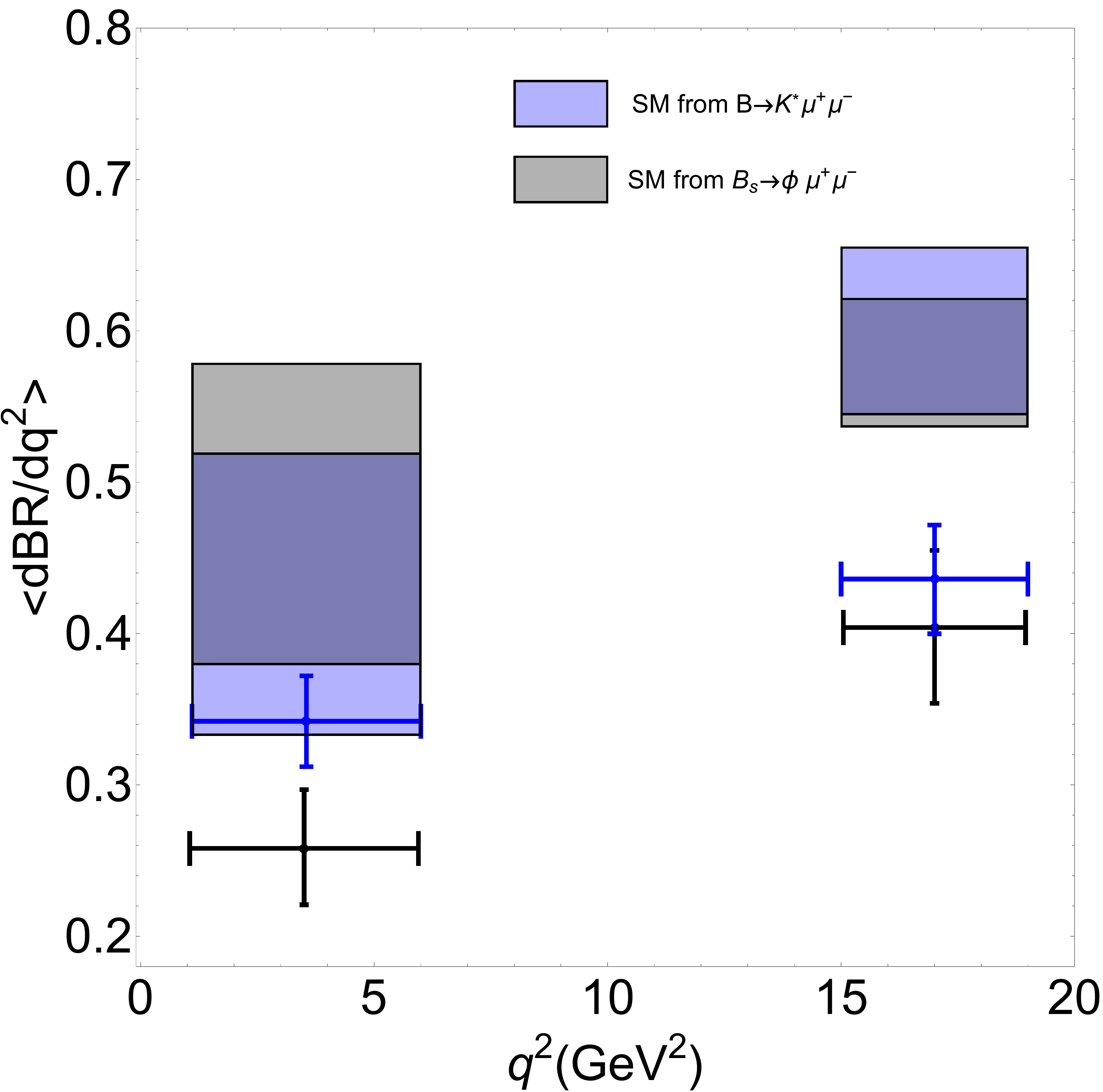}}
\caption{{Theoretical predictions for ${\cal B}(B_s\to \phi\mu^{+}\mu^{-})$ and ${\cal B}(B\to K^{*}\mu^{+}\mu^{-})$ within the SM along with their corresponding experimental measurement. The results at large recoil
are presented here only for illustrative purposes and are based on the form factors presented in Ref.~\cite{Straub:2015ica} (these results are not used in our global analyses). The results at low recoil
are indeed used in Ref.~\cite{Descotes-Genon:2015uva} and are based on available lattice QCD inputs for the form factors.
 }}\label{PlotsBRs}
\end{center}
\end{figure}

Contrary to the case of ${\cal B}({B \to K^*\mu^+\mu^-})$, there are no computations available using the B-meson light-cone sum rules of Refs.~\cite{Khodjamirian:2010vf,Gubernari:2018wyi} for $B_s\to\phi\mu^{+}\mu^{-}$, and one must rely on the estimates given in Ref.~\cite{Straub:2015ica}. One can see in Fig.~\ref{PlotsBRs} that at low recoil, where lattice form factors are used, the prediction for ${\cal B}(B \to K^*\mu^+\mu^-)$ is expected to be slightly larger than ${\cal B}(B_s \to\phi\mu^+\mu^-)$ and indeed data (with large error bars) follows the same trend. On the contrary, in the large-recoil region where the light-cone sum rules results of Ref.~\cite{Straub:2015ica} are used, the SM predictions lead to a larger value for ${\cal B}(B_s \to\phi\mu^+\mu^-)$ than for ${\cal B}(B \to K^*\mu^+\mu^-)$. Surprisingly, data shows the opposite trend, which may come from a statistical fluctuation of the data leading to an inversion of the experimental measurements of both modes at large recoil. Alternatively, this issue may signal a problem in the theoretical prediction of the form factors of Ref.~\cite{Straub:2015ica}. Firstly, these predictions are obtained by combining results in different kinematic regions (light-cone sum rules and lattice QCD) which do not fully agree with each other when they are extrapolated: the fit
to a common parametrisation over the whole kinematic space leads to a fit with uncertainties that may be artificially small due to these incompatibilities of the inputs. Moreover, the choice of the $z$-parametrisation~\cite{Khodjamirian:2010vf,Straub:2015ica} used to describe the  form factors over the whole kinematic range has interesting properties of convergence, but it may in some cases lead to potential unitarity violations~\cite{Gonzalez-Solis:2018ooo}.

Finally, another issue that specifically affects  ${\cal B}(B_s \to\phi\mu^+\mu^-)$ is the $B_s$-$\bar{B}_s$ mixing. As it is well known, $B_s$-$\bar{B}_s$ mixing implies that the time evolution of the $B_s$ meson before its decay involves two mass states with different widths that are linear combinations of the flavour states $B_s$ and $\bar{B}_s$. The current measurements performed at LHCb are integrated over time, and the neat effect of the evolution between the two mass states is a correction of $O(\Delta\Gamma_s/\Gamma_s)$ in the relation between the theoretical computation of the branching ratio and its measurement~\cite{DescotesGenon:2011pb,DeBruyn:2012wj,DeBruyn:2012jp,Descotes-Genon:2015hea}. This effect is taken into account in the global fit~\cite{Descotes-Genon:2015uva} as an additional source of uncertainty for the theoretical estimate of the branching ratios. 

The experimental efficiencies should also be corrected for this effect, which depend on the CP-asymmetry $A_{\Delta\Gamma}$ that can also be affected by NP contributions. It should thus be kept free within a large range in the absence of measurements. Neglecting this effect and assuming a SM value for this asymmetry may lead to an underestimation of some systematics on the efficiencies. For instance, Ref.~\cite{Dettori:2018bwt} showed that this issue can lead to an additional systematic effect of $10\%$ in the  $B_s\to\mu^{+}\mu^{-}$ systematics. The impact on efficiencies from NP effects was indeed considered in Ref.~\cite{Aaij:2015esa} for $B_s\to\phi\mu^{+}\mu^{-}$ by varying $\C{9\mu}$ in the underlying physics model used to compute signal efficiencies, leading to a much smaller effect in this case (of a few percent, in line with back-of-the-envelope estimates).

\subsection{Tensions between large and low recoil in angular observables}
\label{sec:understanding}
 
We discuss for the first time here a rather different type of tension, concerning the $B\to K^*\mu^{+}\mu^{-}$ angular observables at large and low recoil. On the one hand, we observe that branching ratios exhibit the same discrepancy pattern between theory and experiment at low and large recoil~\footnote{This is true for all $b\to s\ell\ell$ modes, apart from the decay $\Lambda_b^0 \to \Lambda \mu^+\mu^-$, where the experimental errors at low recoil are very large and the normalisation chosen prevents further interpretation~\cite{Aaij:2015xza,Aaij:2018gwm}.}. On the other hand, the current deviations at LHCb in $P_5^\prime$ require NP contributions with opposite sign in the two kinematic regions.  Indeed, the pull between the SM value and the LHCb experimental measurement in $\langle P_5^\prime \rangle_{[15,19]}$ has the opposite sign (albeit the significance is only 1.2$\sigma$) w.r.t. its  large-recoil bins, in particular $\langle P'_5\rangle_{[4,6]}$ and $\langle P'_5\rangle_{[6,8]}$. This very slight tension is not there in the case of the Belle data where same-sign deviations are observed, even though the error bars are rather large in this case.

\begin{figure}[t]
\begin{subfigure}{.5\textwidth}
  \centering
  \includegraphics[width=\linewidth]{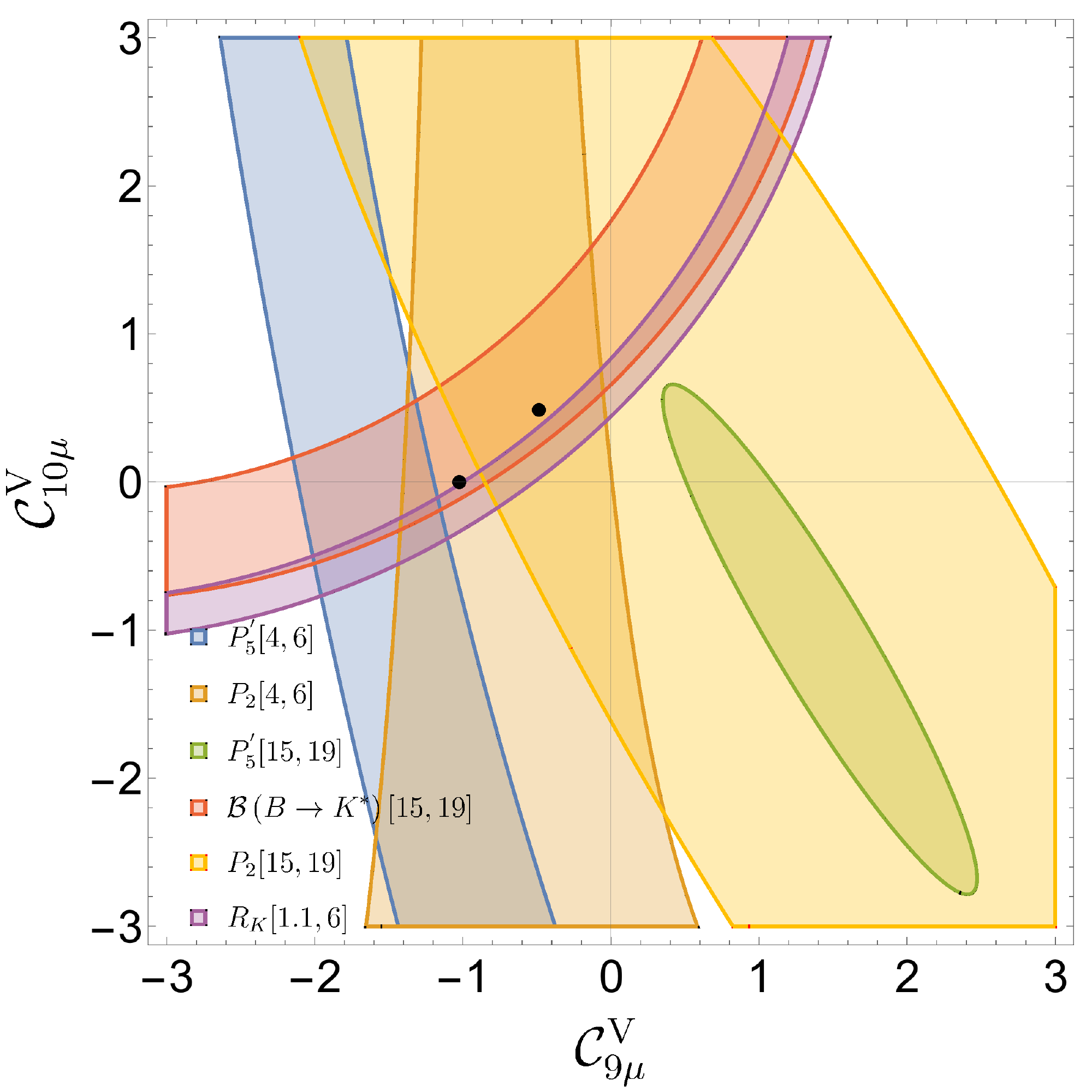}
  \label{fig:overlap 1}
\end{subfigure}%
\begin{subfigure}{.5\textwidth}
  \centering
  \includegraphics[width=\linewidth]{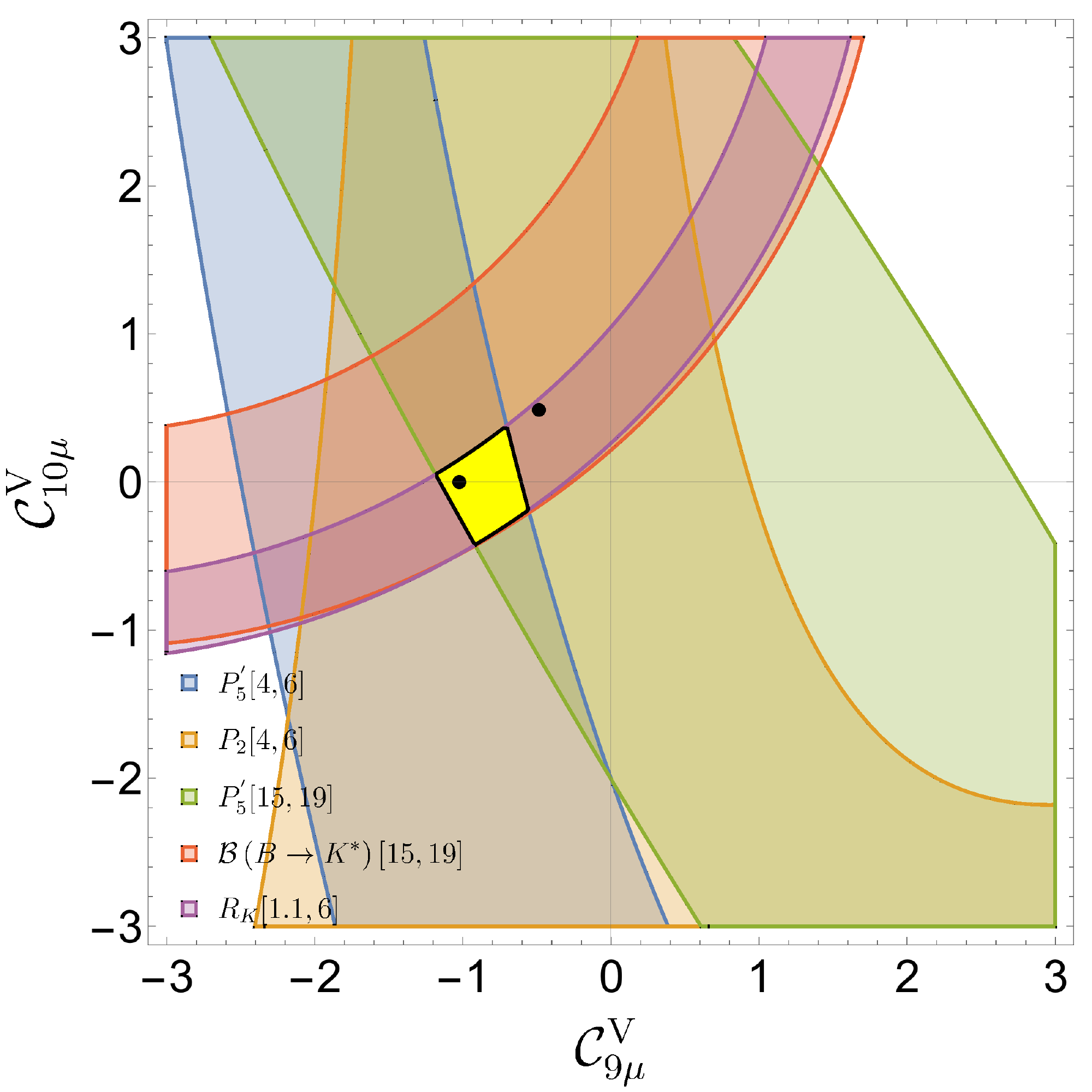}
  \label{fig:overlap 7}
\end{subfigure}
  \caption{68.3\% (left) and 95\% (right) CL solutions regions for the observables discussed in the main text in the $(\C{9\mu}^{\rm V},\C{10\mu}^{\rm V})$ plane. The yellow region corresponds to the overlap region. $\langle P_2\rangle_{[15,19]}$ is only shown in the left panel.}
  \label{fig:overlapSM}
\end{figure}

For the purposes of illustration, let us consider the NP scenario where there is no LFU contribution and NP occurs only in $\C{9\mu}^{\rm V}$ and $\C{10\mu}^{\rm V}$.
This is illustrated in Fig.~\ref{fig:overlapSM} where the constraints for these observables (as well as other relevant observables that will be listed below) are shown at 68.3\% (left) and 95\% (right) CL. One can notice their milder sensitivity to $\C{10\mu}^{\rm V}$.
$\langle P'_5\rangle_{[4,6]}$ (blue region) would prefer a negative $\C{9\mu}^{\rm V}$ while $\langle P'_5\rangle_{[15,19]}$ (green region) would favour a positive $\C{9\mu}^{\rm V}$ at 68.3\% CL. 

Black dots indicate the particular solutions $(-1.02,0)$ and $(-0.45,0.45)$ corresponding to the best-fit points of the 1D favoured scenarios in Ref.~\cite{Alguero:2019ptt} (hypotheses I and II of the present article). We also indicate the constraints from $\langle P_2\rangle_{[4,6]}$,  $\langle P_2\rangle_{[15,19]}$, and $\langle R_K\rangle_{[1.1,6]}$, $\langle{\cal B}(B^{0}\to K^{*0}\mu^{+}\mu^{-})\rangle_{[15,19]}$
since  we believe that they are representative of the set of observables driving our global fit~\footnote{$P_1$ and $P_4^\prime$ observables are known to behave in a more SM-like way than the ones selected here, thus providing weaker constraints.}. The former pair of observables ($\langle P_2\rangle_{[4,6]}$,  $\langle P_2\rangle_{[15,19]}$) has a large overlap region  compatible with the SM while the latter one ($\langle R_K\rangle_{[1.1,6]}$, $\langle{\cal B}(B^{0}\to K^{*0}\mu^{+}\mu^{-})\rangle_{[15,19]}$) overlaps far from the SM point. While $P'_5$ and $R_K$ strongly constrain NP solutions, the $P_2$ bins are weakly constraining. Finally, the yellow region in the right panel in Fig.~\ref{fig:overlapSM} is the overlap of the regions from the five observables obtained after considering the data regions at 95\% CL. 

In summary, an interesting tension between low- and large-recoil regions for $P'_5$  is observed  at the 2-sigma level, favouring $\C{9\mu}$ contributions of different signs in the two kinematic regions. Although not statistically significant, this inner tension seems to require either different sources of NP or a shift in the data once more statistics is added.

\section{Potential of $R_K$ (and $Q_5$) to disentangle NP hypotheses} \label{sec:disentangling}

In this section we discuss the potential impact of the prospective measurements of $\langle R_K\rangle_{[1.1,6]}$ and $\langle Q_5\rangle_{[1.1,6]}$ on the global fits in order to distinguish NP hypotheses. We perform the following illustrative exercise: we vary the experimental values of $\langle R_K\rangle_{[1.1,6]}$ and $\langle Q_5\rangle_{[1.1,6]}$ within suitable ranges, and we perform fits according to these values taken as actual measurements. 
First only $\langle R_K\rangle_{[1.1,6]}$ is allowed to vary before we consider the combined impact of $\langle R_K\rangle_{[1.1,6]}$ and $\langle Q_5\rangle_{[1.1,6]}$. The `pseudo-data' for $\langle R_K\rangle_{[1.1,6]}$  takes into account the expected increase in statistics soon available for this observable. For this exercise we take as the prospective experimental error for $\langle R_K\rangle_{[1.1,6]}$ $+0.044$ : this corresponds to a reduction by $\sqrt{2}$ of the statistical uncertainty in Ref.~\cite{Aaij:2019wad}. Indeed, according to the latter reference, this would amount to the inclusion of the data sets of 2017 and 2018 which are said to have the same statistical power as the combined data set of Run 1, 2015 and 2016.

For each fit (corresponding to a given hypothesis and set of data), both the pull of the hypothesis w.r.t. the SM ($\text{Pull}_\text{SM}$) and the best-fit-point (b.f.p) are computed, which we plot as functions of either $\langle R_K\rangle_{[1.1,6]}$ or $\langle Q_5\rangle_{[1.1,6]}$. Before discussing the results of our analysis, we first state our assumptions:

\begin{itemize}
\item[$\blacktriangleright$] We follow the same approach as in Refs.~\cite{Descotes-Genon:2013wba,Descotes-Genon:2015uva,Capdevila:2017bsm,Alguero:2018nvb,Alguero:2019ptt}. We consider a set of $b\to s\ell\ell$ observables measured by different experiments, we determine the experimental and theoretical correlation matrices between these observables assuming Gaussian distributions. Under a given NP hypothesis (generally described by one or two parameters added to specific short-distance Wilson coefficients), we build a $\chi^2$ from these observables and their correlation matrices that is used to extract the best-fit points and the confidence intervals of the NP parameters, as well as the pull of the NP hypothesis with respect to the SM, within a frequentist framework. 

\item[$\blacktriangleright$] We consider two different kinds of fits with different subsets of observables~\cite{Alguero:2019ptt}: on one side, the global fit (or Fit ``All'', to all 178 available observables) and on the other one, the LFUV fit, where only the observables measuring LFUV are included (plus constraints coming from radiative decays, leading to 20 observables). When several experiments have measured the same observable, we do not average the results but we include all these measurements in the $\chi^2$ taking into account their (theoretical) correlations.

\item[$\blacktriangleright$] Any variation of the experimental value of $\langle R_K\rangle_{[1.1,6]}$ could manifest itself also  in a change in the branching ratios $\mathcal{B}(B^+\to K^+\mu^+\mu^-)$ and/or $\mathcal{B}(B^+\to K^+e^+e^-)$. However, the update of $\langle R_K\rangle_{[1.1,6]}$ in Ref.~\cite{Aaij:2019wad} has not led to significant changes in these branching ratios, and we will assume that this will also occur in the forthcoming updates, so that we modify only the value of $\langle R_K\rangle_{[1.1,6]}$

\item[$\blacktriangleright$] $\langle R_K\rangle_{[1.1,6]}$ is freely varied within a $2\sigma$ range from its current experimental value. It represents a good compromise between a high coverage of the true value and a span compatible with our computational means. $\langle Q_5\rangle_{[1.1,6]}$ is varied within the range $[-0.5, 1.0]$ in order to ensure that we scan over values corresponding to the most relevant NP scenarios (see Fig. 2 of Ref.~\cite{Alguero:2018nvb}).

\item[$\blacktriangleright$] With the increased statistics available at Run 2, it will be possible for experiments to provide more precise determinations of key observables. Therefore, besides the reduction in the error of  $\langle R_K\rangle_{[1.1,6]}$, we assume
a guesstimated uncertainty of order 0.1 for $\langle Q_5\rangle_{[1.1,6]}$.
\end{itemize}

The purpose of this analysis is not to provide precise determinations of the pull of the SM
and the b.f.p.s for different values of $\langle R_K\rangle_{[1.1,6]}$ and $\langle Q_5\rangle_{[1.1,6]}$ but rather to gain qualitative knowledge on how experimental measurements of these two observables will drive the analyses. This is particularly true for the b.f.p. plots that will provide only the central value but not the confidence intervals obtained for the NP contributions.

\begin{figure}[h]
\centering
\begin{subfigure}{.48\textwidth}
  \centering
  \includegraphics[width=0.92\linewidth]{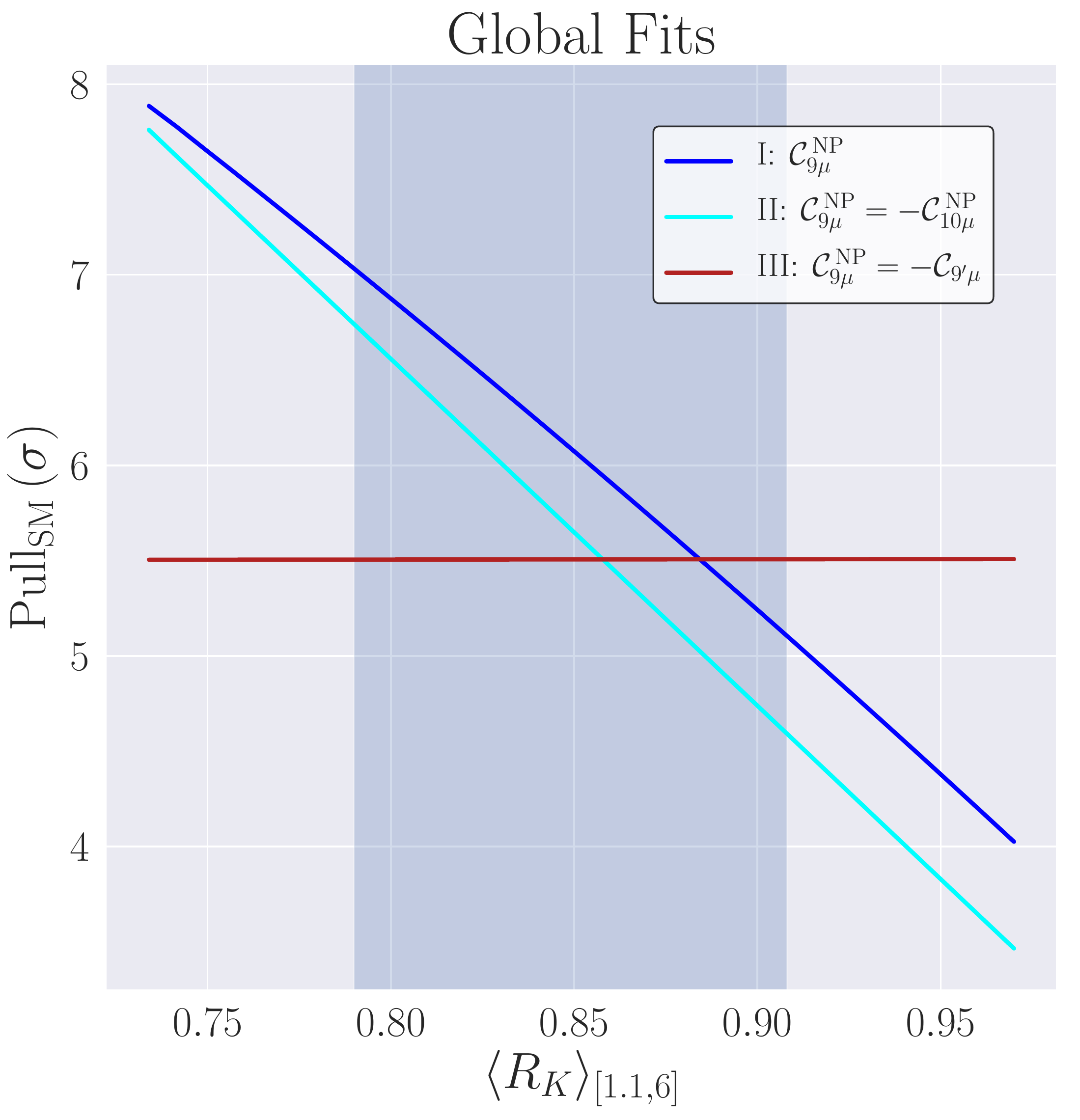}
\end{subfigure}
\begin{subfigure}{.48\textwidth}
  \centering
  \includegraphics[width=0.92\linewidth]{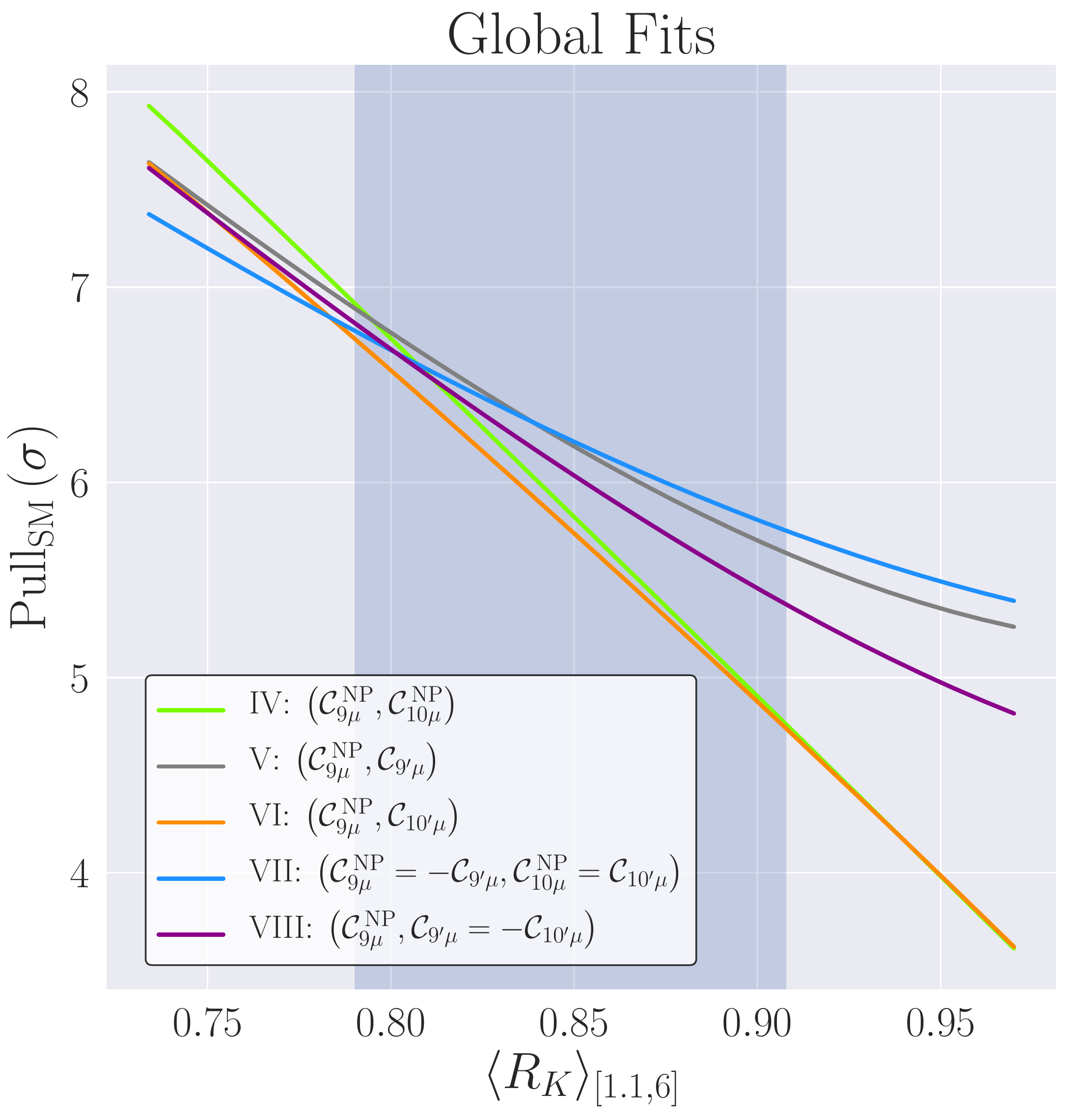}
\end{subfigure}
\begin{subfigure}{.48\textwidth}
  \centering
\includegraphics[width=0.92\textwidth]{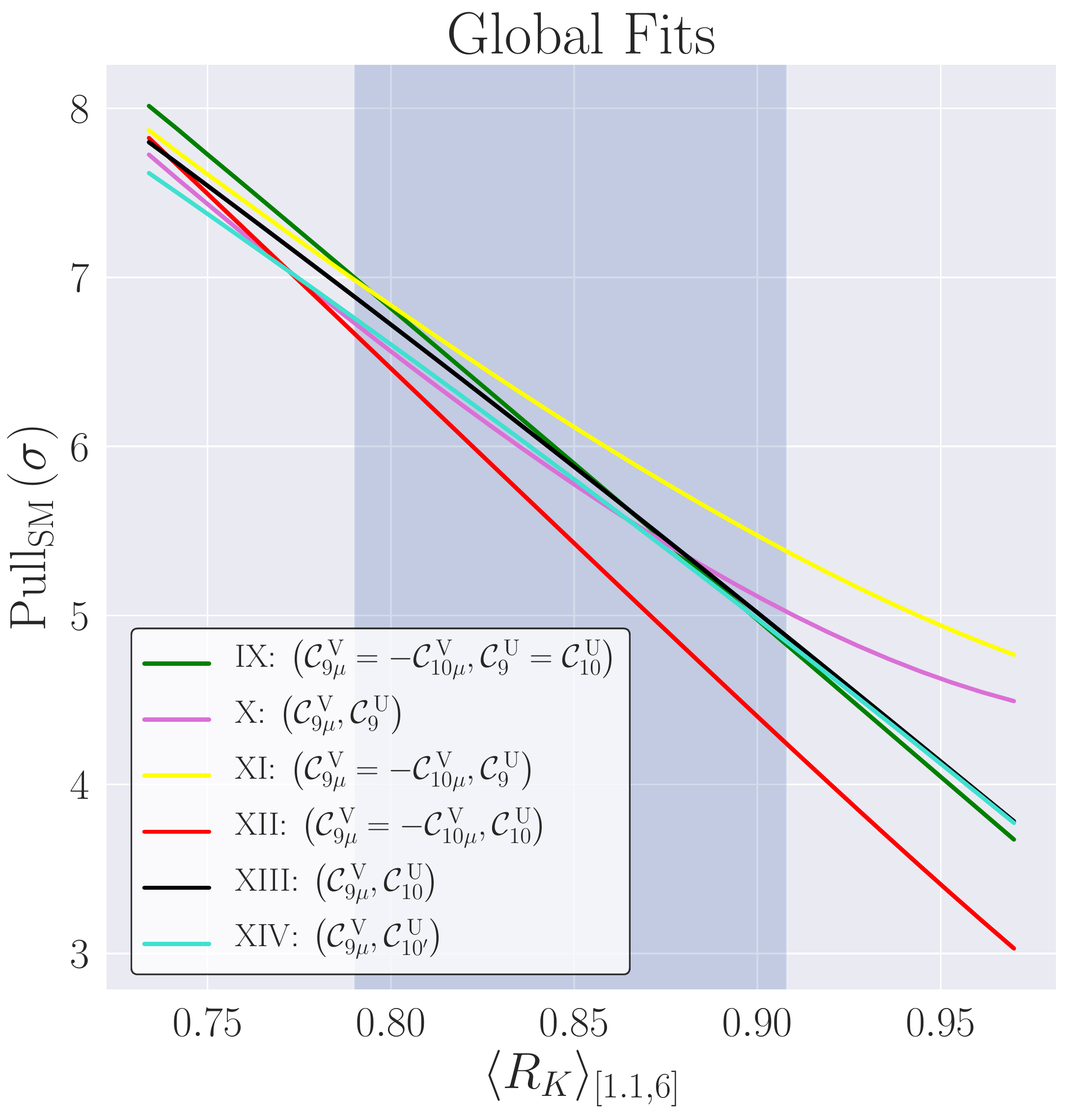}
\end{subfigure}
 \caption{Global fit: Impact of the central value of $\langle R_K\rangle_{[1.1,6]}$ on the $\text{Pull}_\text{SM}$ of the NP scenarios under consideration.}         
\label{fig:GlobalFitPlots_a}
\end{figure}

\clearpage

\subsection{Global Fits}

Figure~\ref{fig:GlobalFitPlots_a} displays the outcome of the global fit (or fit ``All'', involving 178 observables) for the pulls with respect of the SM, assuming different experimental central values of $\langle R_K\rangle_{[1.1,6]}$ and different NP hypotheses varied according to the procedure described above. In App.~\ref{sec:appendixbfp}, Figure~\ref{fig:GlobalFitPlots_b} displays a similar result for the best-fit point of each NP hypothesis. The shaded vertical band in the plots of Figures~\ref{fig:GlobalFitPlots_a} and \ref{fig:GlobalFitPlots_b} highlights the current experimental $1\sigma$ confidence interval for the LHCb average of $\langle R_K\rangle_{[1.1,6]}$.

Figure~\ref{fig:GlobalFitPlots_a} illustrates the relevance of $\langle R_K\rangle_{[1.1,6]}$ on the global fits. For all the NP scenarios considered, except for Hyp. III, $\C{9\mu}^\text{V}=-\C{9'\mu}^\text{V}$, we observe that their corresponding $\text{Pull}_\text{SM}$ undergoes a $\sim 3-4\sigma$ variation from one end of the range of variation of $\langle R_K\rangle_{[1.1,6]}$ to the other. If we restrict the variation of $\langle R_K\rangle_{[1.1,6]}$ to only $1\sigma$, one can see differences of $\sim 2\sigma$ between the two extremes, as expected from the linearity of $\text{Pull}_\text{SM}$ on $\langle R_K\rangle_{[1.1,6]}$ seen in the plots.

The flatness of the $\text{Pull}_\text{SM}$ under the hypothesis III, $\C{9\mu}^\text{V}=-\C{9'\mu}^\text{V}$, can be easily understood. The theoretical prediction of $\langle R_K\rangle_{[1.1,6]}$ is insensitive to the value of $\C{9\mu}^\text{V}=-\C{9'\mu}^\text{V}$, so that it remains constant and equal to 1 to a very high accuracy. Therefore the difference between the theoretical and experimental values of $\langle R_K\rangle_{[1.1,6]}$ does not play any role in the minimisation of the $\chi^2$ function. As a consequence, the b.f.p. is determined using the other observables of the fit, regardless of the experimental value for $\langle R_K\rangle_{[1.1,6]}$ (see Fig.~\ref{fig:GlobalFitPlots_b}), and the contribution of $\langle R_K\rangle_{[1.1,6]}$ cancels in the $\Delta\chi^2=\chi^2_\text{SM}-\chi^2_\text{min}$ statistic. This explains the observed flat curve for 
the $\text{Pull}_\text{SM}$, up to small variations linked to the numerical minimisation of the $\chi^2$ function.

The results in Figure~\ref{fig:GlobalFitPlots_a} show that, for most of the values of $\langle R_K\rangle_{[1.1,6]}$ scanned, it is not possible to fully disentangle all the NP scenarios, with the exception of Hyp III: $\C{9\mu}^\text{V}=-\C{9'\mu}^\text{V}$. However, large values of $\langle R_K\rangle_{[1.1,6]}$ (around 0.90 or above) provide the potential to disentangle some of the NP LFUV scenarios. Many scenarios get their significances down to the range $\sim 3.8\sigma-4.8\sigma$, apart from
scenarios with right-handed currents like Hyps. V, VII, VIII. Indeed, if a new measurement of $\langle R_K\rangle_{[1.1,6]}$ is found in better agreement with its SM prediction, this favours right-handed currents for $\C{9'\mu}^{\rm V}$ cancelling the contribution for $\C{9\mu}^{\rm V}$, but there is still an important number of other tensions (i.e. $R_{K^*}$, $P_{5\mu}^{\prime}$ and $\mathcal{B}(B_s\to\phi\mu^+\mu^-)$) that require NP contributions in order to be explained. Large values of $\langle R_K\rangle_{[1.1,6]}$ would help also to distinguish among NP scenarios featuring both LFUV and LFU NP, separating Hyps. X and XI from the others).

\begin{figure}[t]
\begin{subfigure}{.5\textwidth}
  \centering
  \includegraphics[width=0.92\linewidth]{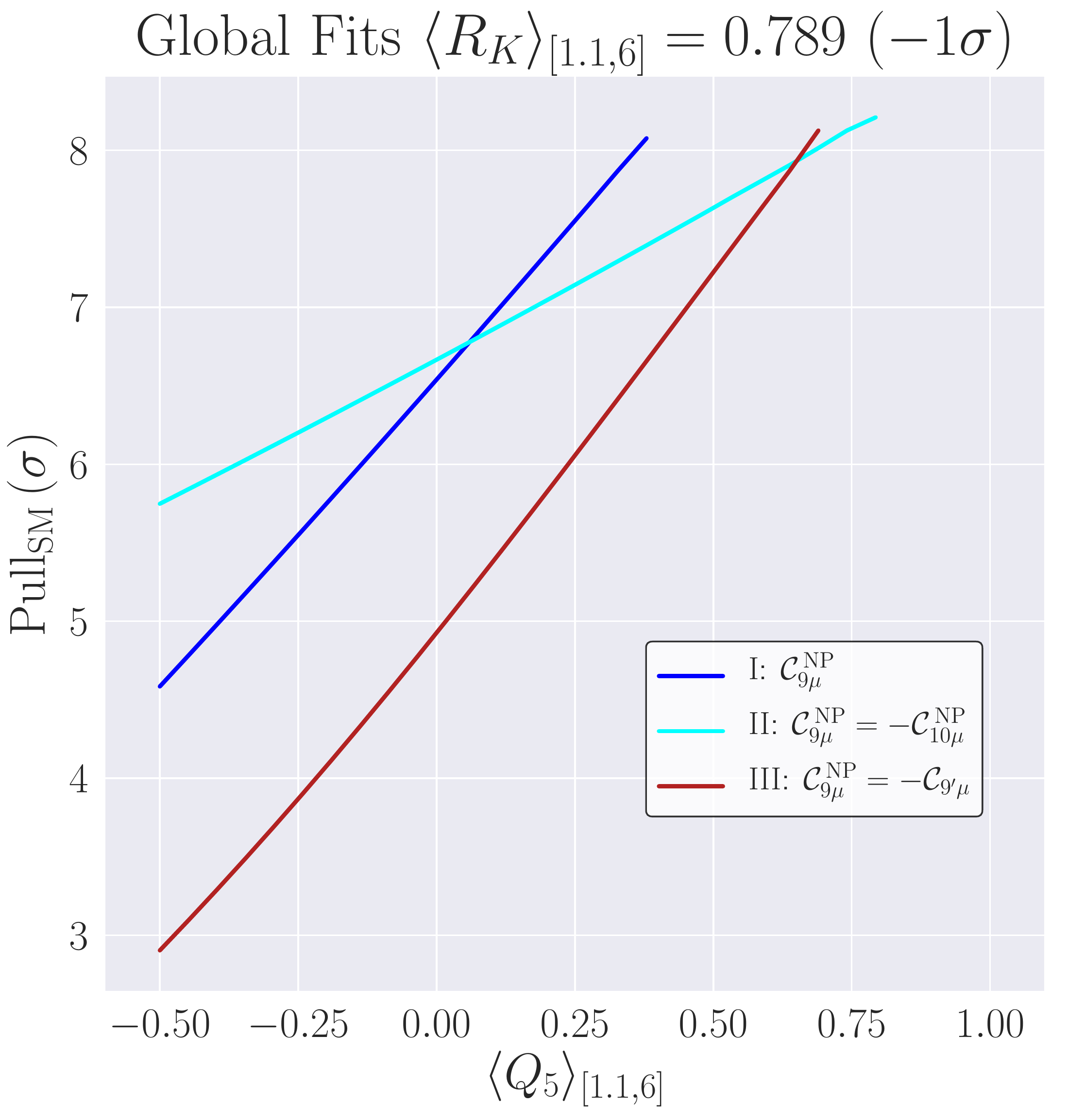}
\end{subfigure}%
\begin{subfigure}{.5\textwidth}
  \centering
  \includegraphics[width=0.92\linewidth]{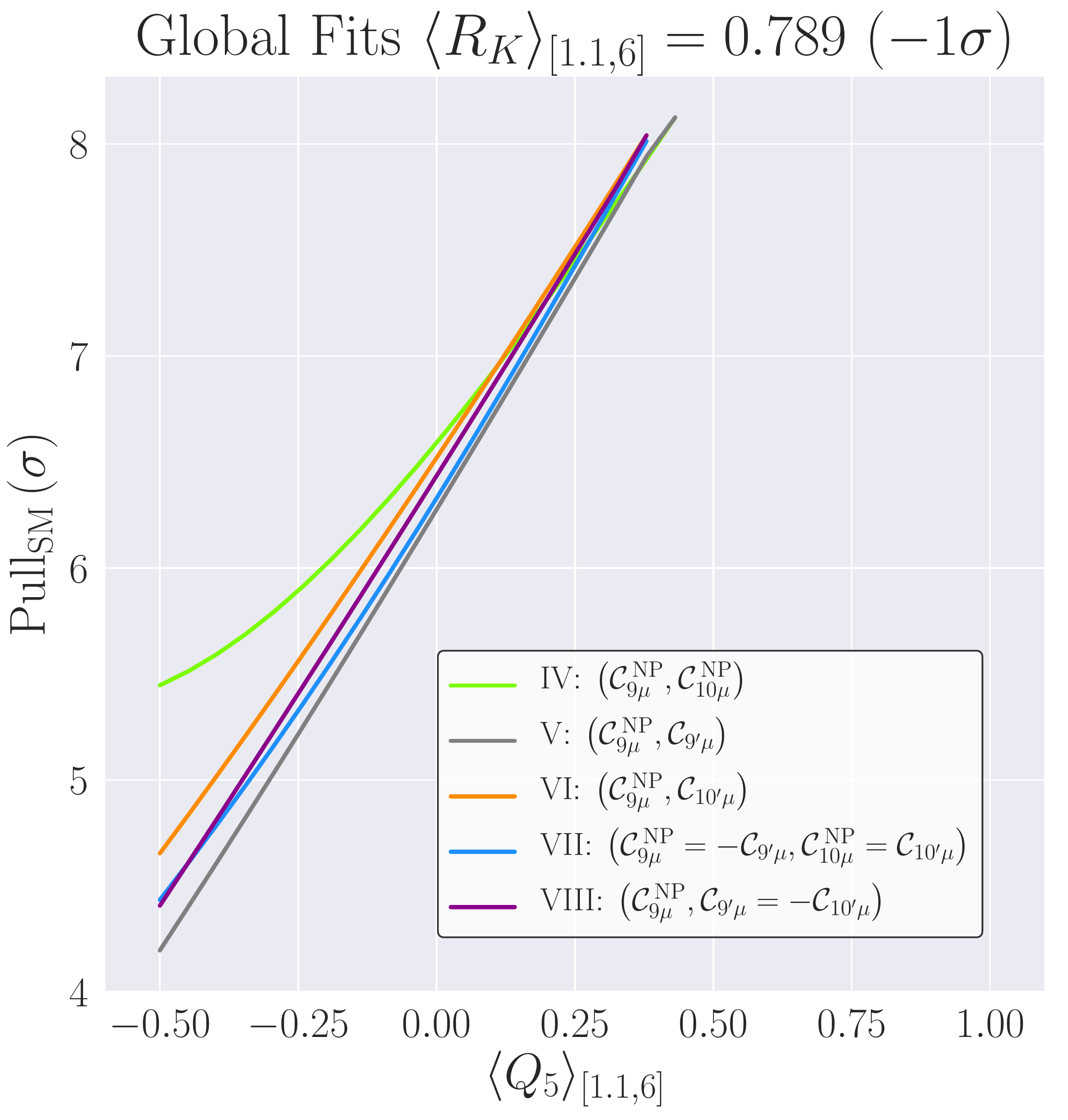}
\end{subfigure}
\begin{subfigure}{.5\textwidth}
  \centering
  \includegraphics[width=0.92\linewidth]{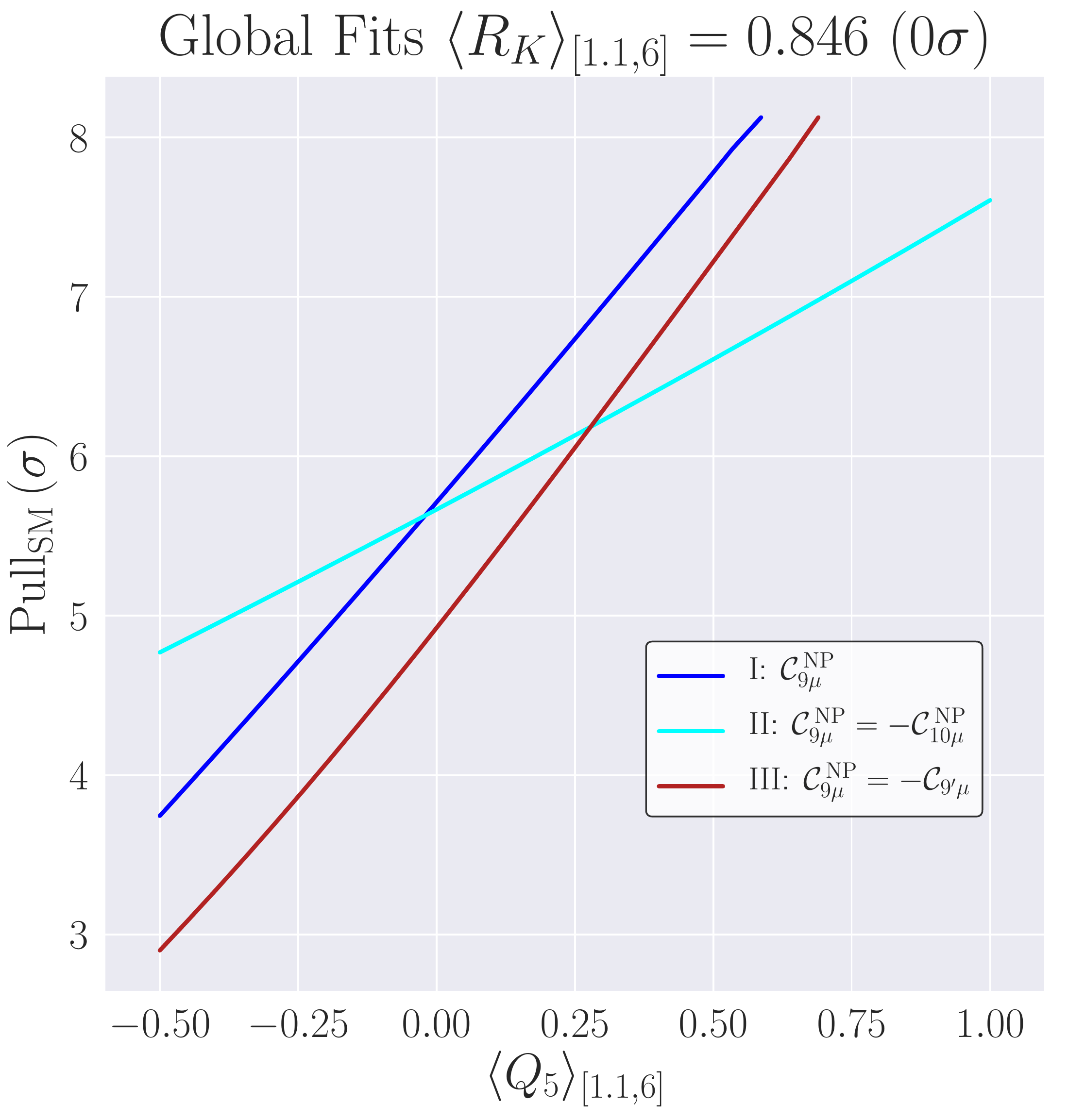}
\end{subfigure}%
\begin{subfigure}{.5\textwidth}
  \centering
  \includegraphics[width=0.92\linewidth]{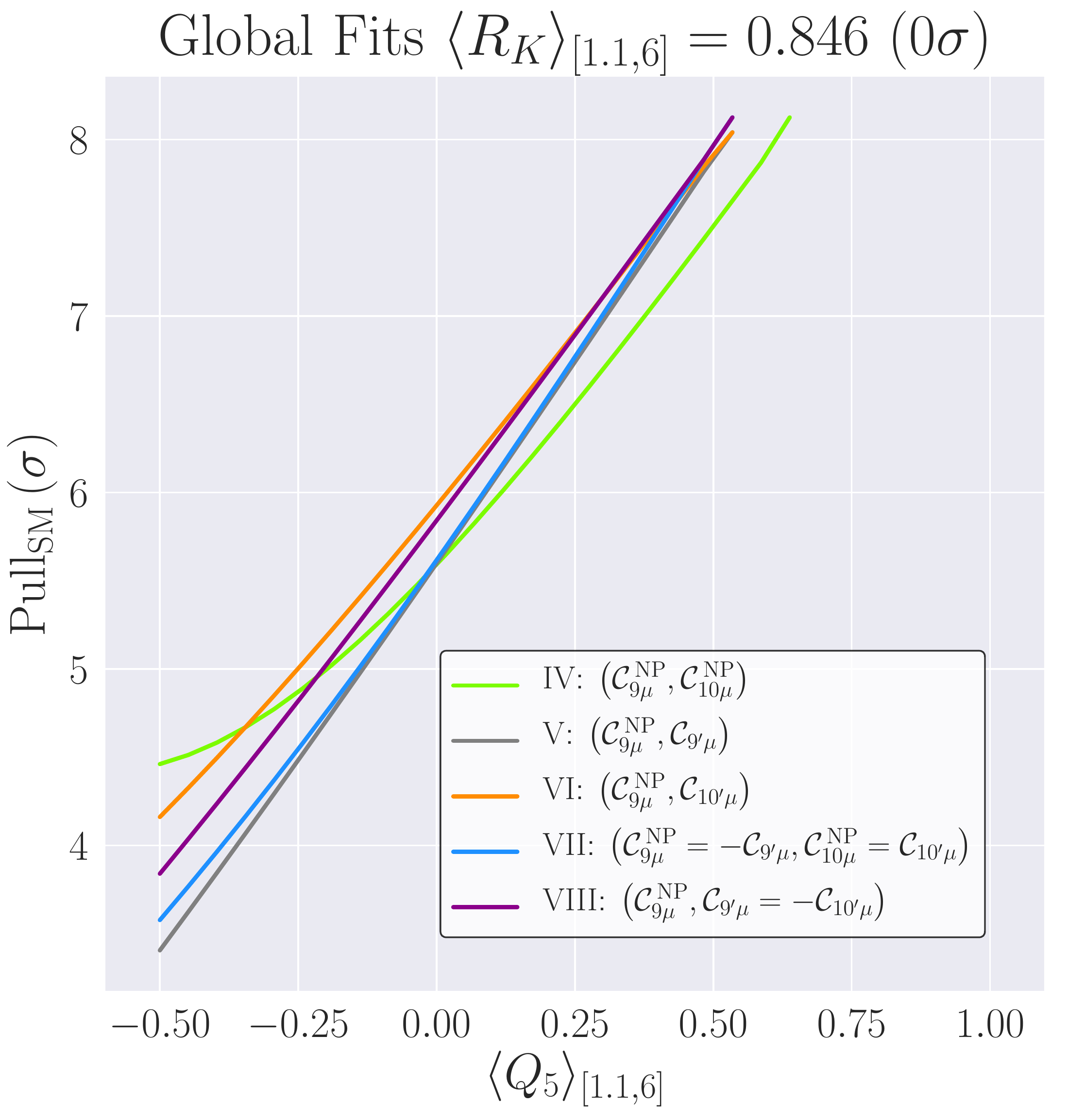}
\end{subfigure}
\begin{subfigure}{.5\textwidth}
  \centering
  \includegraphics[width=0.92\linewidth]{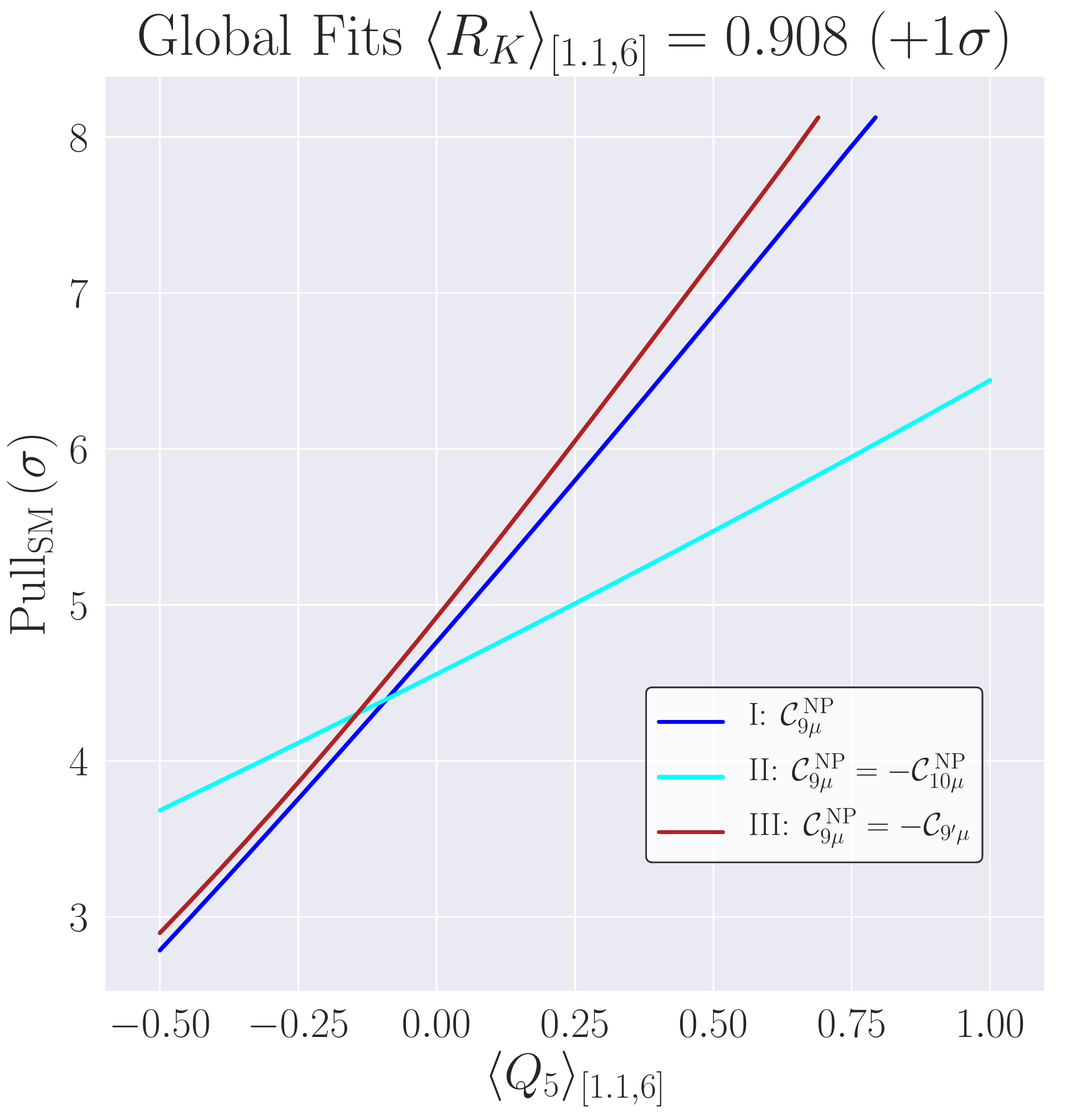}
\end{subfigure}%
\begin{subfigure}{.5\textwidth}
  \centering
  \includegraphics[width=0.92\linewidth]{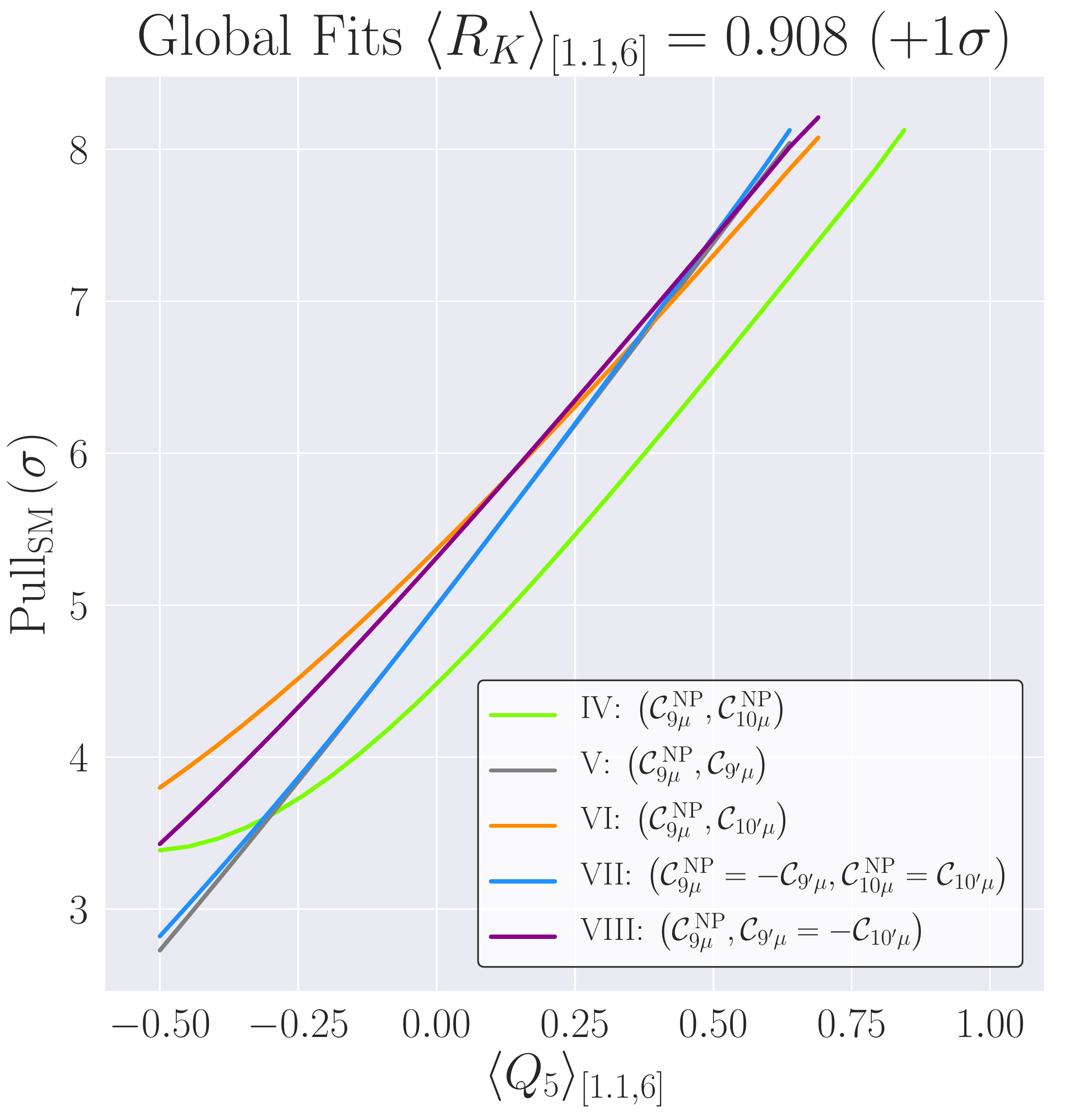}
\end{subfigure}
\caption{Global fit: Impact of $\langle Q_5\rangle_{[1.1,6]}$ on the $\text{Pull}_\text{SM}$ of the NP scenarios under consideration for different values of $\langle R_K\rangle_{[1.1,6]}$.}    
\label{fig:GlobalFitPlotsQ5_1a}
\end{figure}

\begin{figure}[t]
  \centering
\begin{subfigure}{.5\textwidth}
  \centering
  \includegraphics[width=0.92\linewidth]{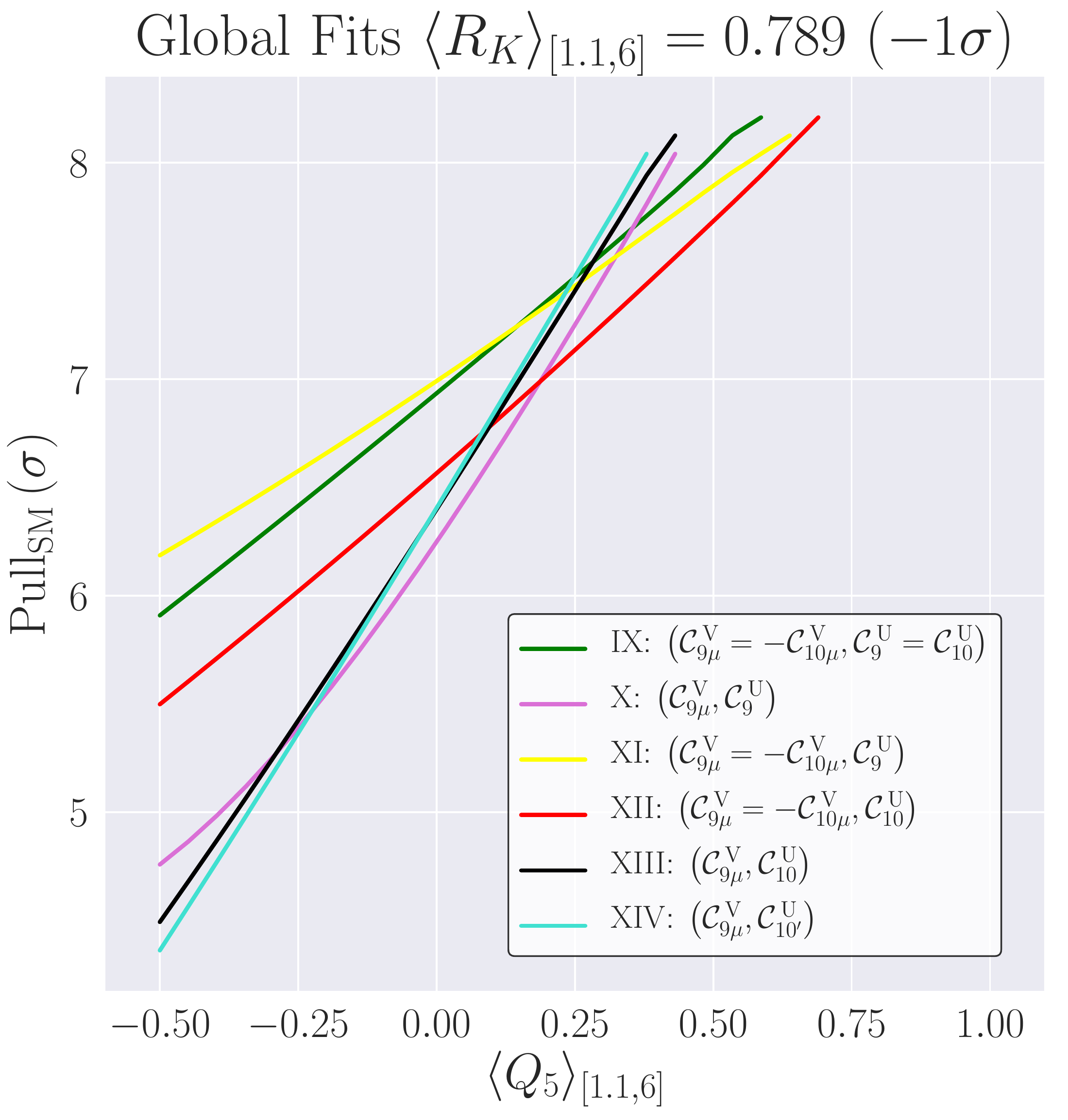}
\end{subfigure}
\begin{subfigure}{.5\textwidth}
  \centering
  \includegraphics[width=0.92\linewidth]{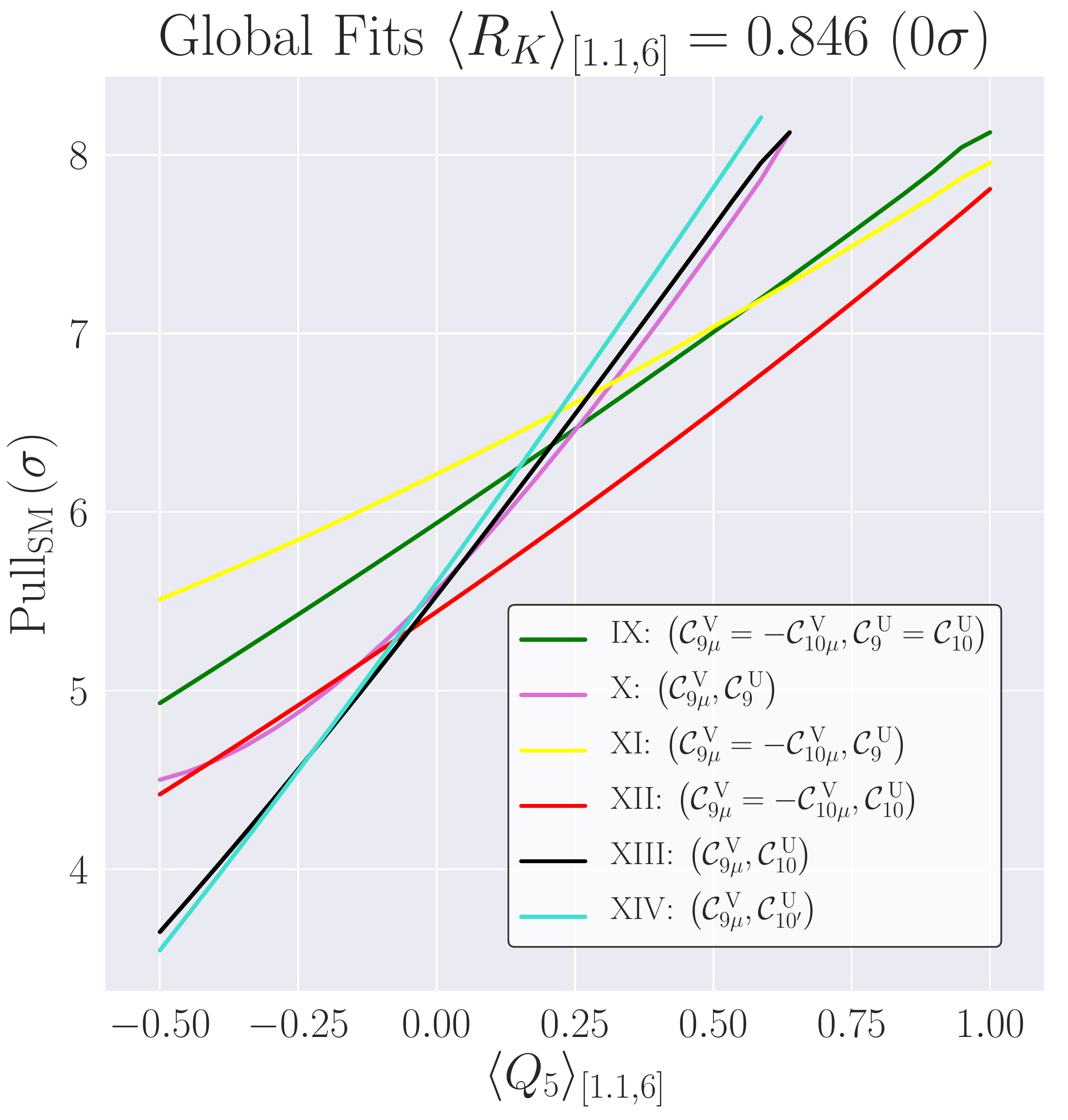}
\end{subfigure}
\begin{subfigure}{.5\textwidth}
\centering
\includegraphics[width=0.92\linewidth]{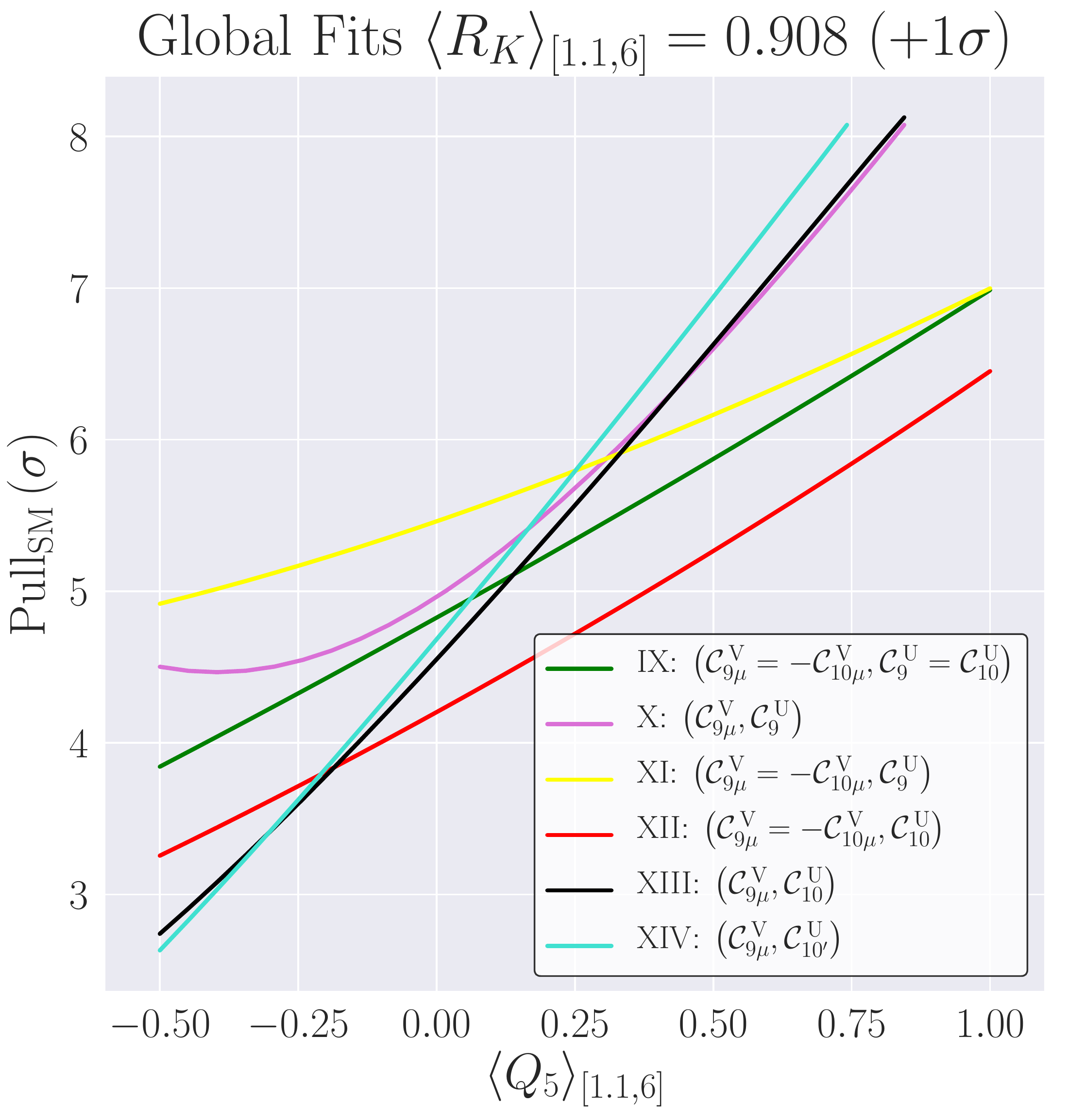}
\end{subfigure}%
\caption{Global fit: Impact of $\langle Q_5\rangle_{[1.1,6]}$ on the $\text{Pull}_\text{SM}$ of the NP scenarios under consideration for different values of $\langle R_K\rangle_{[1.1,6]}$.}    
\label{fig:GlobalFitPlotsQ5_1b}
\end{figure}
\clearpage

We then study the combined influence of $\langle R_K\rangle_{[1.1,6]}$ and $\langle Q_5\rangle_{[1.1,6]}$. The value of $\langle Q_5\rangle_{[1.1,6]}$ is varied as explained above and we repeat the analysis for three different values of $\langle R_K\rangle_{[1.1,6]}$: its current experimental value and the ends of its $1\sigma$ range. Figs.~\ref{fig:GlobalFitPlotsQ5_1a} and \ref{fig:GlobalFitPlotsQ5_1b} show how the $\text{Pull}_\text{SM}$ varies with different experimental values of $\langle Q_5\rangle_{[1.1,6]}$ and $\langle R_K\rangle_{[1.1,6]}$. In App.~\ref{sec:appendixbfp}
 Figs.~\ref{fig:GlobalFitPlotsQ5_2a}, \ref{fig:GlobalFitPlotsQ5_2b} and \ref{fig:GlobalFitPlotsQ5_2c}
show how b.f.p.s are affected, with a rather large variety of behaviours.

We observe that most of the hypotheses see their Pull$_{\rm SM}$ increase for larger values  of $\langle Q_5\rangle_{[1.1,6]}$. One can separate the discussion according to the value of $\langle R_K\rangle_{[1.1,6]}$

\begin{itemize}
\item[$\blacktriangleright$] $\langle R_K\rangle_{[1.1,6]} \simeq 0.8$: The pulls are larger than in the current case. The rather low value of $R_K$ disfavours in general scenarios with right-handed currents with respect to scenarios involving only SM vector operators. A large value of $\langle Q_5\rangle_{[1.1,6]}$ (close to 1) favours $\C{9\mu}^{\rm NP}$, whereas a low value supports NP in both $\C{9\mu}^{\rm NP}$ and $\C{10\mu}^{\rm NP}$ either from  LFUV NP only (Hyps. II, IV) or from a combination of LFU and LFUV NP (Hyp. XI).

\item[$\blacktriangleright$] $\langle R_K\rangle_{[1.1,6]} \simeq 0.85$: The new determination of $\langle R_K\rangle_{[1.1,6]}$ is then nominally close to its current experimental value but with smaller errors. Therefore, the values for the b.f.p.s are numerically similar to the b.f.p.s reported in Ref.~\cite{Capdevila:2017bsm,Alguero:2019ptt}. On one hand, low and large values of $\langle Q_5\rangle_{[1.1,6]}$ provide both a rather clear separation between Hyps. II, IX, X, XI  and Hyps. I, III, XIII, XIV. On the other hand, it does not help to separate a set of hypotheses with LFUV NP only (IV to VIII).

\item[$\blacktriangleright$] $\langle R_K\rangle_{[1.1,6]} \simeq 0.9$: The pulls are lower with respect to their present values. Large values of  $\langle Q_5\rangle_{[1.1,6]}$ allow one to disfavour Hyp II, and low and large values of $\langle Q_5\rangle_{[1.1,6]}$ provide still both a rather clear separation among hypotheses combining LFU and LFUV NP (Hyps. IX, X, XI on one hand and Hyps. XIII, XIV on the other). However, it does not help to separate a set of hypotheses with LFUV NP only (IV to VIII).
\end{itemize}

In summary, if the update of $\langle R_K\rangle_{[1.1,6]}$ is larger than its current value, it can help distinguishing among various NP hypotheses, with a preference for hypotheses involving right-handed currents in $\C{9'\mu}^{\rm V}$. A value of $\langle R_K\rangle_{[1.1,6]}$ similar or smaller than the current value  would clearly disfavour the hypothesis  $\C{9\mu}^\text{V}=-\C{9'\mu}^{\rm V}$, but many other NP hypotheses (with LFUV only or with a combination of LFU and LFUV) cannot be separated.
The observable $\langle Q_5\rangle_{[1.1,6]}$ is an excellent candidate to separate among some of these possibilities. Depending on the situation, low and/or large values of this observables provide a good separation in terms of pulls.

\begin{figure}[h]
\centering
\begin{subfigure}{.5\textwidth}
  \centering
  \includegraphics[width=0.92\linewidth]{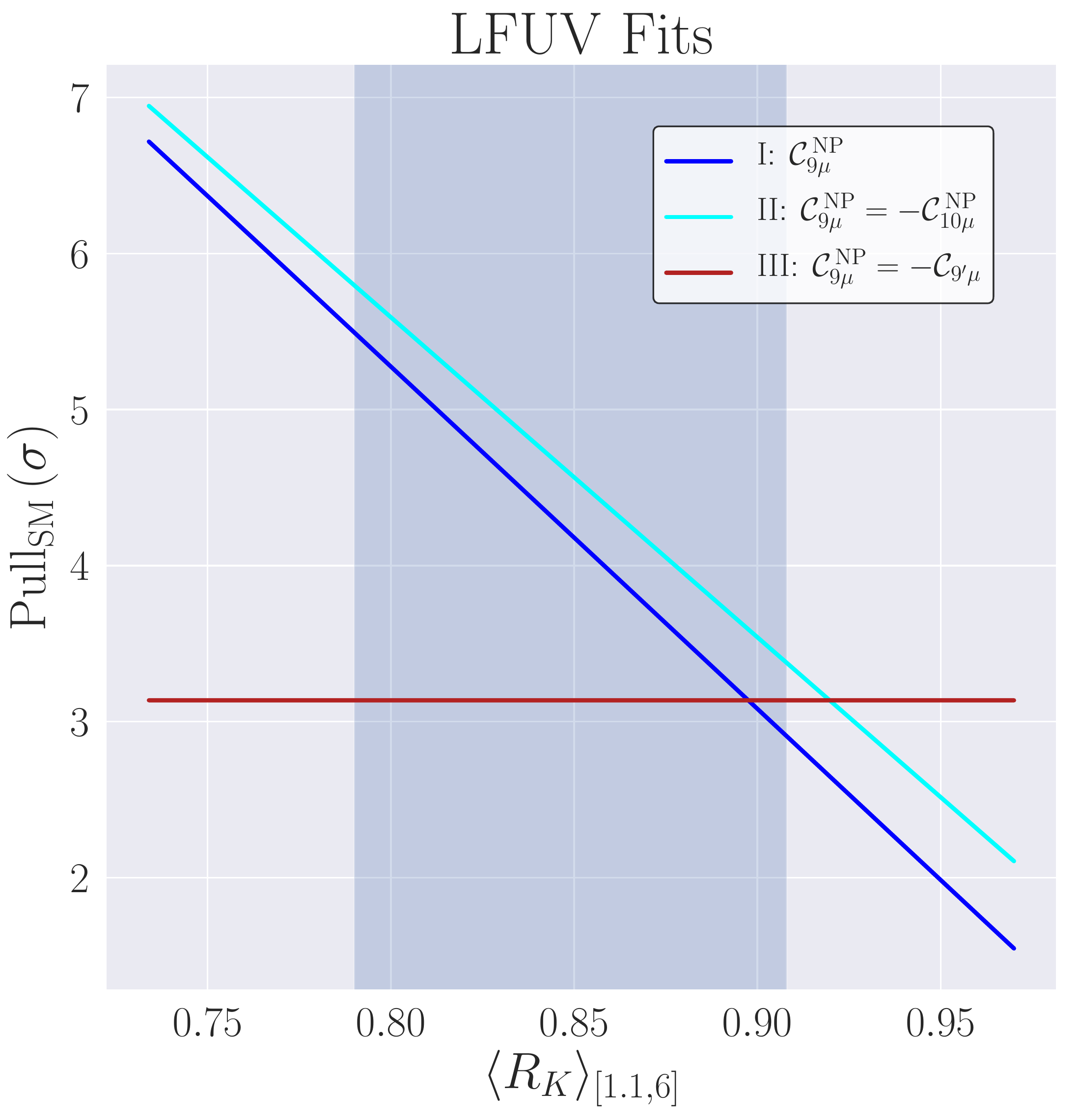}
\end{subfigure}
\begin{subfigure}{.5\textwidth}
  \centering
  \includegraphics[width=0.92\linewidth]{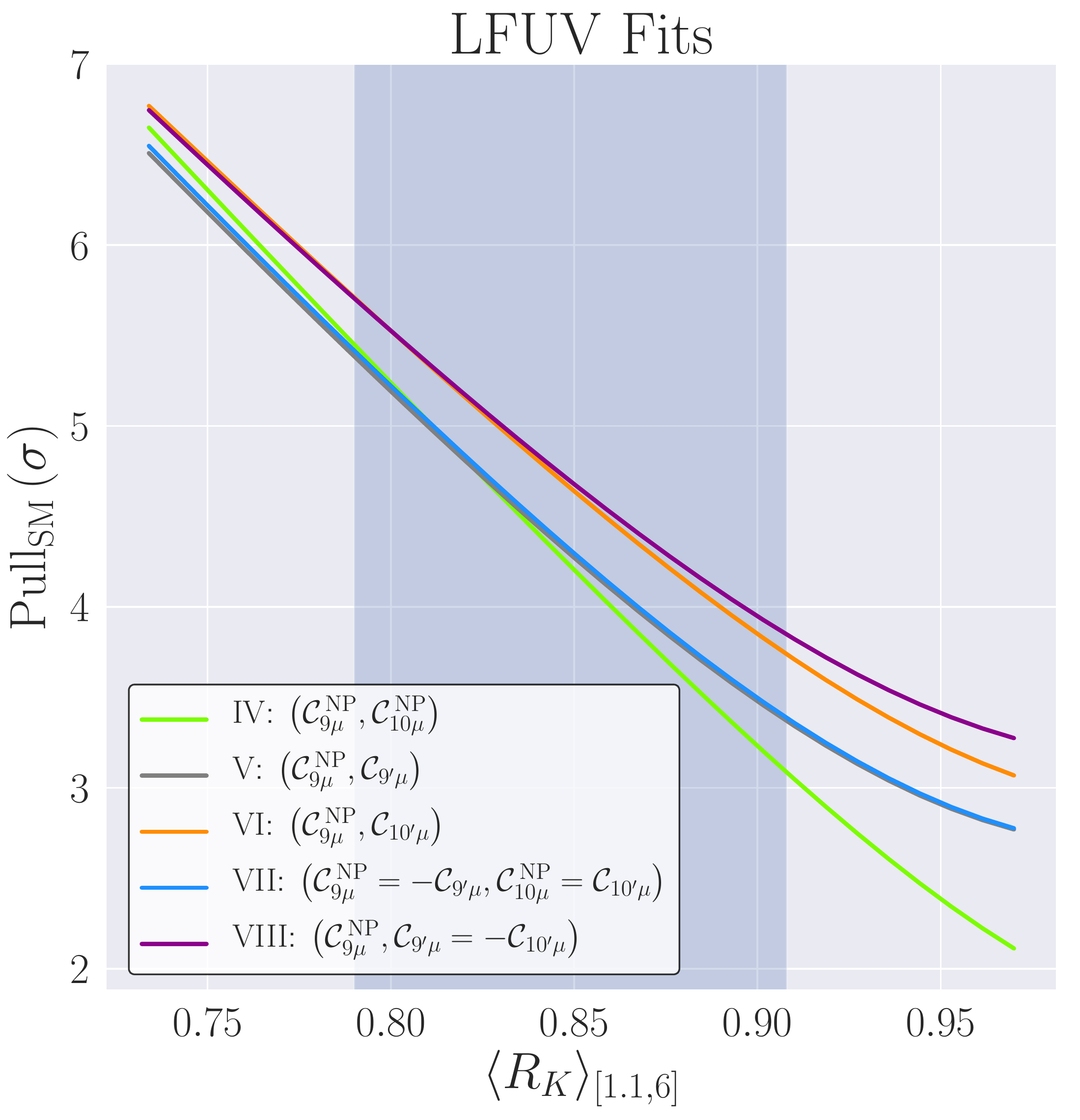}
\end{subfigure}
\begin{subfigure}{.5\textwidth}
\centering
\includegraphics[width=0.92\textwidth]{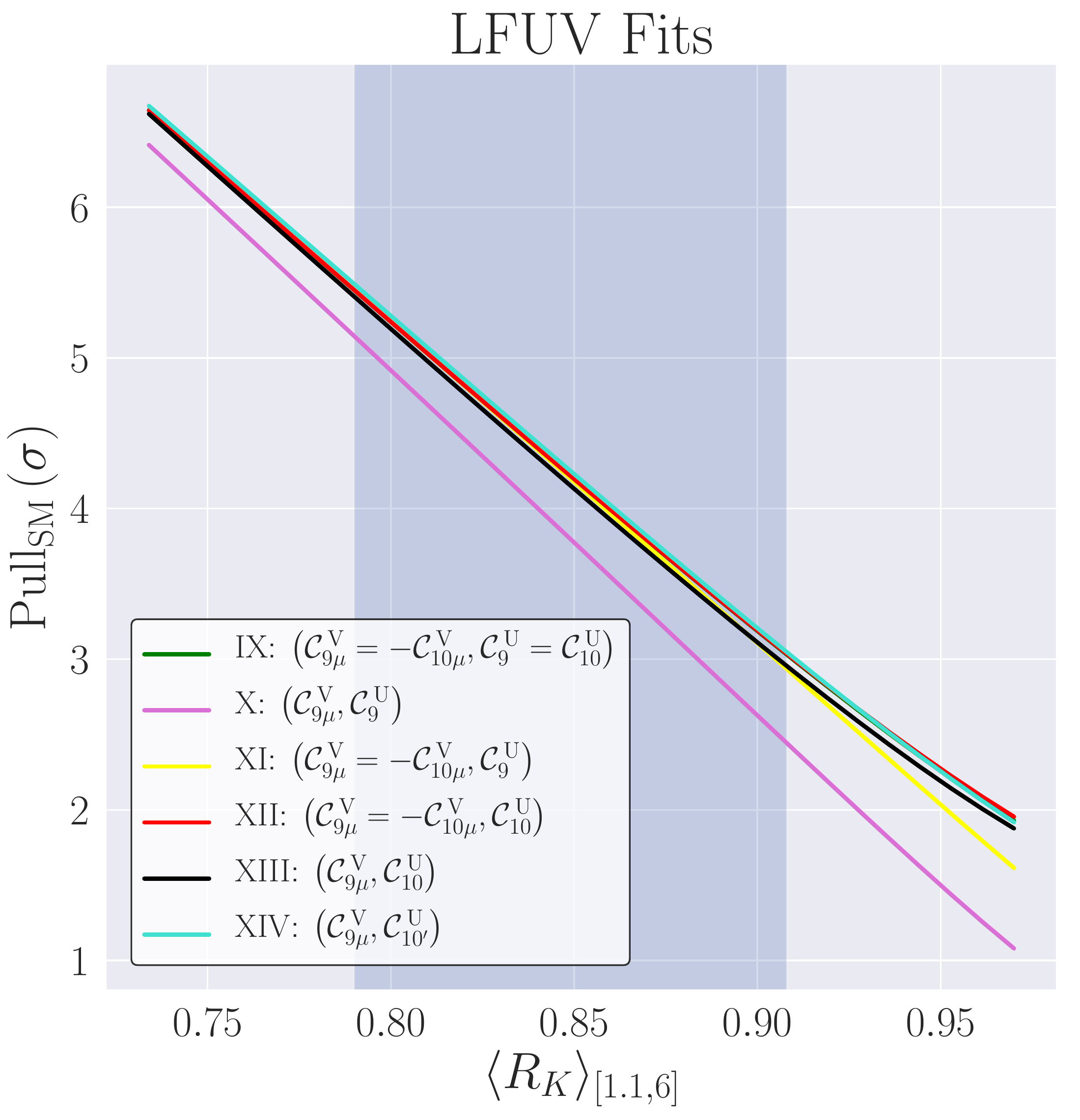}
\end{subfigure}

\caption{LFUV fit: Impact of the central value of $\langle R_K\rangle_{[1.1,6]}$ on the $\text{Pull}_\text{SM}$ of the NP scenarios under consideration.}       
  \label{fig:LFUVFitPlots_a}
\end{figure}

\clearpage

\subsection{LFUV fits}

It is also interesting to address the impact of the observables $\langle R_K\rangle_{[1.1,6]}$ and $\langle Q_5\rangle_{[1.1,6]}$ on the LFUV fits,
which currently include all the $R_K$ and $R_{K^*}$ measurements, the measurements of $Q_i$ $(i=4,5)$ by the Belle collaboration,
all the $b\to s\gamma$ observables available, as well as $\mathcal{B}(B\to X_s\mu^+\mu^-)$ and $\mathcal{B}(B_s\to \mu^+\mu^-)$~\cite{Iwasaki:2005sy,Lees:2013nxa,Aaij:2017vad}. We follow the same guidelines as for the global fits.

\begin{figure}[h]
\begin{subfigure}{.5\textwidth}
  \centering
  \includegraphics[width=0.92\linewidth]{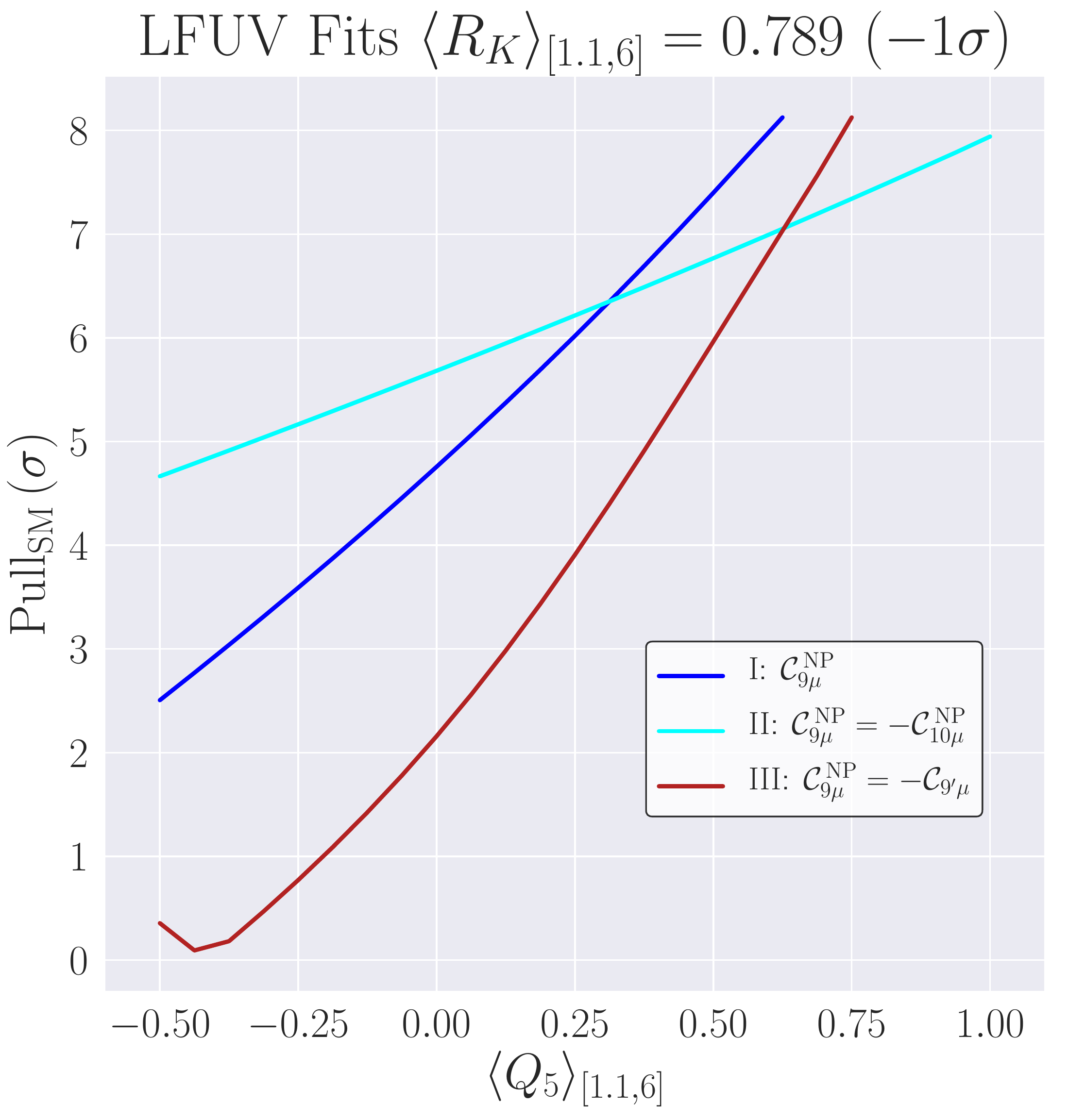}
\end{subfigure}%
\begin{subfigure}{.5\textwidth}
  \centering
  \includegraphics[width=0.92\linewidth]{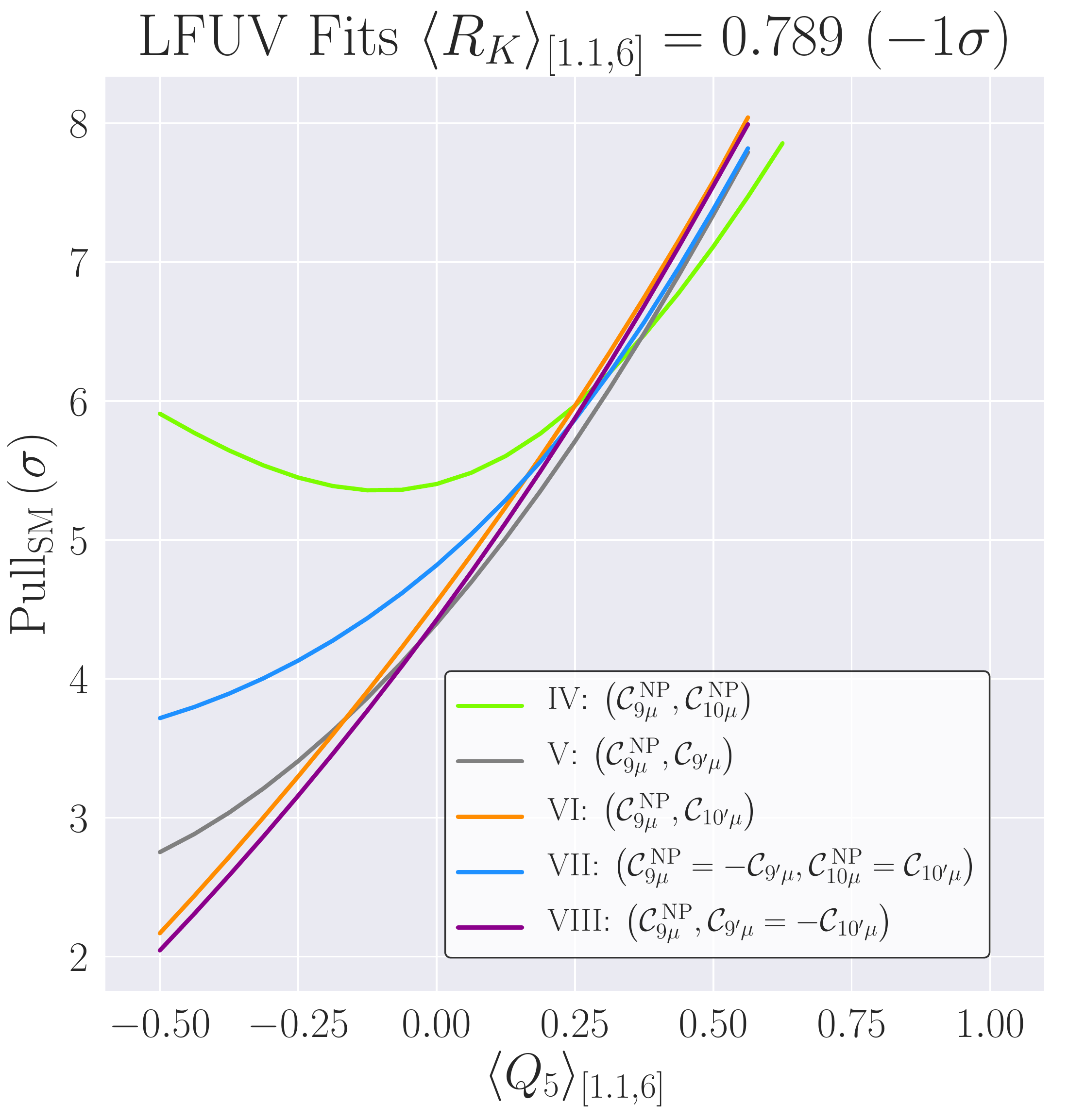}
\end{subfigure}
\begin{subfigure}{.5\textwidth}
  \centering
  \includegraphics[width=0.92\linewidth]{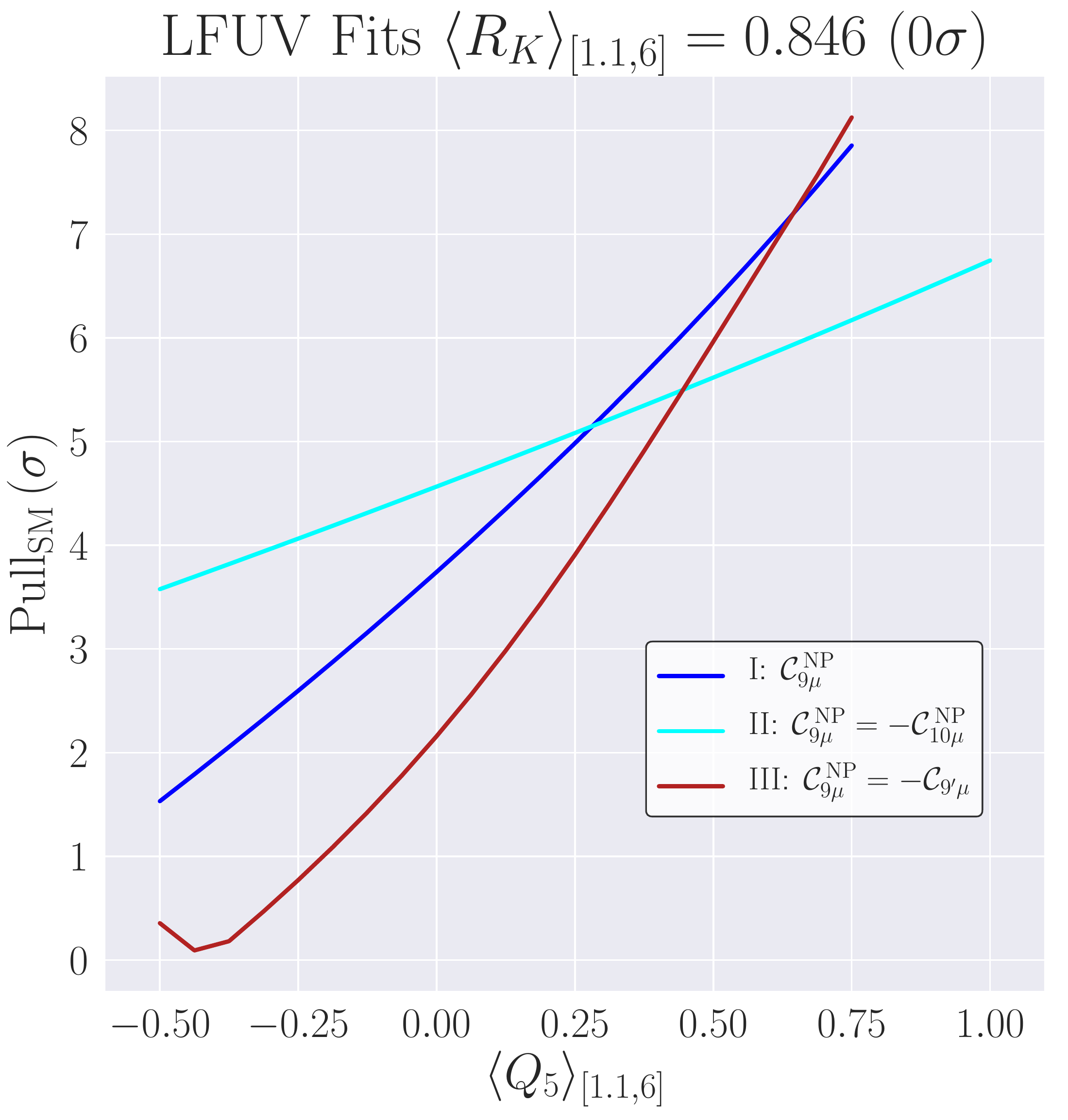}
\end{subfigure}%
\begin{subfigure}{.5\textwidth}
  \centering
  \includegraphics[width=0.92\linewidth]{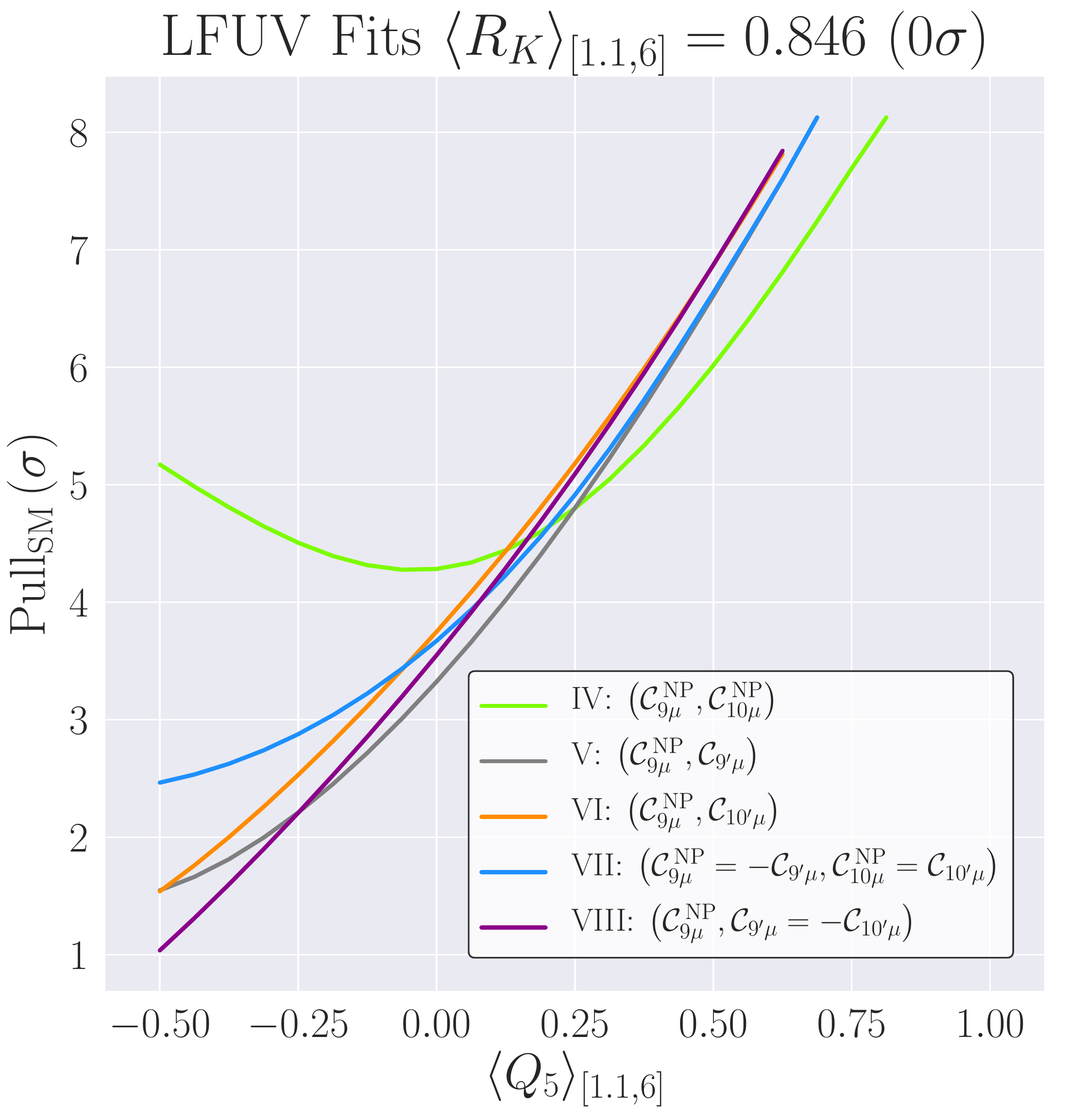}
\end{subfigure}
\begin{subfigure}{.5\textwidth}
  \centering
  \includegraphics[width=0.92\linewidth]{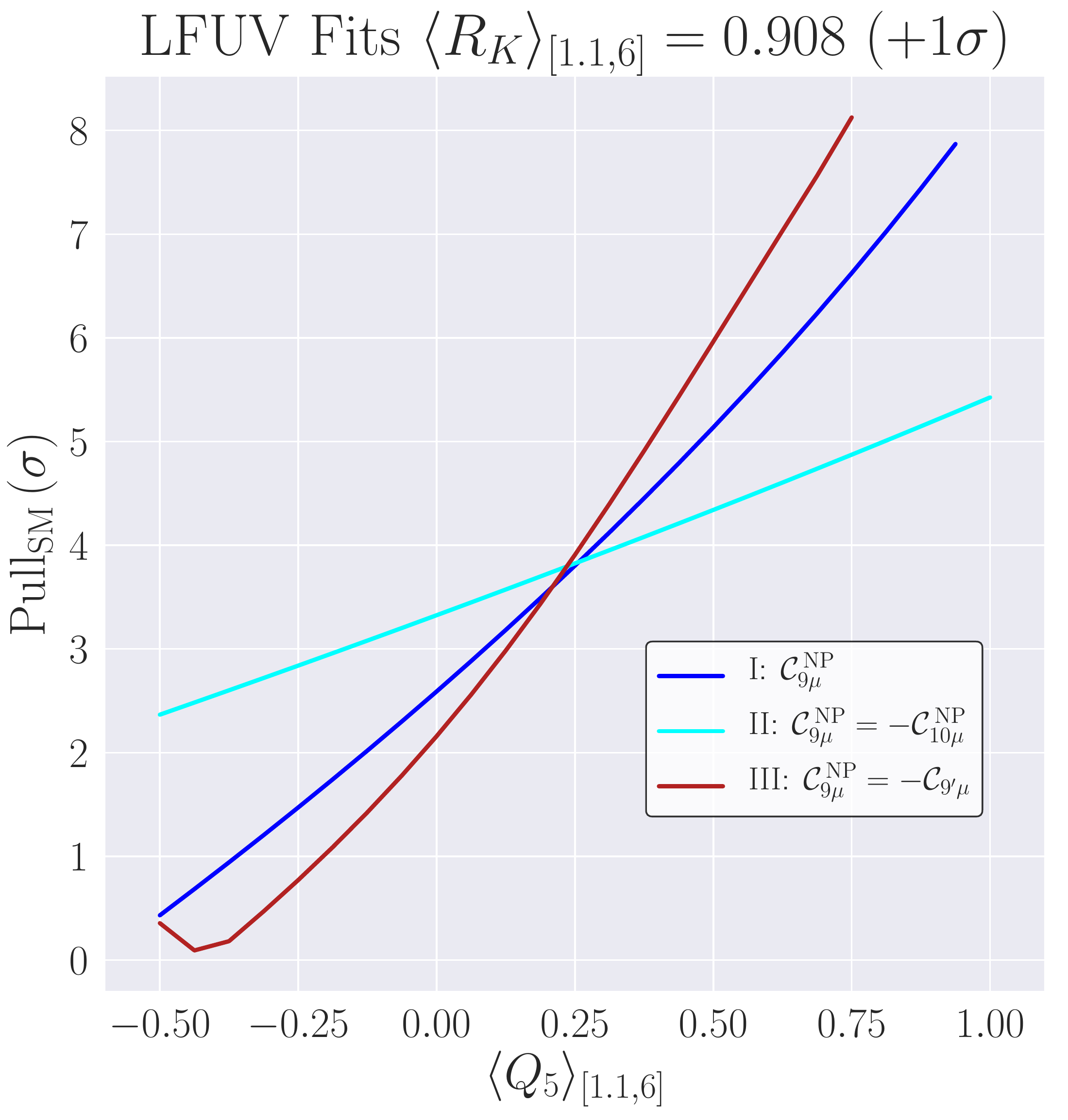}
\end{subfigure}%
\begin{subfigure}{.5\textwidth}
  \centering
  \includegraphics[width=0.92\linewidth]{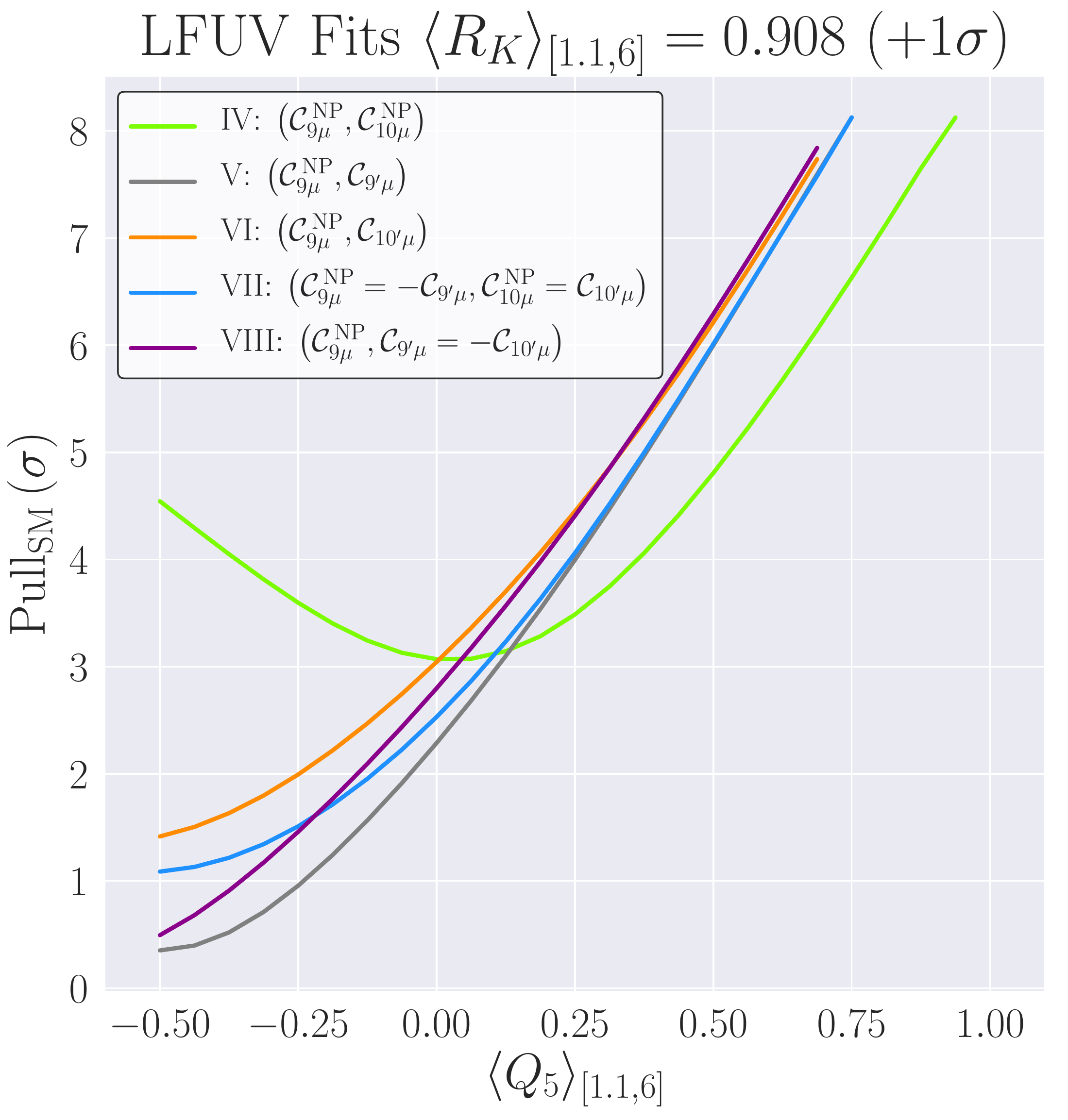}
\end{subfigure}
\caption{LFUV fit: Impact of $\langle Q_5\rangle_{[1.1,6]}$ for different values of $\langle R_K\rangle_{[1.1,6]}$ on the $\text{Pull}_\text{SM}$ of the NP hypotheses under consideration.}         
\label{fig:LFUVFitPlots1}
\end{figure}

\begin{figure}[h]
\centering
\begin{subfigure}{.5\textwidth}
  \centering
  \includegraphics[width=0.92\linewidth]{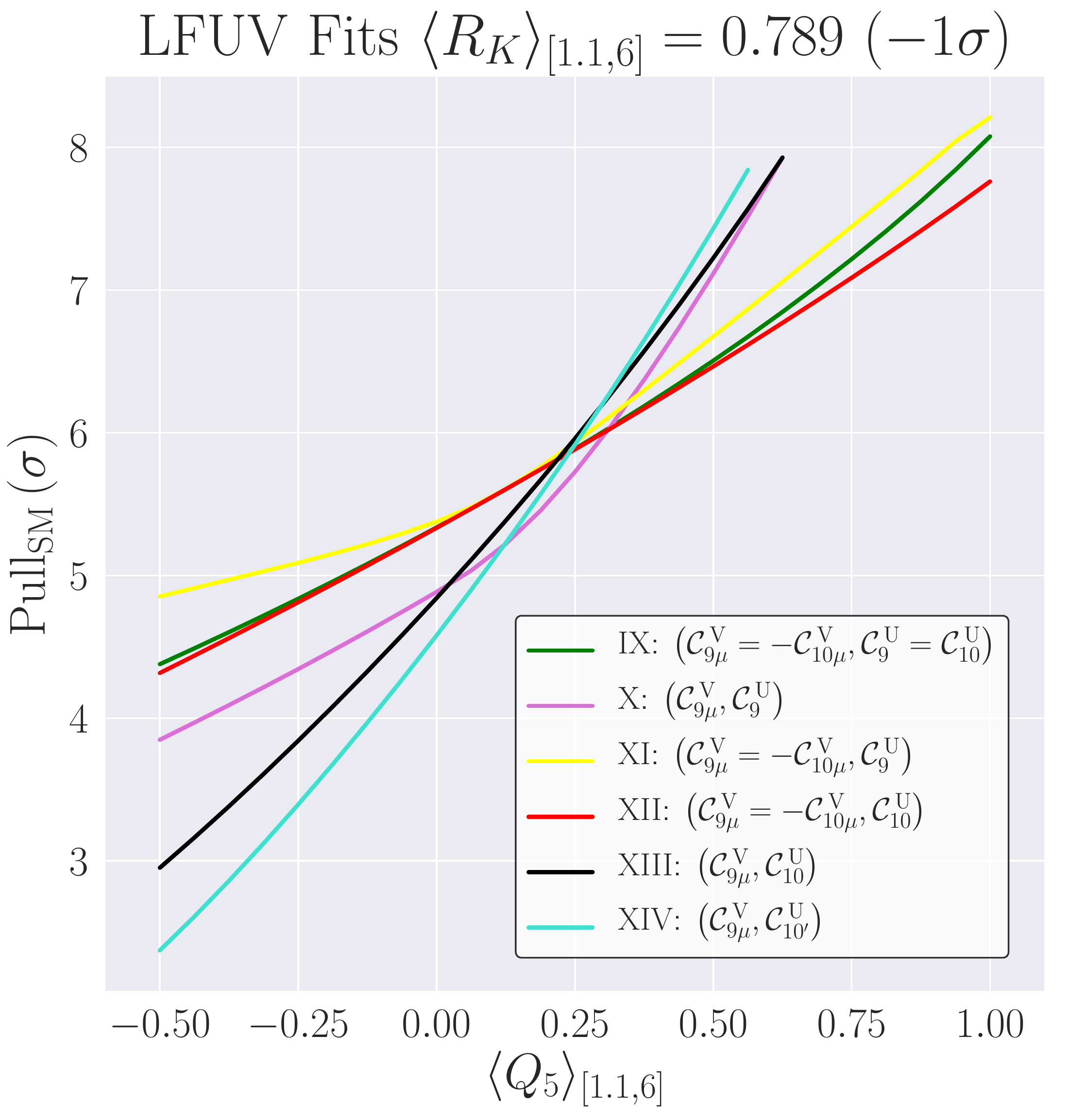}
\end{subfigure}%

\begin{subfigure}{.5\textwidth}
  \centering
  \includegraphics[width=0.92\linewidth]{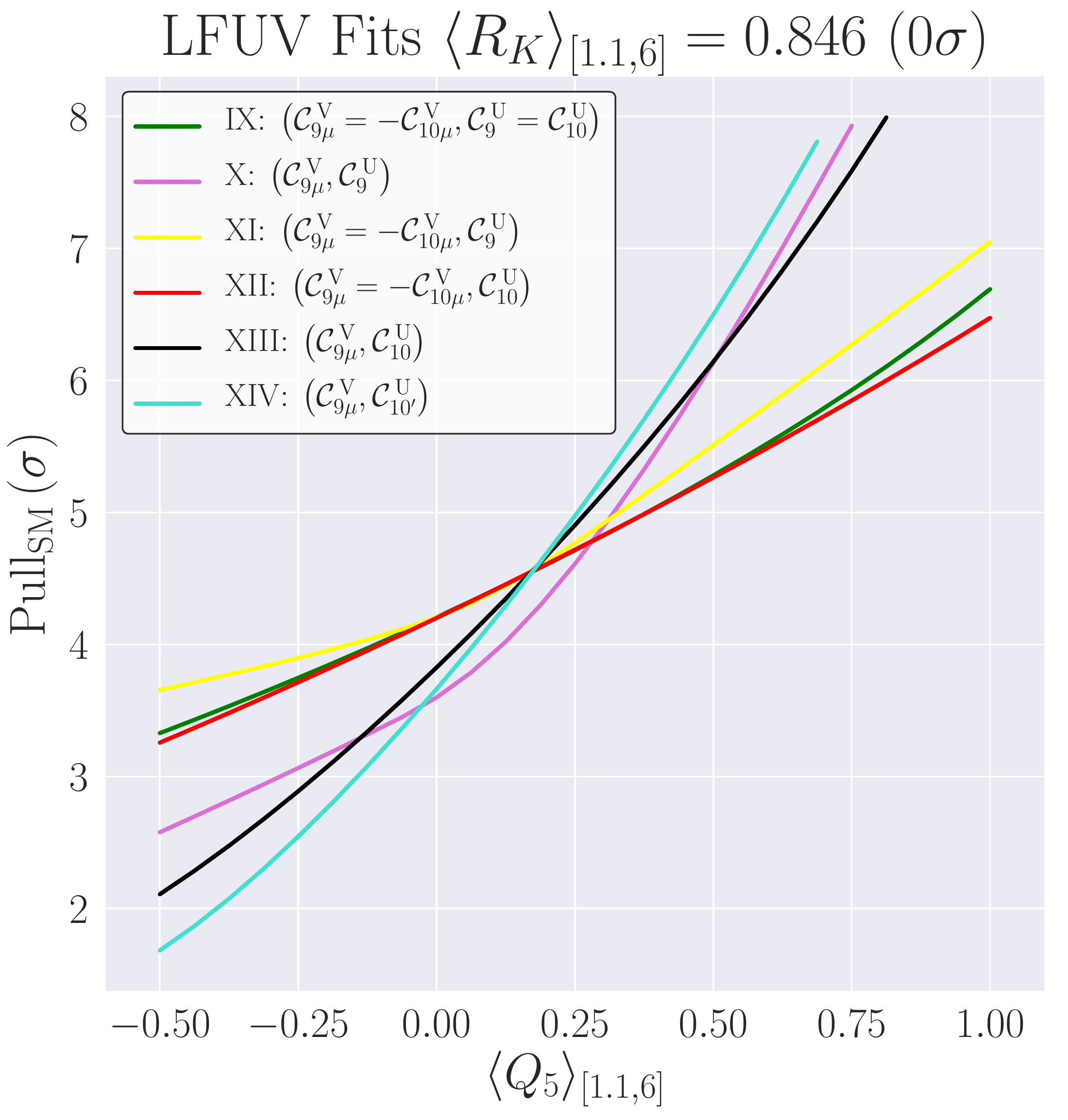}
\end{subfigure}

\begin{subfigure}{.5\textwidth}
\centering
\includegraphics[width=0.92\textwidth]{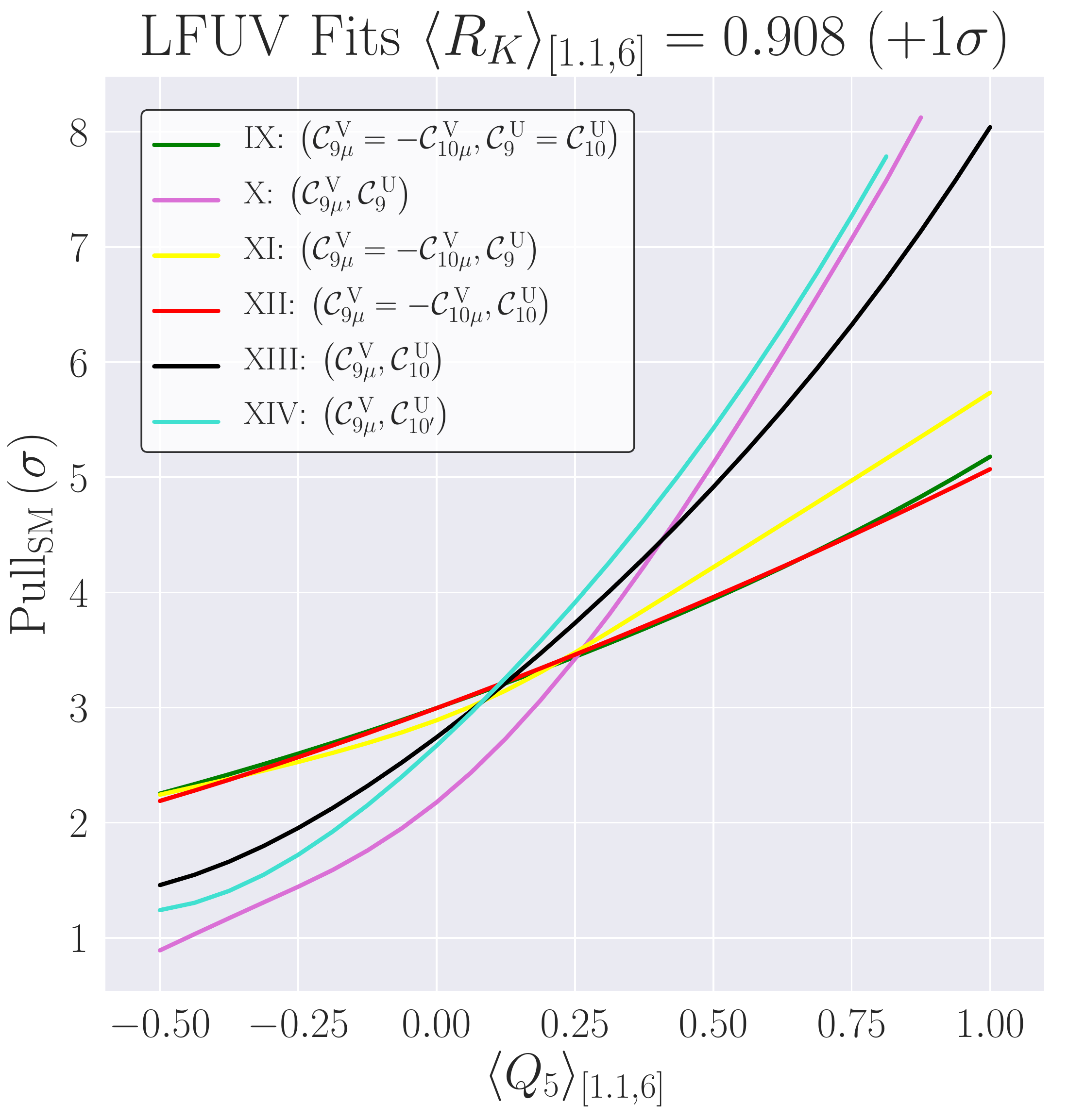}
\end{subfigure}
\caption{LFUV fit: Impact of $\langle Q_5\rangle_{[1.1,6]}$ for different values of $\langle R_K\rangle_{[1.1,6]}$ on the $\text{Pull}_\text{SM}$ for the NP hypotheses under consideration.}         
\label{fig:LFUVFitPlots2}
\end{figure}

We show in Figs.~\ref{fig:LFUVFitPlots_a} the impact of an update of $\langle R_K\rangle_{[1.1,6]}$ on the pull with respect to the Standard Model, whereas the
role played by $\langle Q_5\rangle_{[1.1,6]}$ is indicated by Figs.~\ref{fig:LFUVFitPlots1} and \ref{fig:LFUVFitPlots2}.
One can also plot the variation of the best-fit point as a function of $\langle R_K\rangle_{[1.1,6]}$ and
$\langle Q_5\rangle_{[1.1,6]}$, leading to plots similar to the case of the global fit, but the information gained this way is not very illuminating and we refrain from showing them.

Most of the features observed in the global fit are also observed in the LFUV fit, although with lower pulls. For different values of $\langle R_K\rangle_{[1.1,6]}$, the separation among hypotheses can be improved if one measures either low or large values of $\langle Q_5\rangle_{[1.1,6]}$. One can in particular notice that the separation between the LFUV NP hypotheses (IV to VIII) seems easier for low values of $\langle Q_5\rangle_{[1.1,6]}$ compared to the fit ``All''. This means that the observables in the LFUV fit are more sensitive to details of this scenario, but this gets compensated by other observables in the fit ``All'' so that the sensitivity is reduced in the more complete fit.

LFUV fits lead us to draw similar conclusions to the ones extracted from the global fits. However they exhibit a stronger clustering of the pulls, especially if only $\langle R_K\rangle_{[1.1,6]}$ is used to discriminate them.
Moreover, if we take a given hypothesis with LFUV-NP only, and consider hypotheses obtained by adding further LFU-NP contributions, we obtain very similar pulls.
Therefore,  the only way to distinguish among different LFU-NP hypotheses consists in performing the global fits and having access to both muonic and electronic branching ratios.
The addition of $\langle Q_5\rangle_{[1.1,6]}$  
allows for strategies that enable the discrimination of various scenarios in a similar way to the case of the global fits.

\clearpage

\section{Pulls of individual observables}\label{sec:pulls}

Discerning among different NP hypotheses using the significance of the ${\rm Pull}_{\rm SM}$ proved difficult using the current
set of data. 
In Sec.~\ref{sec:disentangling} we have discussed how new data could possibly improve the situation and allow us to disentangle the different scenarios. We discuss here a complementary approach based on
analysing the current picture for each hypothesis using the correlated pull of individual observables.

We consider the \textit{individual pull of an observable}, which can be written for the $i^{\rm th}$ observable as:
\begin{equation}
{\rm pull}_i^{\rm obs} = \sqrt{ \chi^2_{\rm min} - \chi^2_{{\rm \, min \, w/o \, obs \, }i} }\, \, , \label{eq:pulls}
\end{equation}
\noindent
where $\chi^2_{\rm min}$ and $\chi^2_{{\rm \, min \, w/o \, obs \, }i}$ are the minimal values of the $\chi^2$ with and without the $i^{\rm th}$ observable
\begin{align}
\nonumber
\chi^2&=\sum_{ij}({\cal O}^{th}-{\cal O}^{exp})_{i}V^{-1}_{ij}({\cal O}^{th}-{\cal O}^{exp})_{j}\, , \\
\chi^2_{{\rm \, w/o \, obs \, }i}&=\sum_{i\neq j}({\cal O}^{th}-{\cal O}^{exp})_{i}V^{-1}_{ij}({\cal O}^{th}-{\cal O}^{exp})_{j}\, ,\label{eq:chi2}
\end{align}
\noindent
where ${\cal O}^{th}$ and ${\cal O}^{exp}$ are the theoretical prediction and the experimental value for the observable respectively, and $V^{-1}_{ij}$ corresponds to the covariance matrix element $\{i,j\}$. 

This definition, already used in Refs.~\cite{Lenz:2010gu,Charles:2011va,Lenz:2012az,Charles:2016qtt}, differs from the definition often adopted e.g. in the context of electroweak precision observables~\cite{:2005ema}: this naive pull is defined as the difference between the experimental value and the theoretical value at the b.f.p., normalised by the uncertainty. As discussed in Ref.~\cite{demortier}, this definition should be refined. The pull of an observable can be considered as the assessment of the impact of the additional external constraint to the fit given by this observable, which must be assessed using a pull involving the central values and the uncertainties with and without the constraint given by the observable. It can be easily shown that the definition of Ref.~\cite{demortier} is equivalent to our definition. Moreover, this definition follows the same approach of pulls as for the comparison of hypotheses and it can be expected to follow a normal law with zero mean and unit width. We stress that it includes correlations, so that it might have a different value from the naive pull in the presence of large correlations among observables (from experimental or theoretical nature).

In the following, we will compare the pulls, as defined in Eq.~(\ref{eq:pulls}), under the SM hypothesis and under various NP hypotheses for all observables. However, we will focus mainly on the observables yielding ${\rm pulls}^{\rm obs}$ larger than 1.5. These ${\rm pulls}^{\rm obs}$ with respect to the SM are expected to be reduced under the various NP hypotheses, but some might remain (while the others disappear), providing interesting insights into the role of the various observables within a given NP hypothesis.
 
\begin{figure}[t]
\begin{center}
\includegraphics[width=\textwidth]{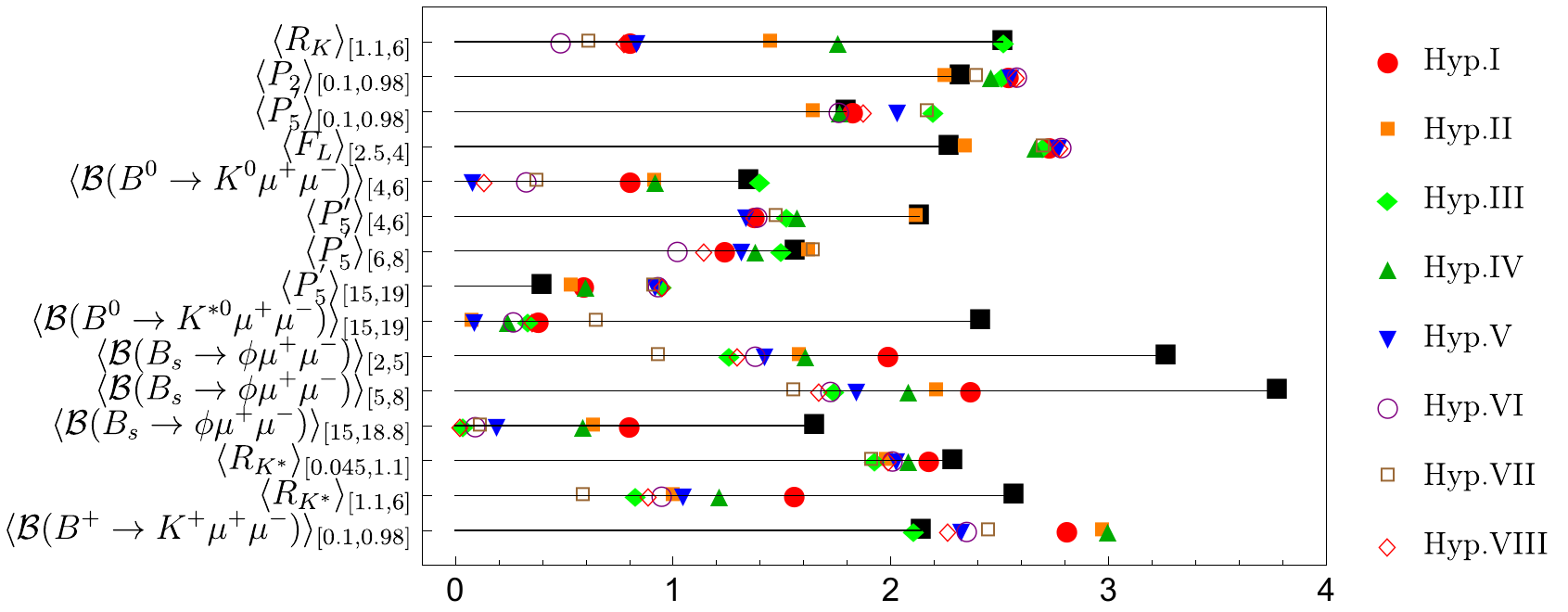}
\caption{${\rm pull}^{\rm obs}$ defined in Eq.~(\ref{eq:pulls}) for each selected observable within different NP hypotheses (from I to VIII) compared to the Standard Model (black squares).}
\label{PlotRes1}
\end{center}
\end{figure}

\begin{figure}[t]
\begin{center}
\includegraphics[width=\textwidth]{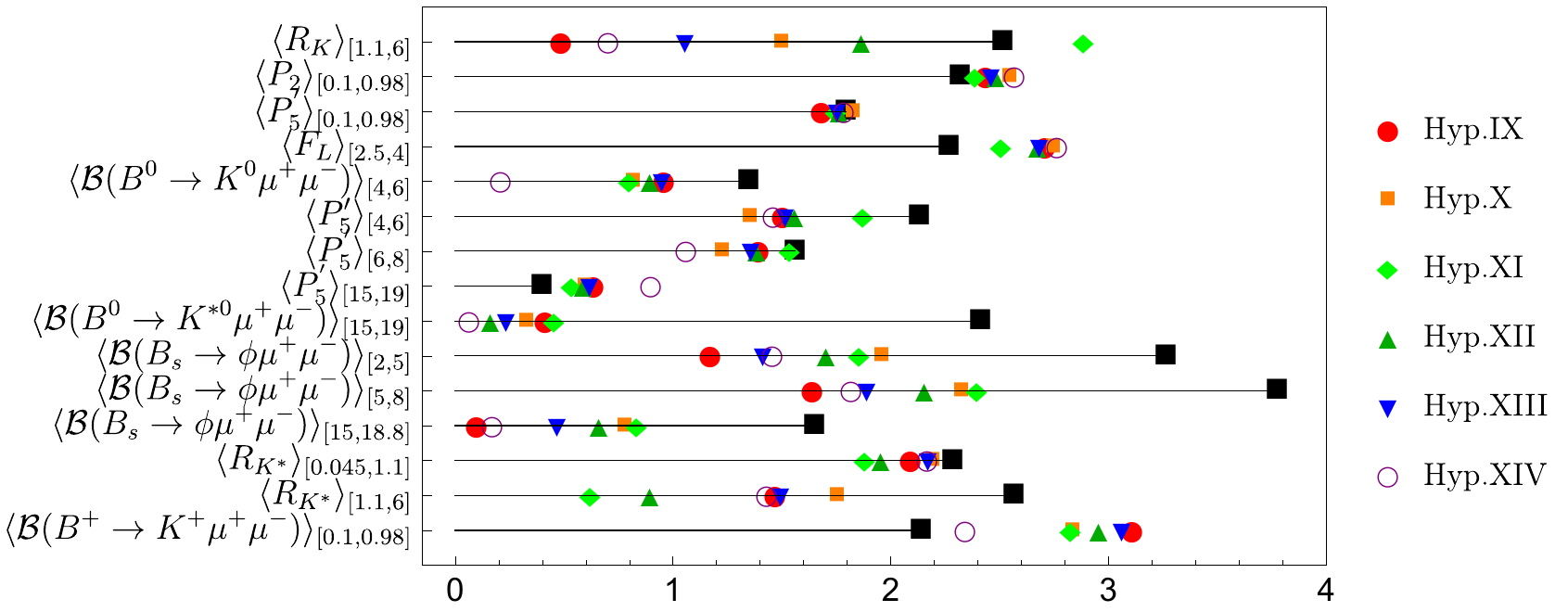}
\caption{${\rm pull}^{\rm obs}$ defined in Eq.~(\ref{eq:pulls}) for each selected observable within different NP hypotheses (from IX to XIV) compared to the Standard Model (black squares).}
\label{PlotRes2}
\end{center}
\end{figure}

The observables under discussion are visualized in Figs.~\ref{PlotRes1} and \ref{PlotRes2}. Black squares represent the ${\rm pull}^{\rm obs}$ of the observable $i$ within the SM computed following Eq.~(\ref{eq:pulls}), while coloured and empty shapes represent the ${\rm pull}^{\rm obs}$ of the same observable under different NP hypotheses.

\subsection{Observables with a large ${\rm pull}_i^{\rm obs}$ within the Standard Model}

Looking at Figs.~\ref{PlotRes1} and \ref{PlotRes2}, the first observable considered is $\langle R_K\rangle_{[1.1,6]}$: it is very sensitive to LFUV NP, it has a large ${\rm pull}^{\rm obs}$ with respect to the SM.
It has also a small ${\rm pull}^{\rm obs}$ in all the NP hypotheses considered except for Hyps.~III and XI. 
Remarkably, it has almost no correlation with any other observable in the fit\footnote{There is little theoretical correlation between  $R_K$ and the rest of the observables as the hadronic form factors, which are the main source of correlated uncertainty among observables, cancel in $R_K$. The experimental correlation between $R_K$ and the $B^{+} \to K^{+}\ell\ell$ branching ratio is not public and therefore assumed to be zero here. We checked the lack of correlation between $\langle R_K\rangle_{[1.1,6]}$ and the rest of the observables at the level of our covariance matrix.} 
\cite{Capdevila:2018jhy} and, as a consequence, the fit must satisfy both $\langle R_K\rangle_{[1.1,6]}$ and the rest of the observables in parallel, leading to a potential tension unless the solution for both sectors of the fit is similar. 

Another observable showing a tension with the SM is $\langle P_2\rangle_{[0.1,0.98]}$. Remarkably, none of the hypotheses considered is able to  reduce the tension of this observable (see Figs.~\ref{PlotRes1} and \ref{PlotRes2}) so it remains poorly explained either within the SM or under the NP hypotheses considered. 
 A similar problem is seen in $\langle P'_5\rangle_{[0.1,0.98]}$ even though for this observable the ${\rm pull}^{\rm obs}$ within the SM is smaller than for  $\langle P_2\rangle_{[0.1,0.98]}$. Including a NP contribution to $\C{7}$ (which affects mainly the first bin) does not improve these two observables as they require contributions of opposite signs. Therefore, the small tensions observed in the first bin for these two observables 
 will require more data to be understood.
  
The polarization fraction $\langle F_L\rangle_{[2.5,4]}$ follows a similar trend as $\langle P'_5\rangle_{[0.1,0.98]}$. Its current measurement shows little deviation with the SM prediction \cite{Descotes-Genon:2015uva}  in any of its bins due to the large theoretical errors, but its ${\rm pull}^{\rm obs}$ is around $\sim~2$ under the SM hypothesis due to the correlations with the rest of the observables involved in the fit.
Moreover, for all NP hypotheses considered its value for the ${\rm pull}^{\rm obs}$ is larger than in the SM case. 

\begin{figure}[h]
\begin{center}
\includegraphics[width=8cm]{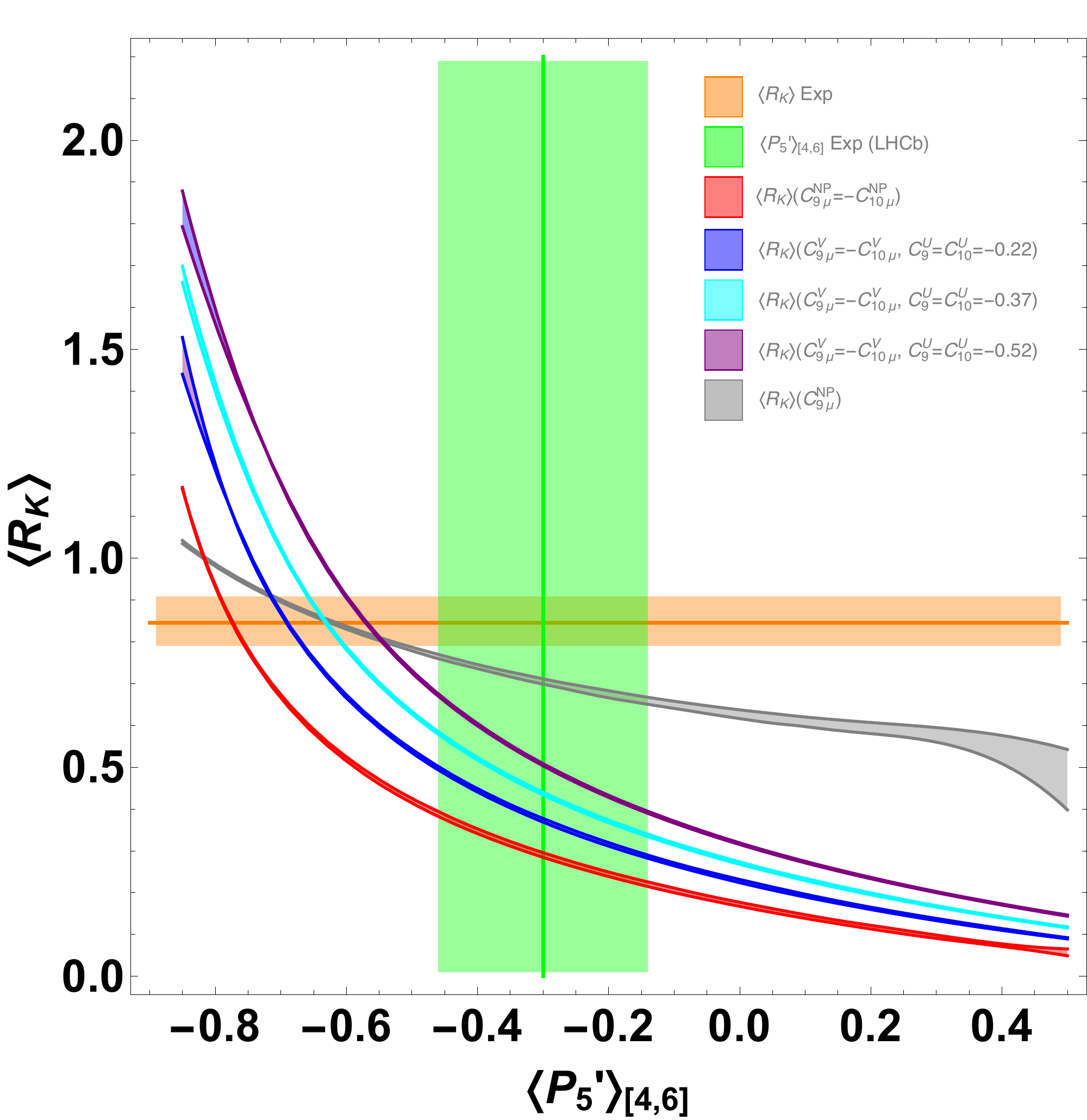}
\caption{$\langle R_K\rangle_{[1.1,6]}$ versus $\langle P_5^\prime \rangle_{[4,6]}$ in three different scenarios: $\C{9\mu}^{\rm V}$ (grey), $\C{9\mu}^{\rm V}=-\C{10\mu}^{\rm V}$ (red), and $\C{9\mu}^{\rm V}=-\C{10\mu}^{\rm V}$, $\C{9}^{\rm U}=\C{10}^{\rm U}$ (three different values of the LFU contributions, blue, light blue and purple).
In each case, the band is obtained by varying the LFU contribution within 1$\sigma$. The current experimental values from LHCb are also indicated (green vertical and orange horizontal bands).}
\label{figrkp5}
\end{center}
\end{figure}
 

In the case of the observables $\langle P'_5\rangle_{[4,6]}$ and $\langle P'_5\rangle_{[6,8]}$ the tension between the SM prediction and the experimental value is somewhat reduced if one introduces NP contributions. The less efficient hypothesis is Hyp.~II, $\C{9\mu}^{\rm NP}=-\C{10\mu}^{\rm NP}$, and one of the best possibilities is Hyp.~I, i.e. a single contribution $\C{9\mu}^{\rm NP}$.  As shown in Fig. \ref{figrkp5}, the latter hypothesis (grey band) better accommodates at the same time both the deviations in $\langle P'_5\rangle_{[4,6]}$ and $\langle R_K\rangle_{[1.1,6]}$ w.r.t. the solution with $\C{9\mu}^{\rm NP}=-\C{10\mu}^{\rm NP}$, although the updated experimental value of $\langle R_K\rangle_{[1.1,6]}$ makes the compatibility only partial, whereas the former hypothesis (red band) is unable to obtain the same agreement. If one favours this scenario $\C{9\mu}^{\rm NP}=-\C{10\mu}^{\rm NP}$, a compatibility level similar to that of $\C{9\mu}^{\rm NP}$ can be obtained by adding a LFU contribution $\C{9}^{\rm U}=\C{10}^{\rm U}$ ~\cite{Alguero:2018nvb} (corresponding to Hyp.~IX), since $\langle R_K\rangle_{[1.1,6]} = 1+0.49 \C{9\mu}^{\rm V}$ while $\langle P'_5\rangle_{[4,6]} = -0.81 - 0.27 \C{9}^{\rm U} - 0.14 \C{9\mu}^{\rm V} +0.07 (\C{9}^{\rm U})^2 +0.05 (\C{9\mu}^{\rm V})^2+0.11\C{9}^{\rm U}\C{9\mu}^{\rm V}$~\cite{Alguero:2018nvb}. This implies that a NP scenario with  $\C{9\mu}^{\rm V}=-\C{10\mu}^{\rm V}$ would then need a LFU universal contribution 
$\C{9}^{\rm U}=\C{10}^{\rm U}$ to be viable. This is illustrated in Fig.~\ref{figrkp5}
where Hyp.~IX is displayed varying the LFU contribution within its $1\sigma$ confidence interval according to the global fit (blue, light blue and purple bands). 

One of the tensions of the fits described in Sec.~\ref{sec:understanding} belongs to the low-recoil region. This corresponds mainly to two observables here: $\langle P'_5\rangle_{[15,19]}$ and $\langle{\cal B} (B^{0} \to K^{*0}\mu^{+}\mu^{-})\rangle_{[15,19]}$. For $\langle P'_5\rangle_{[15,19]}$, including NP contributions leads to an increase  of ${\rm pull}^{\rm obs}$ in all cases, see Figs.~\ref{PlotRes1} and \ref{PlotRes2}. The reduction of this tension would require NP contributions of opposite sign to the ones favoured by the rest of the observables of the fit. On the other hand, the branching ratio $\langle{\cal B} (B^{0} \to K^{*0}\mu^{+}\mu^{-})\rangle_{[15,19]}$ has its ${\rm pull}^{\rm obs}$ significantly reduced under all NP hypotheses. This shows that the correlations between $\langle{\cal B}(B^{0} \to K^{*0}\mu^{+}\mu^{-})\rangle_{[15,19]}$ and the rest of the observables allow for the reduction of tensions under the different favoured NP hypotheses. 

Deviations with respect to the SM are also found for the branching ratio of the $B_s \to \phi\mu^{+}\mu^{-}$ channel. All NP hypotheses yield ${\rm pulls}^{\rm obs}$ 
smaller than their SM counterparts, specially for the low-recoil bin and parallel to its $B \to K^*$ counterpart.
The preferred scenarios considering the three bins of the branching ratio in the $B_s \to \phi\mu^{+}\mu^{-}$ channel from Figs.~\ref{PlotRes1} and \ref{PlotRes2} are those with $\C{9\mu}^{\rm V}=-\C{10\mu}^{\rm V}$. 

Finally, the tension for $\langle R_{K^*}\rangle_{[1.1,6]}$ is also alleviated by introducing a NP contribution. Whereas the large bin $\langle R_{K^*}\rangle_{[1.1,6]}$ gets a significant reduction in its ${\rm pull}^{\rm obs}$ under all NP hypotheses, the first bin $[0.1,0.98]$ is just slightly improved when compared to the SM. This could point towards an inconsistency 
between the first bin and the other bins of $R_{K^*}$ if one takes into account the issues discussed above regarding the first bin of $P_2$ and $P'_5$. We should comment here on
Ref.~\cite{Kumar:2019qbv}, where several NP scenarios are discussed in order to explain the low value of $\langle R_{K^*}\rangle_{[0.1,0.98]}$, including contributions to both $b\to s\mu\mu$ and $b\to see$ channels as well as considering the presence of right-handed currents. It is possible to connect some of the scenarios from \cite{Kumar:2019qbv} with the description of NP in terms of LFU and LFUV contributions proposed in Ref.~\cite{Alguero:2018nvb}. Remarkably, the Wilson coefficients obtained in scenario S10 of \cite{Kumar:2019qbv} (following their notation) are in excellent agreement with the results obtained in \cite{Alguero:2018nvb} for the 4D fit with $\{\C{9\mu}^{\rm V},\C{10\mu}^{\rm V},\C{9}^{\rm U},\C{10}^{\rm U}\}$. 

Hyp.~II works better than Hyp.~I in reducing the ${\rm pull}^{\rm obs}$ of the LFUV observables, as can be easily understood looking at the structure of the observables in terms of their Wilson coefficients provided in Ref.~\cite{Alguero:2018nvb}: $\langle R_{K}\rangle_{[1,6]} = 1 + 0.23 \C{9\mu}^{\rm V} - 0.26 \C{10\mu}^{\rm V}$ while $\langle R_{K^*}\rangle_{[1.1,6]} = 1 + 0.16 \C{9\mu}^{\rm V}-0.29 \C{10\mu}^{\rm V}$. Then, the ratio $({\langle R_{K}\rangle_{[1,6]} -1})/({\langle R_{K^*}\rangle_{[1.1,6]}-1})$  turns out to be $1.4$ in Hyp.~I and 1.1 in Hyp.~II.

\subsection{Observables with no ${\rm pull}_i^{\rm obs}$ with respect to the Standard Model}

The kind of analysis undertaken in the previous subsection can also be conducted starting from observables that currently have very small ${\rm pulls}^{\rm obs}$ with respect to the SM predictions and see how they are affected once a fit under a NP hypothesis is performed. It is to expect their ${\rm pulls}^{\rm obs}$ will grow under NP hypotheses unless they are observables insensitive to the NP hypotheses considered. For this purpose we consider a group of observables with pulls between 0 and 0.2 within the SM.

After a detailed analysis, we found very small ${\rm pulls}^{\rm obs}$
for all of them, except $\langle{\cal B} (B^+\to K^+\mu^{+}\mu^{-})\rangle_{[0.1,0.98]}$,
 under all the NP hypotheses considered, i.e.  they are basically NP insensitive. In other words, these observables could be removed from the fit and the rest of the observables would remain unaffected. It is remarkable that again an observable taken in the first bin, $\langle{\cal B} (B^+\to K^+\mu^{+}\mu^{-})\rangle_{[0.1,0.98]}$, has a worse ${\rm pull}^{\rm obs}$ under all NP hypotheses than under the SM. 
This goes in the same direction as the issues observed
 in the first bin of observables of the $B\to K^{*}\mu^{+}\mu^{-}$ channel such as $P_2$, $P'_5$ and $R_{K^*}$, which points out the need of a better understanding of the low-$q^2$ region.

\section{Sensitivity of observables to different Wilson coefficients}\label{sec:param}

A detailed scrutiny of the dependence in Wilson coefficients of $b\to s\ell\ell$ observables is useful in order to understand tensions in the global fit under various NP hypotheses. The complexity of the calculation prevents such systematic study, but we will rely on an approximate parametrisation of the observables in order to identify the sensitivity of interesting observables to different Wilson coefficients/NP hypotheses.

\subsection{Observables and parametrisations}

We use our current programs~\cite{Descotes-Genon:2015uva} to generate predictions according to a grid in the space of Wilson parameters and fit the proposed parametrisations to this set of pseudo-data.  We collect the coefficients in the vector
\begin{equation}
X=\left\{\C{9\mu}^\text{NP},\C{10\mu}^\text{NP}, \C{9'\mu},\C{10'\mu},\C{9e}^\text{NP},\C{10e}^\text{NP}, \C{9'e},\C{10'e}\right\}
\end{equation}
For simplicity, we consider only real NP contributions to the Wilson coefficients $\C{9,9',10,10'}$ for muon and for electron operators: we assume that we can neglect NP contributions to $\C{7}$, $\C{7'}$ (based on $b\to s\gamma$ observables in good agreement with the SM) as well as to scalar, pseudoscalar and tensor operators (based on the outcome of the general global fits performed on $b\to s\ell\ell$ observables which do not favour large contributions to these Wilson coefficients. We also neglect imaginary contributions for the Wilson coefficients (as there are no signals of NP sources of CP violation from the corresponding $b\to s\ell\ell$ observables).    
We consider only the set of LHCb observables present in the global fit of Ref.~\cite{Descotes-Genon:2015uva} with the corresponding binning in $q^2$, as well as the Belle observables $Q_4$ and $Q_5$. We generate points along a grid for all $\C{i\ell}^\text{NP}$ ($\ell=\mu, e$) in the range $[-1,1]$ for $i=10,9',10'$, but $[-2,0]$ for $i=9$, computing both central values and theoretical uncertainties for all the observables of interest. 
We can divide the set of observables into three different groups: LFD observables, LFUV differences and LFUV ratios, with three different approximate quadratic parametrisations:
\begin{itemize}
\item LFD observables: this category includes branching ratios and (optimized and averaged) angular observables \cite{Matias:2012xw,Altmannshofer:2008dz} governed by a lepton-specific $b \to s\ell\ell$ transition. We use a general quadratic parametrisation in the NP contributions to the Wilson coefficients ($i,j=9,10,9',10'$):
\begin{equation}
\mathcal{O} = \alpha + \sum_i\beta_i \C{i\ell}^\text{NP} + \sum_{i\geq j} \gamma_{ij}\C{i\ell}^\text{NP}\C{j\ell}^\text{NP}
\end{equation}
\item LFUV differences: $Q_{1,2,4,5}$ and $\delta S_{5,6s}$ observables that measure differences in angular observables between muonic and electronic channels \cite{Capdevila:2016ivx},
\begin{equation}
Q_i=P_{i\mu}^{(')}-P_{ie}^{(')\, ,} \qquad \delta S_i=S_{i\mu}-S_{ie}
\end{equation} 
In this case we use the following parametrisation ($i,j=9,10,9',10'$)
\begin{align}
\mathcal{O} _{\mu}-\mathcal{O} _{e}
&=\alpha_\mu-\alpha_e+\sum_i\left(\beta_{i\mu}\,\mathcal{C}_{i\mu}^\text{NP}-\beta_{ie}\,\mathcal{C}_{ie}^\text{NP}\right)\\
&\quad+\sum_{i\geq j}\left(\gamma_{ij\mu}\,\mathcal{C}_{i\mu}^\text{NP}\mathcal{C}_{j\mu}^\text{NP}-\gamma_{ije}\,\mathcal{C}_{ie}^\text{NP}\mathcal{C}_{je}^\text{NP}\right)\nonumber
\end{align}
\item LFUV ratios: these type of observables are defined as ratios of integrated branching fractions involving muonic (numerator) and electronic (denominator) final states and we use ($i,j=9,10,9',10'$):
\begin{equation}
\dfrac{\mathcal{O}_\mu}{\mathcal{O}_e}=\dfrac{\alpha_\mu + \sum_i\beta_{i\mu} \C{i\mu}^\text{NP} + \sum_{i\geq j} \gamma_{ij\mu}\C{i\mu}^\text{NP}\C{j\mu}^\text{NP}}{\alpha_e + \sum_i\beta_{ie} \C{ie}^\text{NP} + \sum_{i\geq j} \gamma_{ije}\C{ie}^\text{NP}\C{je}^\text{NP}}
\end{equation}
\end{itemize}

Not all observables can be parametrised in a fully satisfying way over this grid through such a quadratic approximation. The difference is at most of 0.3 for all observables (central value and uncertainty) over the whole range of scan.
$P_i(B\to K^*\mu^{+}\mu^{-})$ observables turn out to be more difficult to fit with a purely quadratic parametrisation, 
with differences between the approximate parametrisation and the exact values above
0.15 in some parts of the scan range (this could be expected due to their normalisation, leading
to rapidly varying functions when the Wilson coefficients are changed). However, even this approximate parametrisation is able to catch a few interesting aspects of the sensitivity to Wilson coefficients.

\subsection{Observables as conics}

Let us start the discussion with the $b\to s\mu\mu$ observables, with a quadratic parametrisation that can be seen as a conic in the space of 
\begin{equation}
{\mathcal O}=a+(X-X_0)^T.M.(X-X_0)
\end{equation}
The symmetric matrix $M$ can be diagonalised as $M=R^T.\Delta.R$
where $R$ is an orthogonal (rotation) matrix, and $\Delta$ is a diagonal matrix with eigenvalues ordered in (absolute) size. We can define the (rotated) $Y$-basis
\begin{equation}
Y=R.(X-X_0)\qquad {\mathcal O}=a+R^T.\Delta.R
\end{equation}
The largest eigenvalues of $\Delta$ correspond to directions in the $Y$-basis that are able to change the value of ${\mathcal O}$ in the most significant way. These directions correspond to linear combinations of Wilson coefficients to which the observable ${\mathcal O}$ is the most sensitive.

We can consider the directions corresponding to the largest eigenvalues in absolute value (``dominant ones'') and we can determine whether some directions are similar for different observables. This will indicate that some observables share the same sensitivity to NP for their dominant directions.
We perform this comparison considering only the dominant eigenvalue as well as the eigenvalues that are at most 80\% of this eigenvalue~\footnote{Let us add that the comparison of eigenvalues can be performed only for a given observable (possibly in different bins). The rescaling of an observable by an arbitrary factor will affect the absolute values of the eigenvalues but not their relative importance.}, and we consider ``parallel'' directions that correspond to pairs of vectors with an angle with a cosine of 0.9 or more (in absolute value). We perform this comparison for all the bins of the different observables. Due to the large number of directions to consider, we give our conclusions in a compact way, where each observable mentioned means ``some of the bins of this observable''. We can observe several classes of observables exhibiting dominant parallel directions:
\begin{itemize}
\item observables which are expected to exhibit similar directions due to flavour symmetries
\begin{eqnarray}
&&\hspace*{-0.5cm}{\cal B}(B^0\to K^{*0}),{\cal B}(B^+\to K^{*+}),{\cal B}(B_s\to \phi)
\qquad
{\cal B}(B^0\to K^{0}),{\cal B}(B^+\to K^{+})
\\
&&\hspace*{-0.5cm} P_i(B^0\to K^{*0}), P_i(B_s\to \phi)\qquad (i=1,4,6,F_L)
\end{eqnarray}

\item observables from the same process
\begin{eqnarray}
&&{\cal B}(B^0\to K^{*0}),P_i(B^0\to K^{*0})\qquad (i=1,3,4,5,6,8,F_L)\\
&&{\cal B}(B_s\to \phi),P_i(B_s\to \phi)\qquad (i=1,4,6,F_L)\\
&&P_i(B_s\to \phi),P_j(B_s\to \phi)\qquad (\{i,j\}=\{F_L,1\},\{F_L,4\},\{1,4\}\\
&&P_i(B^0\to K^{*0}),P_j(B^0\to K^{*0})\\
&&\quad (\{i,j\}
  =\{F_L,1\},\{F_L,2\},\{F_L,4\},\{F_L,8\},\\
&&\quad\qquad\qquad\{1,2\},\{1,3\},\{1,4\},\{1,5\},\{1,8\},\\
&&\quad\qquad\qquad \{2,4\},\{2,5\},\{2,6\},\{2,8\},\{3,4\},\{3,8\},\{4,8\},\{6,7\})
\end{eqnarray}
\item observables with unexpected correlations
\begin{eqnarray}
&&\hspace*{-0.5cm}{\cal B}(B^+\to K^{+}),{\cal B}(B^0\to K^{0}),P_1(B^0\to K^{*0}),P_1(B_s\to\phi),P_3(B^0\to K^{*0})
\end{eqnarray}
\end{itemize}

If we now restrict the analysis to observables that are currently deviating by more than 2$\sigma$, we can easily consider a wider range of directions in each case (going down to eigenvalues that are 30\% of the maximal ones), we can provide a slightly more precise description of the correlations
\begin{itemize}
\item between observables as expected from flavour symmetries: 
\begin{eqnarray}
&&{\cal B}(B^0\to K^{*0})[15,19],{\cal B}(B^+\to K^{*+})[15,19],{\cal B}(B_s\to \phi)[15,18.8]
\end{eqnarray}
\item between different close bins of the same observable
\begin{eqnarray}
&&P_5^\prime(B^0\to K^{*0})[4,6],P_5^\prime(B^0\to K^{*0})[6,8]\\
&&{\cal B}(B_s\to\phi)[2,5],{\cal B}(B_s\to\phi)[5,8]
\end{eqnarray}
\item between observables that are not obviously correlated
\begin{eqnarray}
&&{\cal B}(B_s\to\phi)[2,5],{\cal B}(B_s\to\phi)[5,8],P_5^\prime(B^0\to K^{*0})[4,6],\\
&&\qquad {\cal B}(B^0\to K^{*0})[15,19],{\cal B}(B^+\to K^{*+})[15,19]
\end{eqnarray}
\end{itemize}
The same analysis carried out for LFUV quantities involves the same correlations, which is to be expected as they involve ratio or differences of the same observables considered in the muon and the electron cases.

Another comment is in order. We have performed our analysis of the directions in the general case where NP is allowed in ${\cal C}_{9,10,9',10'}$ and we have studied the dominant directions for each observable. Scenarios where some of these NP contributions are assumed to be zero correspond to projections of this analysis on specific hyperplanes. Obviously, if directions are similar in the general  ${\cal C}_{9,10,9',10'}$ space, they remain similar after projection on a given hyperplane, but the projection can also lead to additional pairs of observables with (projected) dominant directions that are parallel (although they are not parallel when considered in the whole space).

\subsection{Directions favoured by the anomalies}

Up to now, we have considered only the leading directions (corresponding to  the main axes of the conics) for the quadratic approximation of the theoretical expression of the $b\to s\mu\mu$ observables. This provides information on the sensitivity to specific directions and the fact that some of these directions are common to several observables. However, since we have not compared these predictions to the deviations observed currently, 
this does not provide information on the specific changes in the Wilson coefficients preferred by the global fit. One can perform a global analysis of the constraints~\cite{Capdevila:2018jhy}, but one can also exploit the above parametrisation to gain some insights to understand better the outcome of the global fit.

This can be done in the following way. As indicated above, we can write a $b\to s\mu\mu$ observable in the $Y$-space
as 
\begin{equation}
{\cal O}=a+Y^T.\Delta.Y
\end{equation}
where the largest axis of the conic correspond to the 1st coordinate in the $Y$-space, with smaller and smaller axes correspond to the 2nd, 3rd\ldots coordinates. Each observable tends to favour a shift in $Y$ in order to reduce ${\cal O}-{\cal O}_{\rm exp}$ compared to ${\cal O}_{\rm SM}-{\cal O}_{\rm exp}$. One can represent this problem as the fact that the conic ${\cal O}={\cal O}_{\rm exp}$ does not pass through the SM point $Y_{SM}=-R.X_0$. This distance corresponds to the NP shift to perform in some of the Wilson coefficients. In principle, one could chose any NP shift in the Wilson coefficients to go from the SM point to the conic section  ${\cal O}={\cal O}_{\rm exp}$, but it turns out to be interesting to think in terms of the main axes of the conic section and first to determine the distance between the SM point and the conic section along a given axis $i$. This is obtained by 
\begin{equation}
\delta Y^{(i)}_{j} =\pm \delta_{ij} \sqrt{\frac{{\cal O}_{\rm exp}-a-\sum_{k\neq i} Y_{SM,k}^2}{\Delta_i}}
\end{equation}
which corresponds to a NP in the Wilson coefficients $\delta X^{(i)}=X_0+R^T.\delta Y^{(i)}$.

Let us emphasize that this equation does not necessarily have a solution, as it is not necessarily possible to reach the conic section from the SM in all directions (this is easily seen in the case of an ellipsoid, depending on the position of the SM point, inside, outside and close or outside and far away).
The above equation illustrates two different aspects of the problem: on the one hand, a large shift in ${\cal O}$ is achieved along the 1st coordinate through a smaller shift in $Y$ (and thus in the Wilson coefficients) than along higher coordinates, but on the other hand, the distance between the SM point and the conic section might be easier to bridge along higher coordinates than along the first coordinate. The shifts $\delta Y^{(i)}$ are thus interesting tools to identify the directions preferred by each observable that deviate significantly from the SM.

If we perform this exercise for the observables deviating by more than $2\sigma$ and keeping only dominant directions (at least 30\% of the largest eigenvalue), we obtain the results in Tab.~\ref{tab:distanceconic}. We computed the distances $\delta Y^{(i)}$ taking into account both experimental and theoretical uncertainties, as indicated in the corresponding table. We also indicate the corresponding directions in terms of $X$-coordinates. We show only directions which can accommodate the central value experimentally measured using the  parametrisation that we used.

\begin{table}
\begin{center}
\begin{tabular}{|c|c|c|c|}
\hline
Observable & Direction $i$ & $\delta Y^{(i)}$ & $X_i$\\
\hline
$P_2(B^0\to K^{*0})$ [0.1,0.98] & 1  &  $11.82 \pm 5.89$  &    (0.01,0.05,0.66,0.75)\\
$P_2(B^0\to K^{*0})$ [0.1,0.98] & 2   & $4.71 \pm 3.62$   &    (-0.47,-0.88,0.02,0.04)\\
$P_2(B^0\to K^{*0})$ [0.1,0.98] & 4  &  $7.93 \pm 6.30$  &    (0.88,-0.47,0.02,-0.00)\\
$P_5^\prime(B^0\to K^{*0})$ [4.0,6.0] & 1 &   $1.83 \pm 0.24$  &  (0.94,0.22,-0.26,0.02)\\
$P_5^\prime(B^0\to K^{*0})$ [4.0,6.0] & 2 &   $1.96 \pm 0.28$  &   (-0.29,0.49,-0.57,0.59)\\
$P_5^\prime(B^0\to K^{*0})$ [4.0,6.0] & 4 &   $8.85 \pm 1.82$  &   (-0.05,-0.63,-0.74,-0.22)\\
$P_5^\prime(B^0\to K^{*0})$ [6.0,8.0] & 1 &   $1.95 \pm 0.32$  &   (0.87,0.43,-0.22,-0.04)\\
$P_5^\prime(B^0\to K^{*0})$ [6.0,8.0] & 2 &   $1.84 \pm 0.37$  &   (-0.31,0.40,-0.60,0.62)\\
${\cal B}(B^0\to K^{*0})$ [15.,19.] & 2 &  $ 0.79 \pm 0.15$  & (0.14,-0.69,-0.14,0.70)\\
${\cal B}(B^+\to K^{*+})$ [15.,19.] & 2 &   $1.18 \pm 0.17$  &  (0.14,-0.69,-0.14,0.70)\\
${\cal B}(B_s\to\phi)$ [2.0,5.0] & 2 &   $2.34 \pm 0.63$  &   (-0.03,-0.70,0.02,0.71)\\
${\cal B}(B_s\to\phi)$ [15.,18.8] & 2 &   $0.65 \pm 0.16$ &   (-0.36,0.60,0.36,-0.62)\\
\hline
\end{tabular}
\caption{Distance $\delta Y^{(i)}$ from the SM to the LFD conic ${\cal O}-{\cal O}_{\rm exp}$ following the main axis $i$ of the conic section. The last column indicates the vector corresponding to the main axis $i$ in the basis $X=(\C{9\mu},\C{10\mu},\C{9'\mu},\C{10'\mu})$.} \label{tab:distanceconic}
\end{center}
\end{table}

We see that
\begin{itemize}
\item $P_2$ is affected by large uncertainties that make it difficult to reach definite conclusions
\item the 4 branching ratios prefer a single direction, which is essentially along $\simeq \C{10\mu}-\C{10'\mu}$
\item the two bins in $P_5^\prime$ have similar behaviours, with two directions favoured ($\C{9\mu}$ with a small $\C{10\mu}$ component, or along $\simeq \C{10\mu}-\C{9'\mu}+\C{10'\mu}$)
\end{itemize}
The opposite signs for $\C{10'}$ between the branching ratios and the two bins in $P_5^\prime$ make it difficult to use this specific parameter to improve the agreement of all observables with experiment. A better agreement between theory and data for these observables can thus be reached by performing shifts in $\C{9\mu}$ and $\C{10\mu}$. Naturally, this very qualitative argument does not take into account the remaining observables, which are in good agreement with the SM and constrain also the size of the NP shifts in $\C{9\mu}$ and $\C{10\mu}$. However it is interesting to see that this rough analysis in terms of favoured directions supports the outcome of the global fit.

A similar analysis can be performed in the case of $R_K$ and $R_{K^*}$, but almost all directions are equally favoured in both muon and electron directions, without shedding further light on the preferences for the global fit.

\section{Conclusions}

Over the last few years, the rare $b\to s\ell\ell$ decays have proved particularly interesting, with a large set of deviations from SM expectations. If a first explanation was provided by a large NP contribution to the Wilson coefficient $\C{9\mu}$, the current set of data allows for several different NP scenarios, which feature large NP contributions violating lepton-flavour universality (affecting $b\to s\mu\mu$ but not $b\to s ee$) in connection with 
the measurement of the LFUV ratios $R_K$ and $R_{K^*}$. This does not exhaust the possibilities, and for instance, it is also possible to add NP contributions satisfying lepton-flavour universality (affecting $b\to s\mu\mu$ and $b\to s ee$) in the same way. With this wealth of possibilities, it is important to understand how current and forthcoming data may favour one scenario over others, so that one can pin down the best NP explanation for the whole set of anomalies observed.

We have first considered some of the current inner tensions of the fits. We highlighted the situation of $R_{K^*}$ in the lowest bin, the tension between low and large recoils for $B_s\to\phi\mu^{+}\mu^{-}$, and a similar issue with $B\to K^*\mu^{+}\mu^{-}$ angular observables. We then discussed the impact of forthcoming measurements of $\langle R_K\rangle_{[1.1,6]}$ and $\langle Q_5\rangle_{[1.1,6]}$ to disentangle NP hypotheses. We considered various central values for these two measurements and we assumed some reduction in the experimental uncertainties in order to study how pulls w.r.t. SM and best-fit points would evolve. $\langle R_K\rangle_{[1.1,6]}$ alone proves to have only a limited ability to separate the various NP hypotheses: $\C{9\mu}^{\rm V}=-\C{9'\mu}^{\rm V}$ is the only hypothesis strongly affected. On the other hand, the combination of $\langle R_K\rangle_{[1.1,6]}$ and $\langle Q_5\rangle_{[1.1,6]}$ proves much more efficient to separate various favoured hypotheses, either with only LFUV NP contributions or with both LFUV and LFU contributions.

We then considered two different approaches to understand the current structure of the global fit. Firstly, we discussed the pulls associated with individual observables. Their value can be compared under the SM hypothesis and under various NP hypotheses, indicating whether they are easier to accommodate once NP contributions are added. As could be expected, no single hypothesis is clearly preferred: under each hypothesis, some of the observables get larger pulls and others smaller ones. Secondly, we looked at approximate quadratic parametrisations of the observables with large pulls in the SM. We could then identify directions in the space of Wilson coefficients to which the observables showed a large sensitivity to any shift. Interestingly, the same directions are often favoured by a large set of observables. We also analysed the shifts preferred by the observables showing large tensions with the SM using the same approach of directions.

The overall conclusion of our study is that the current measurements of $b\to s\ell\ell$ observables  are not enough to disentangle various NP hypotheses: the tensions with the SM do not point towards an unambiguous NP scenario, and several compete at the same level, with only LFUV-NP contributions or combining LFUV and LFU contributions. An update of $\langle R_K\rangle_{[1.1,6]}$ could prove interesting in confirming (or not) the existence of LFUV NP contributions, but our study shows that in most cases, this update will not help in lifting the degeneracy among the various NP hypotheses. On the other hand, we have shown that the determination of $\langle Q_5\rangle_{[1.1,6]}$ might bring interesting information that would be complementary to the existing observables and that could help significantly in disentangling the various scenarios (depending on the actual measured value).
The use of extended data sets, the analyses in different experimental environments and the inclusion of additional observables will all prove particularly important in the coming months in order to cross-check and constrain the dynamics at work in $b\to s\ell\ell$ transitions. This will prove essential to identify models of New Physics that could ultimately resolve the anomalies currently observed in the $b$-quark sector.

\begin{acknowledgments}

This work received financial support from the grants CICYT-FEDER-FPA 2014-55613-P and FPA 2017-86989-P [JM, SDG, BC, MA], and from Centro de Excelencia Severo Ochoa SEV-2012-0234 [BC]; from the EU Horizon 2020 program from the grants No. 690575, No. 674896 and No. 692194 [SDG]. J.M. acknowledges financial support from the Catalan ICREA Academia Program. The work of P.M. is supported by the Beatriu de Pin\'os postdoctoral program of the Government of Catalonia's Secretariat for Universities and Research of the Ministry of Economy and Knowledge of Spain.

\end{acknowledgments}

\appendix

\section{Dependence of best-fit points on $\langle R_K\rangle_{[1.1,6]}$ and $\langle Q_5\rangle_{[1.1,6]}$}\label{sec:appendixbfp}

In this appendix, we show how the best-fit point of each NP hypothesis considered varies assuming particular values of
$\langle R_K\rangle_{[1.1,6]}$ and $\langle Q_5\rangle_{[1.1,6]}$. For the two-dimensional hypothesis, we show in solid and dashed lines of the same colour the variations of the two parameters describing the NP contributions of the hypotheses.

\begin{figure}[h]
\begin{subfigure}{.5\textwidth}
  \centering
  \includegraphics[width=0.95\linewidth]{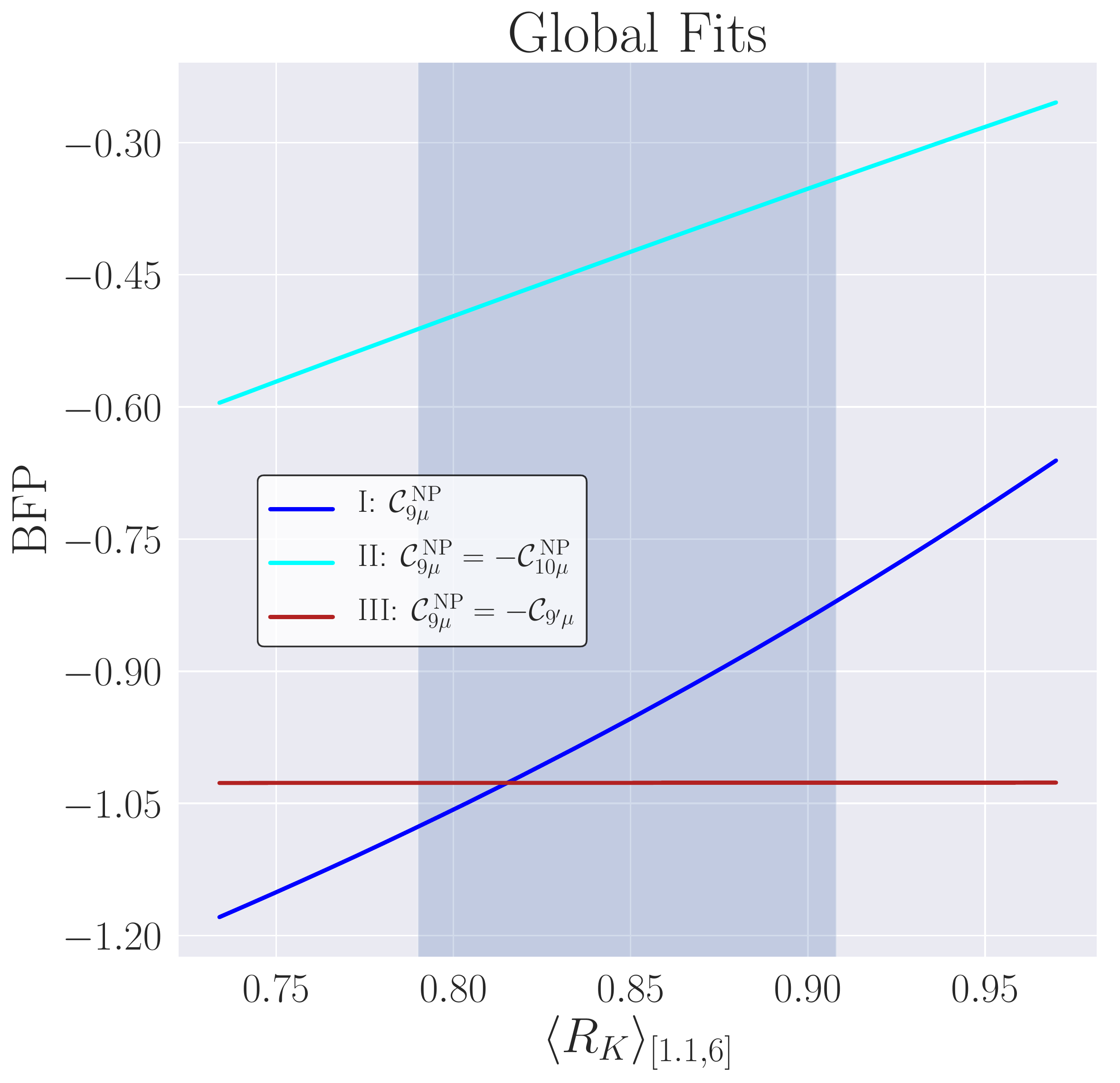}
\end{subfigure}%
\begin{subfigure}{.5\textwidth}
  \centering
  \includegraphics[width=0.95\linewidth]{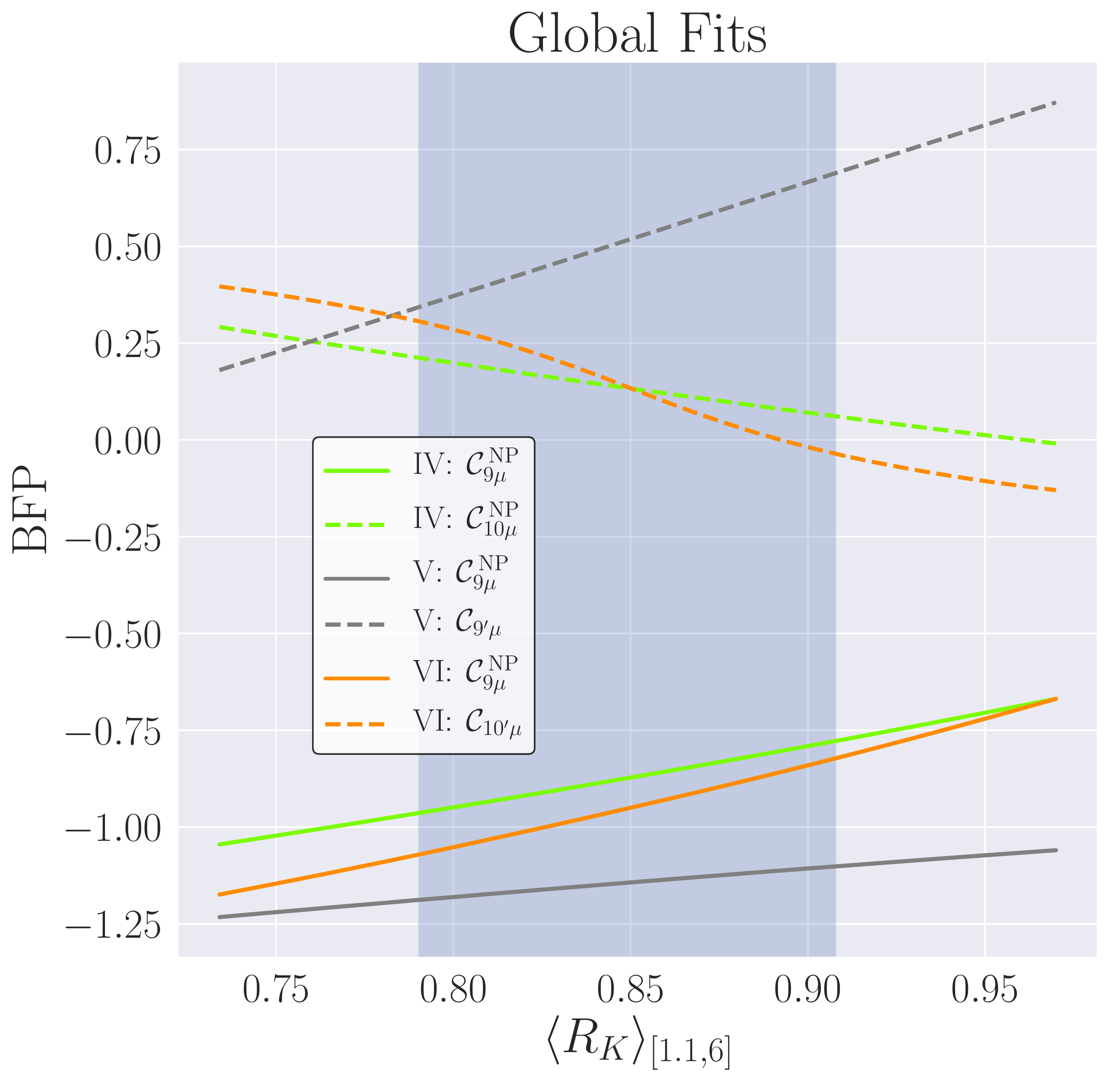}
\end{subfigure}
\begin{subfigure}{.5\textwidth}
  \centering
  \includegraphics[width=0.95\linewidth]{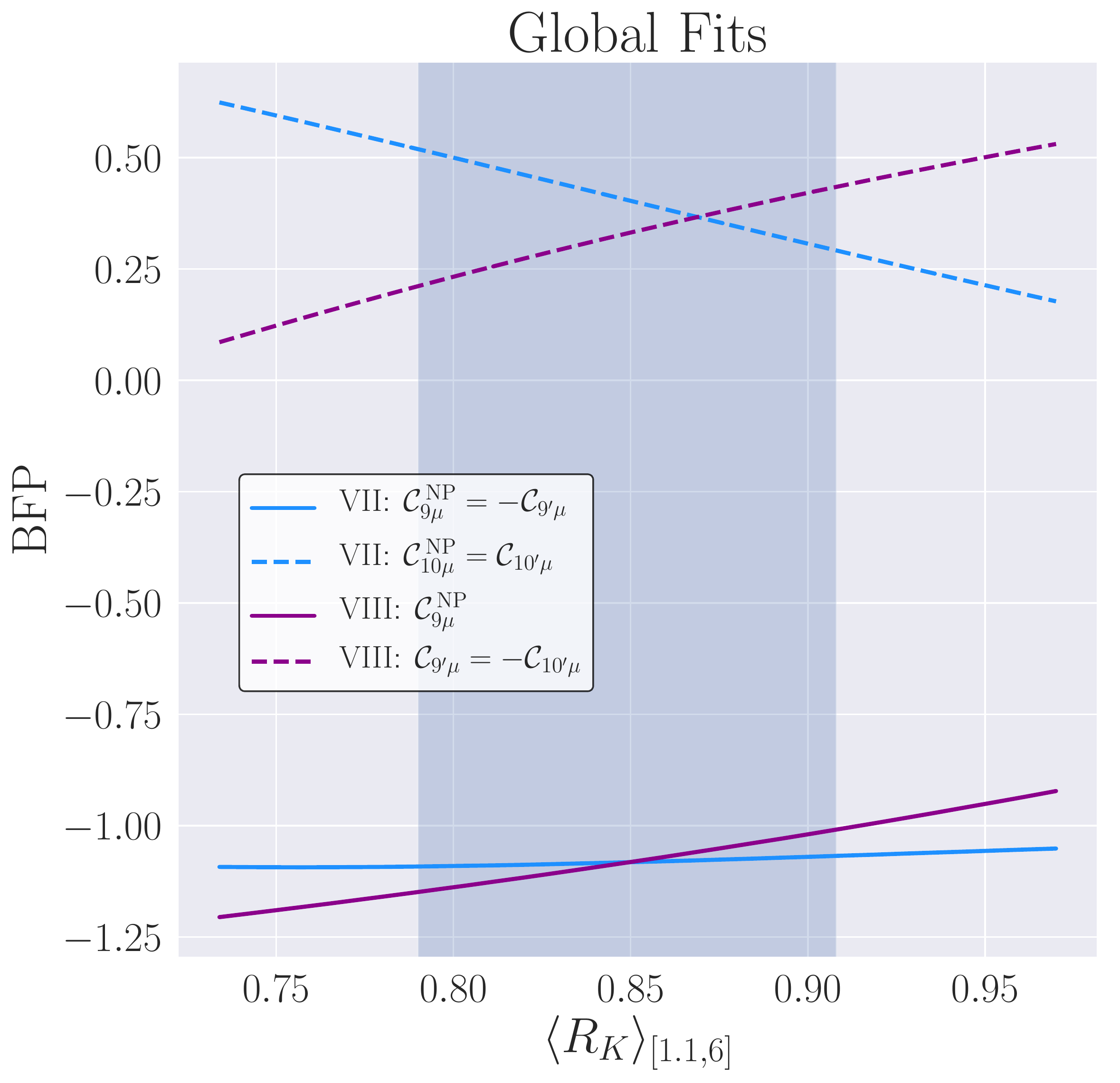}
\end{subfigure}%
\begin{subfigure}{.5\textwidth}
  \centering
  \includegraphics[width=0.95\linewidth]{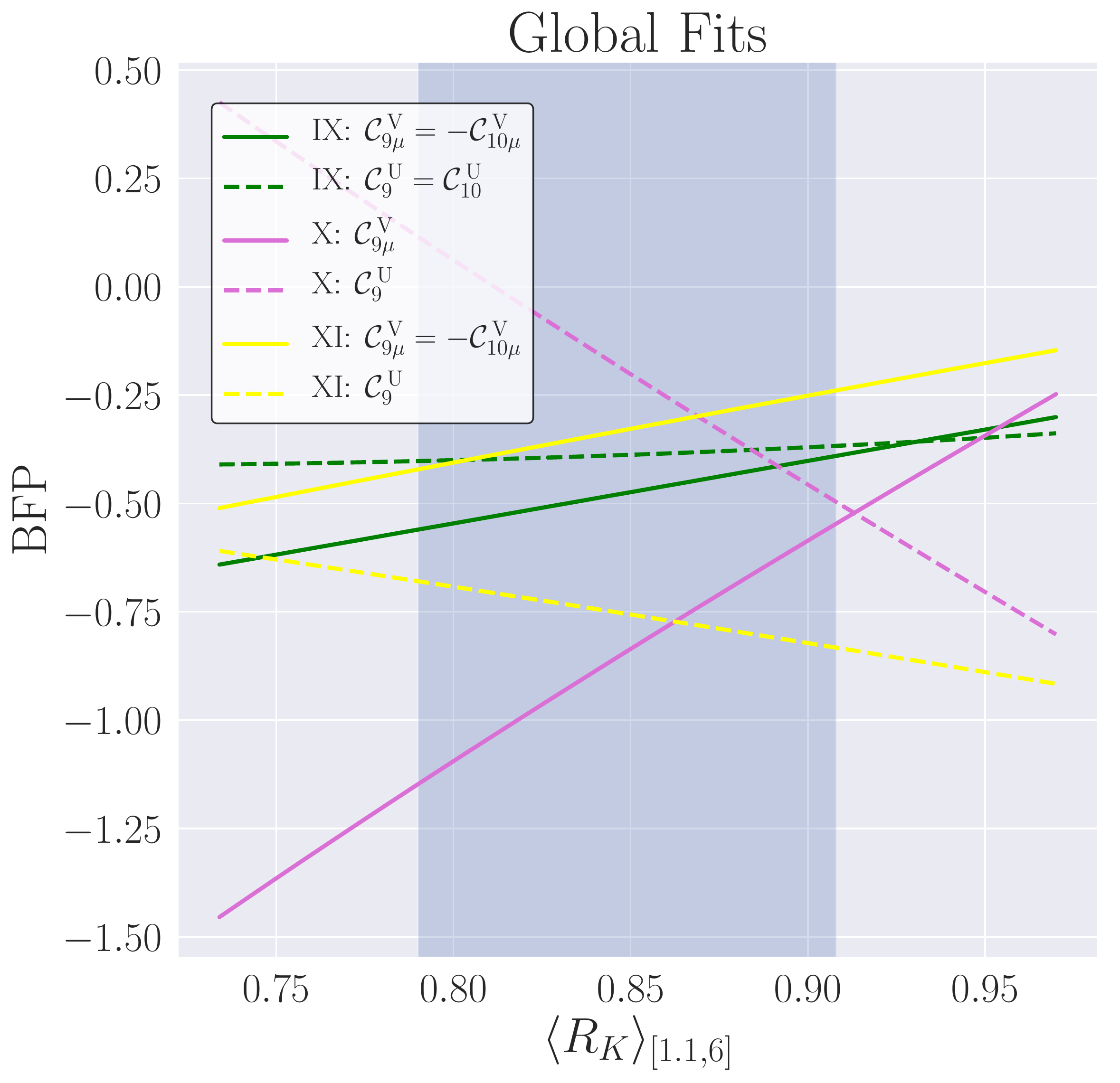}
\end{subfigure}
\centering
\includegraphics[width=0.475\linewidth]{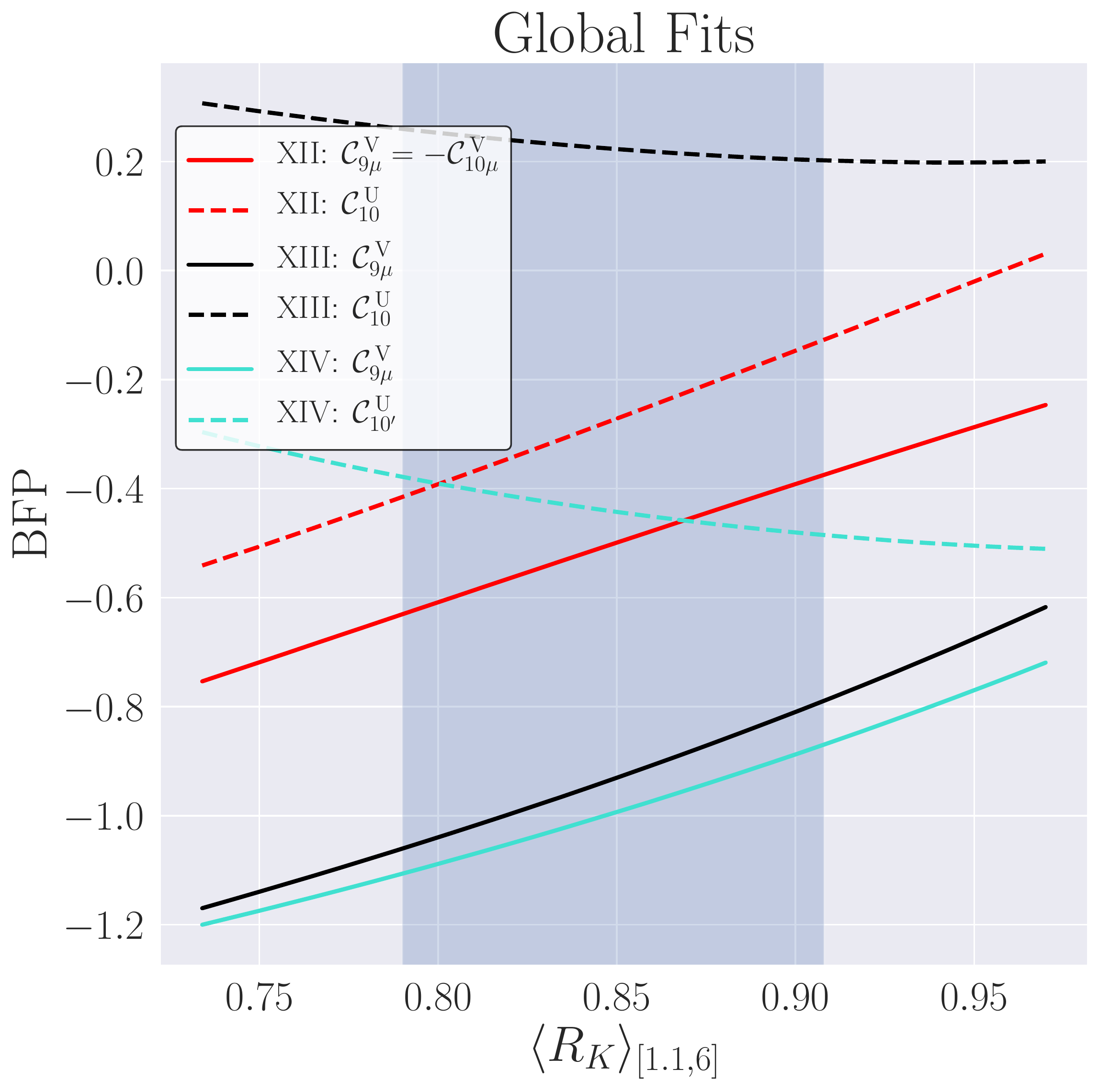}
\caption{Global fit: Impact of the central value of $\langle R_K\rangle_{[1.1,6]}$ on the b.f.p.s of the NP scenarios under consideration.}         
\label{fig:GlobalFitPlots_b}
\end{figure}

\begin{figure}[h]
\begin{subfigure}{.5\textwidth}
  \centering
  \includegraphics[width=0.92\linewidth]{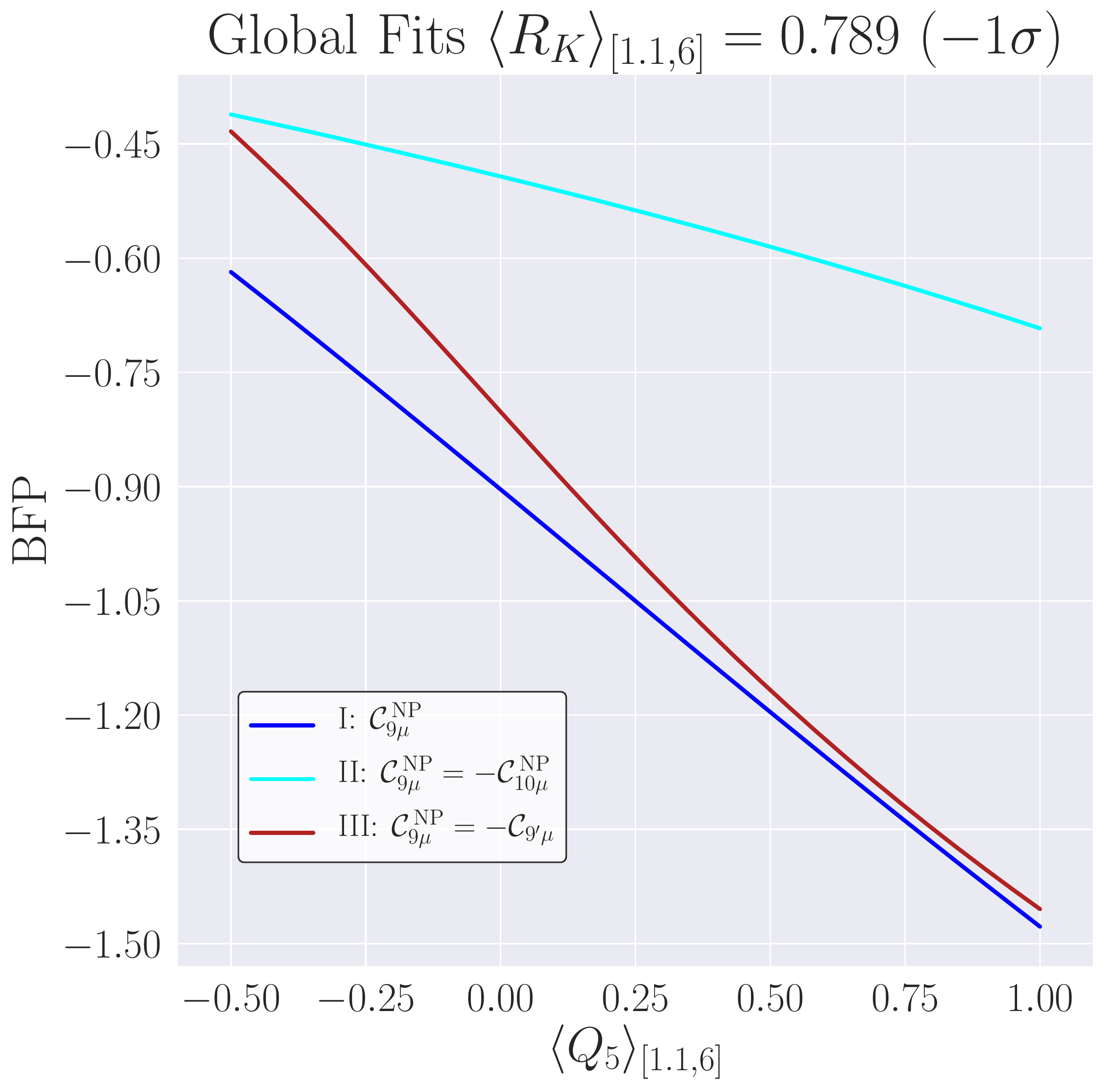}
\end{subfigure}%
\begin{subfigure}{.5\textwidth}
  \centering
  \includegraphics[width=0.92\linewidth]{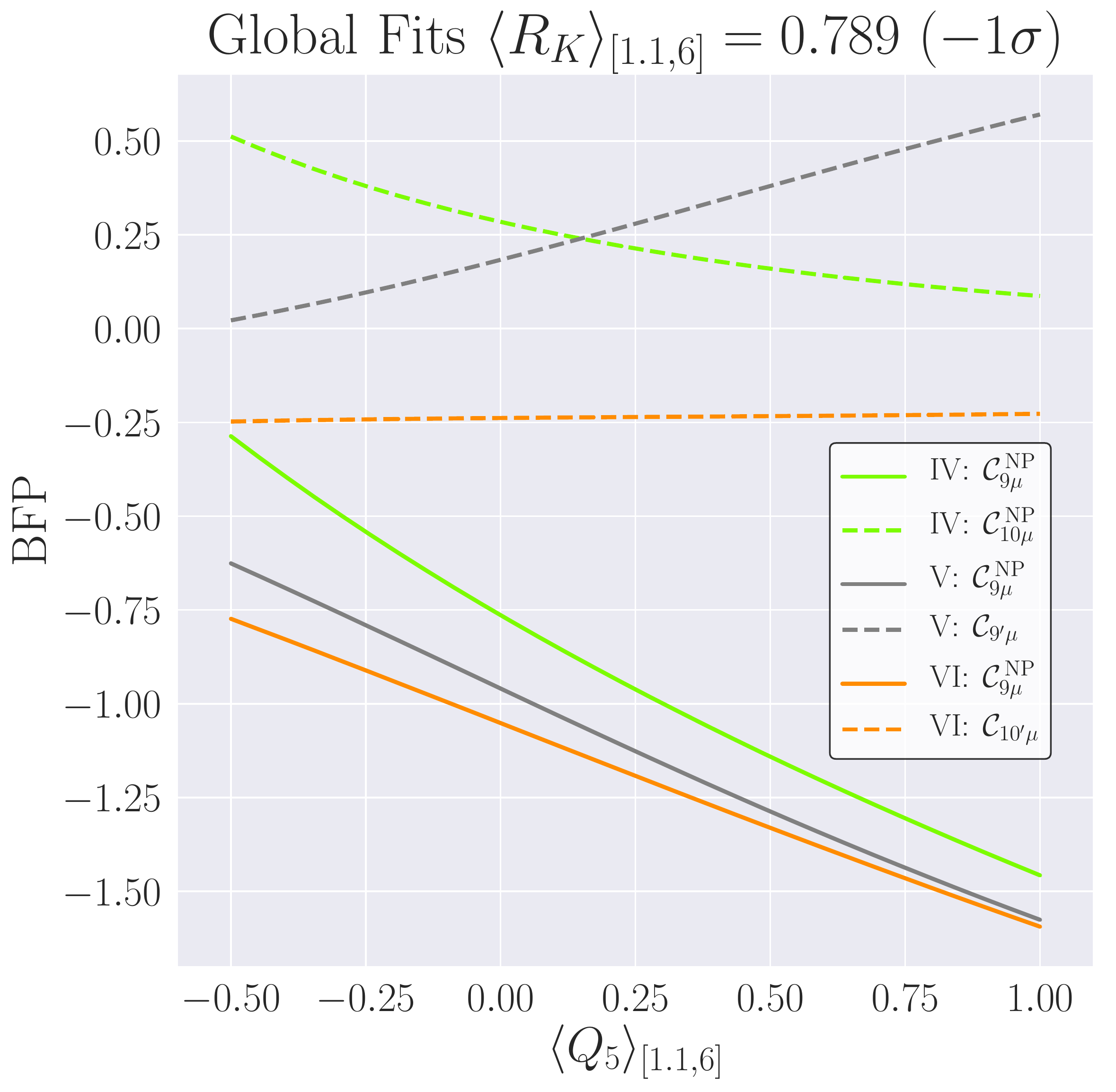}
\end{subfigure}
\begin{subfigure}{.5\textwidth}
  \centering
  \includegraphics[width=0.92\linewidth]{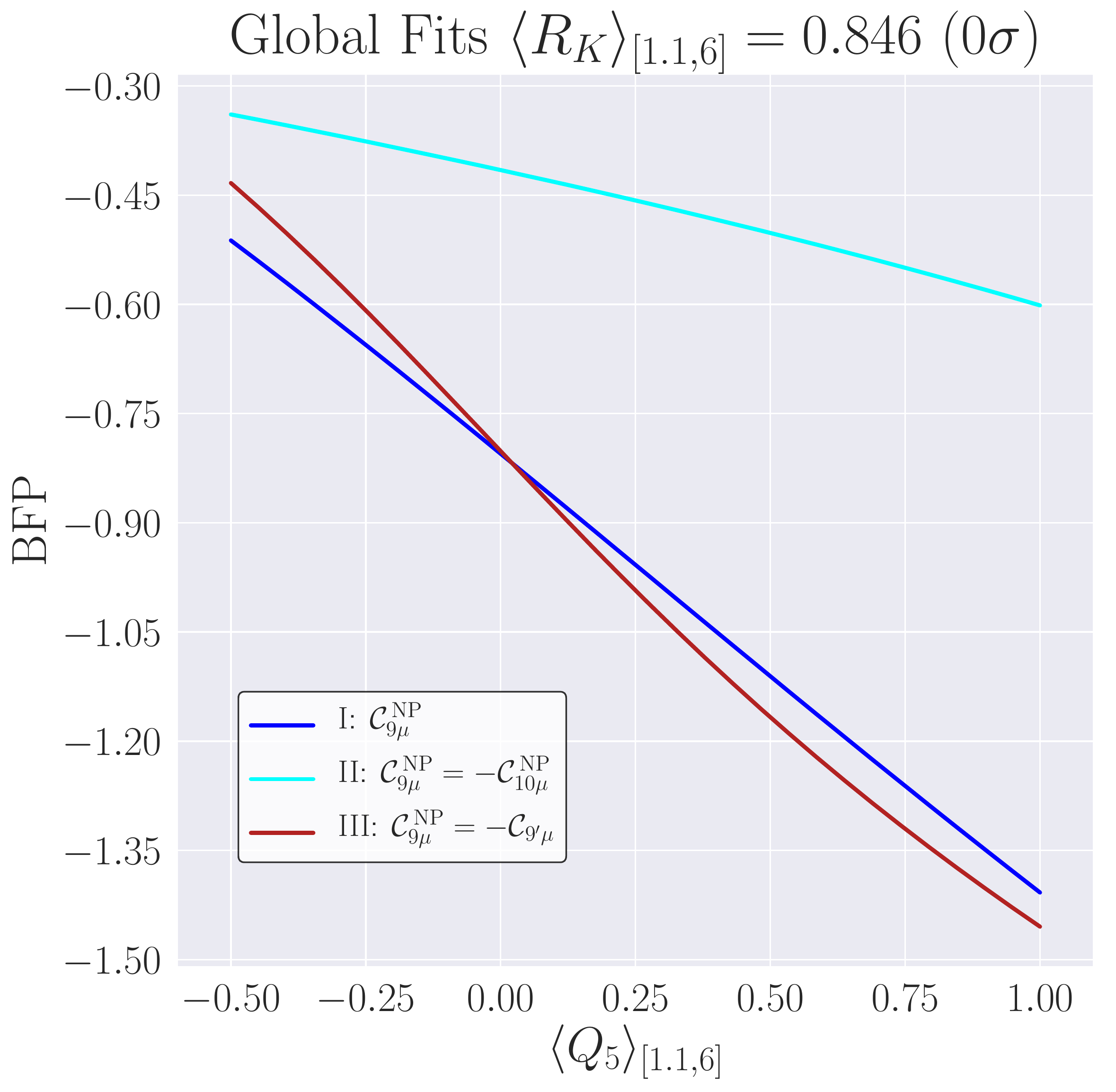}
\end{subfigure}%
\begin{subfigure}{.5\textwidth}
  \centering
  \includegraphics[width=0.92\linewidth]{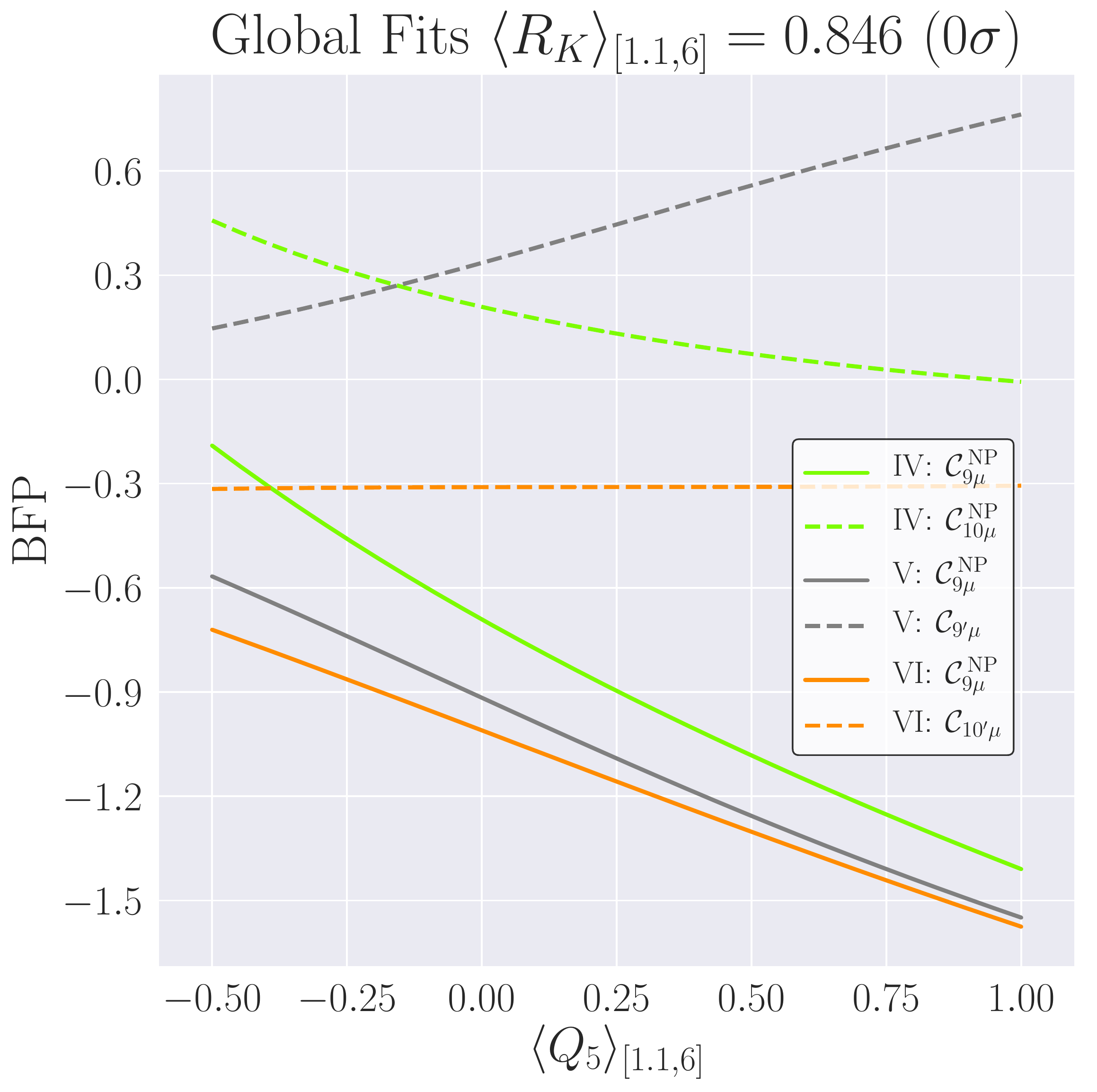}
\end{subfigure}
\begin{subfigure}{.5\textwidth}
  \centering
  \includegraphics[width=0.92\linewidth]{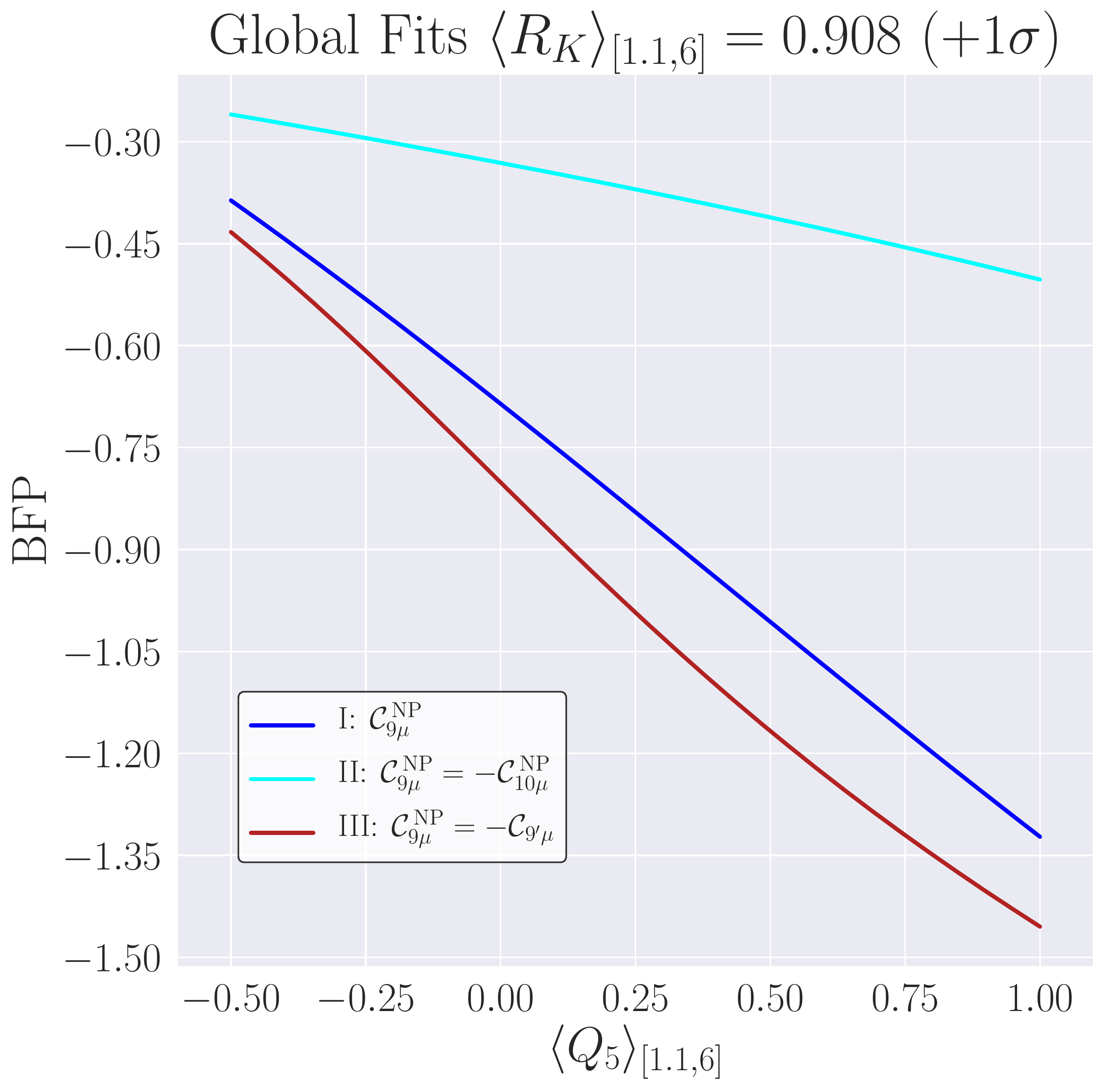}
\end{subfigure}%
\begin{subfigure}{.5\textwidth}
  \centering
  \includegraphics[width=0.92\linewidth]{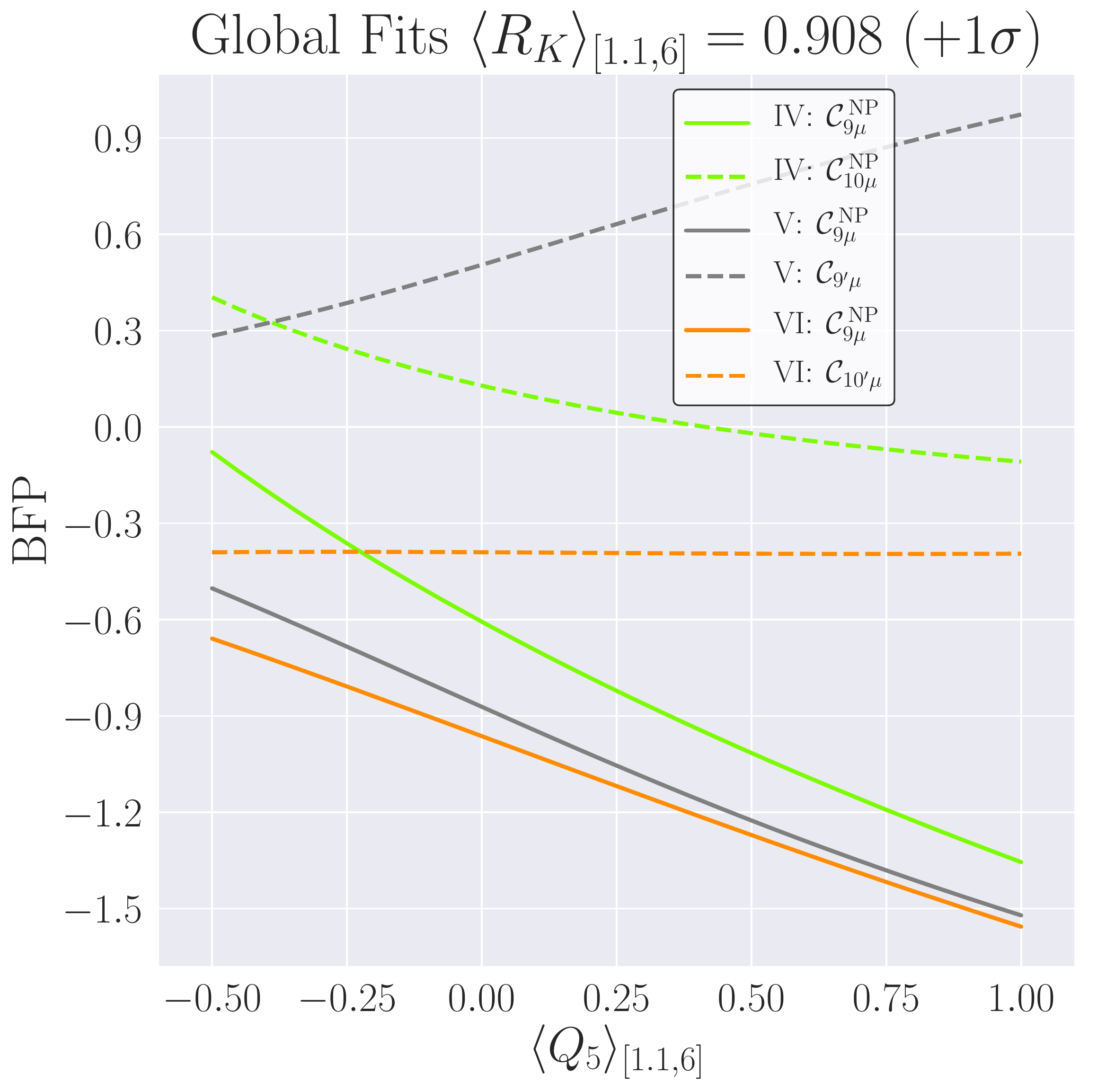}
\end{subfigure}
\caption{Global fit: Impact of $\langle Q_5\rangle_{[1.1,6]}$ on the b.f.p.s of the NP hypotheses under consideration for different values of $\langle R_K\rangle_{[1.1,6]}$.}         
\label{fig:GlobalFitPlotsQ5_2a}
\end{figure}

\begin{figure}[h]
\begin{subfigure}{.5\textwidth}
  \centering
  \includegraphics[width=0.92\linewidth]{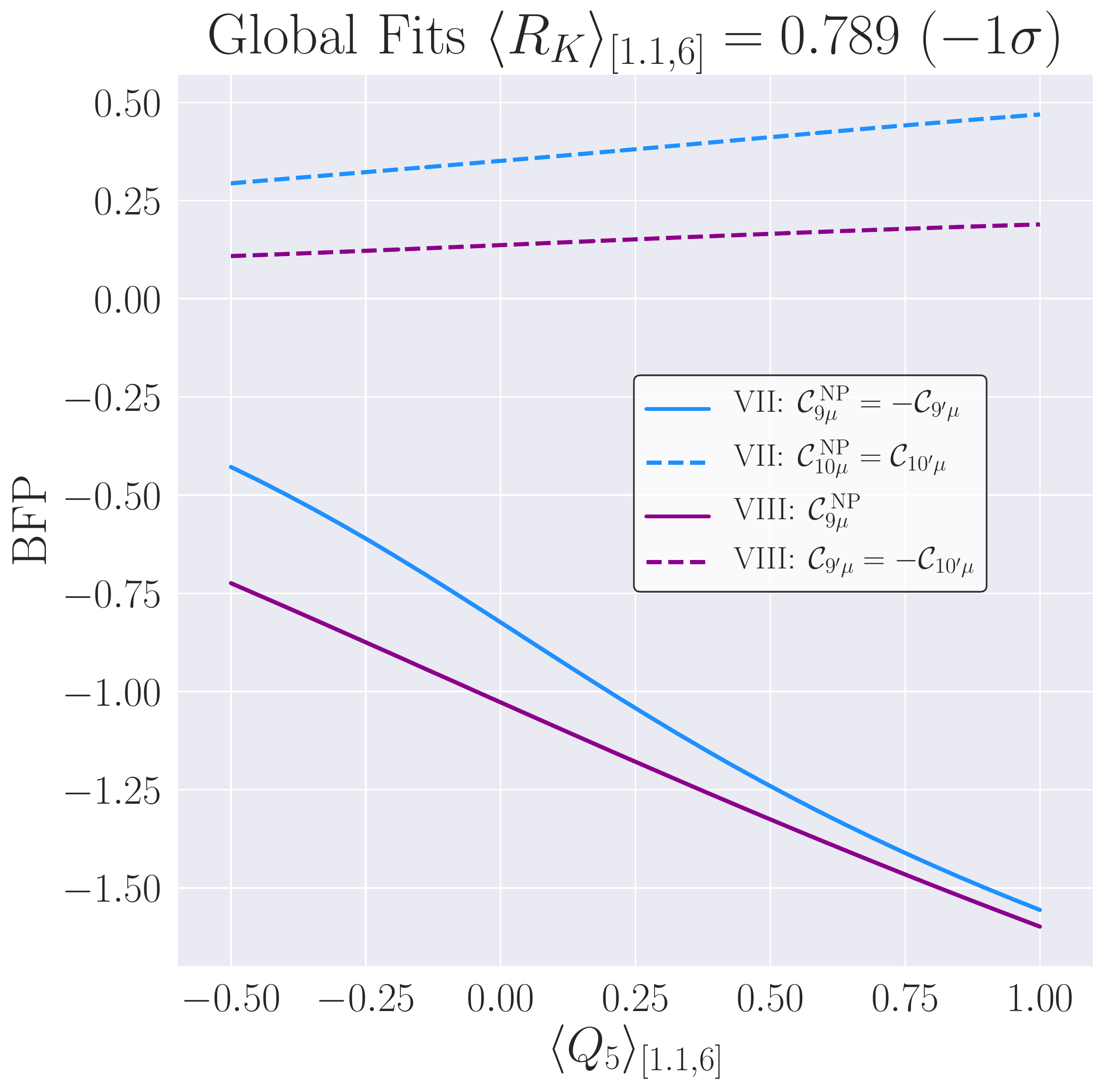}
\end{subfigure}%
\begin{subfigure}{.5\textwidth}
  \centering
  \includegraphics[width=0.92\linewidth]{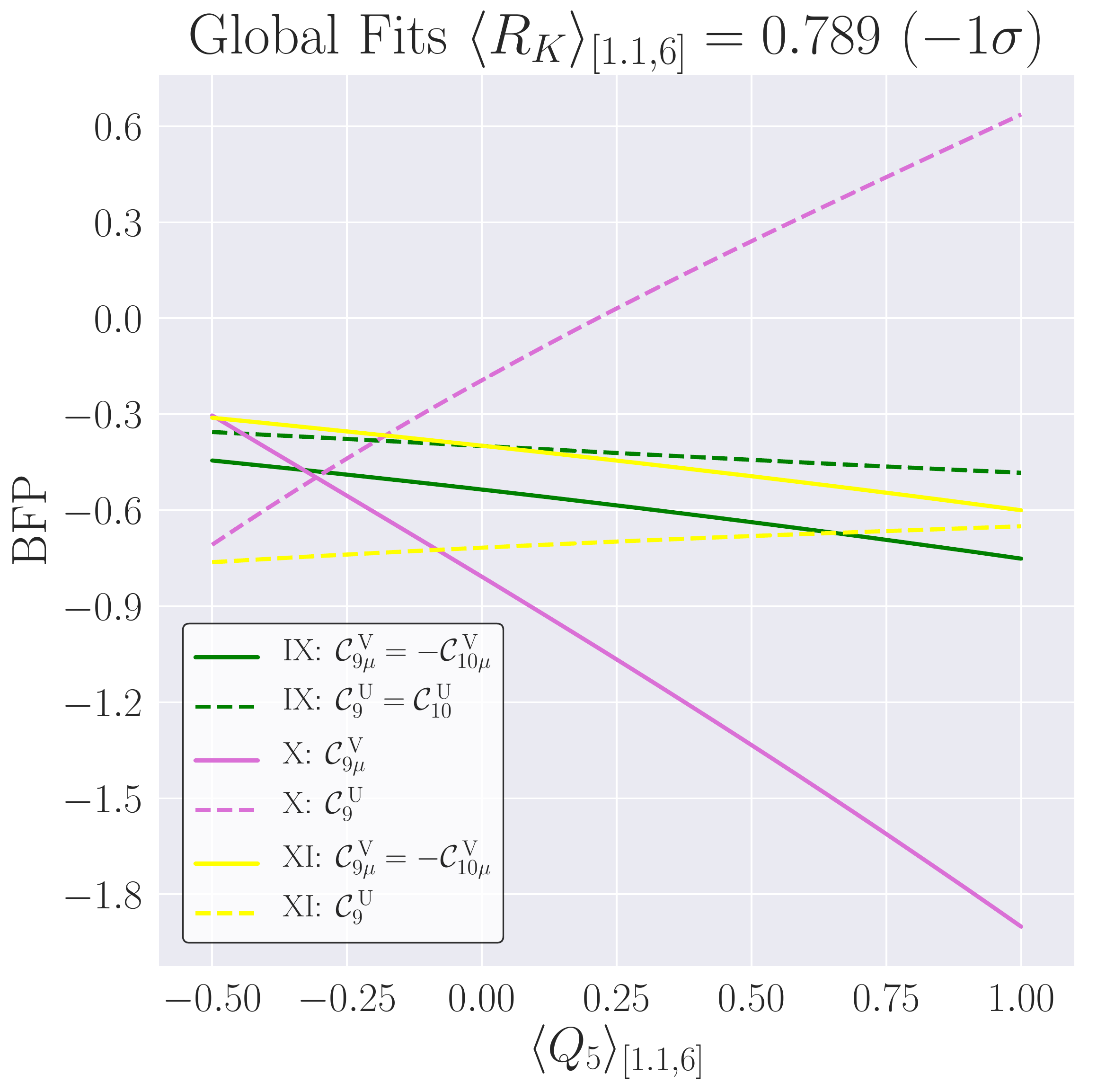}
\end{subfigure}
\begin{subfigure}{.5\textwidth}
  \centering
  \includegraphics[width=0.92\linewidth]{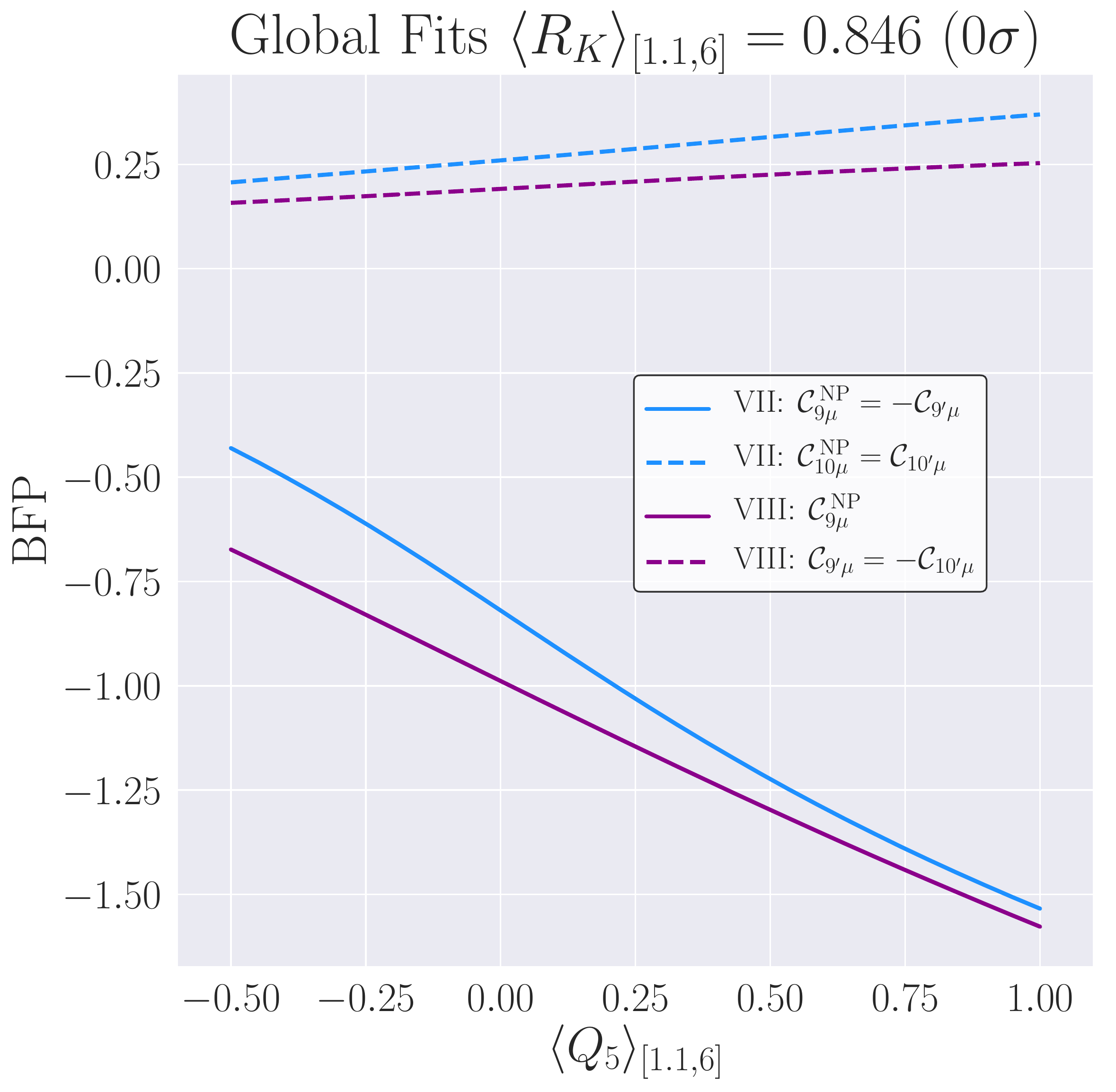}
\end{subfigure}%
\begin{subfigure}{.5\textwidth}
  \centering
  \includegraphics[width=0.92\linewidth]{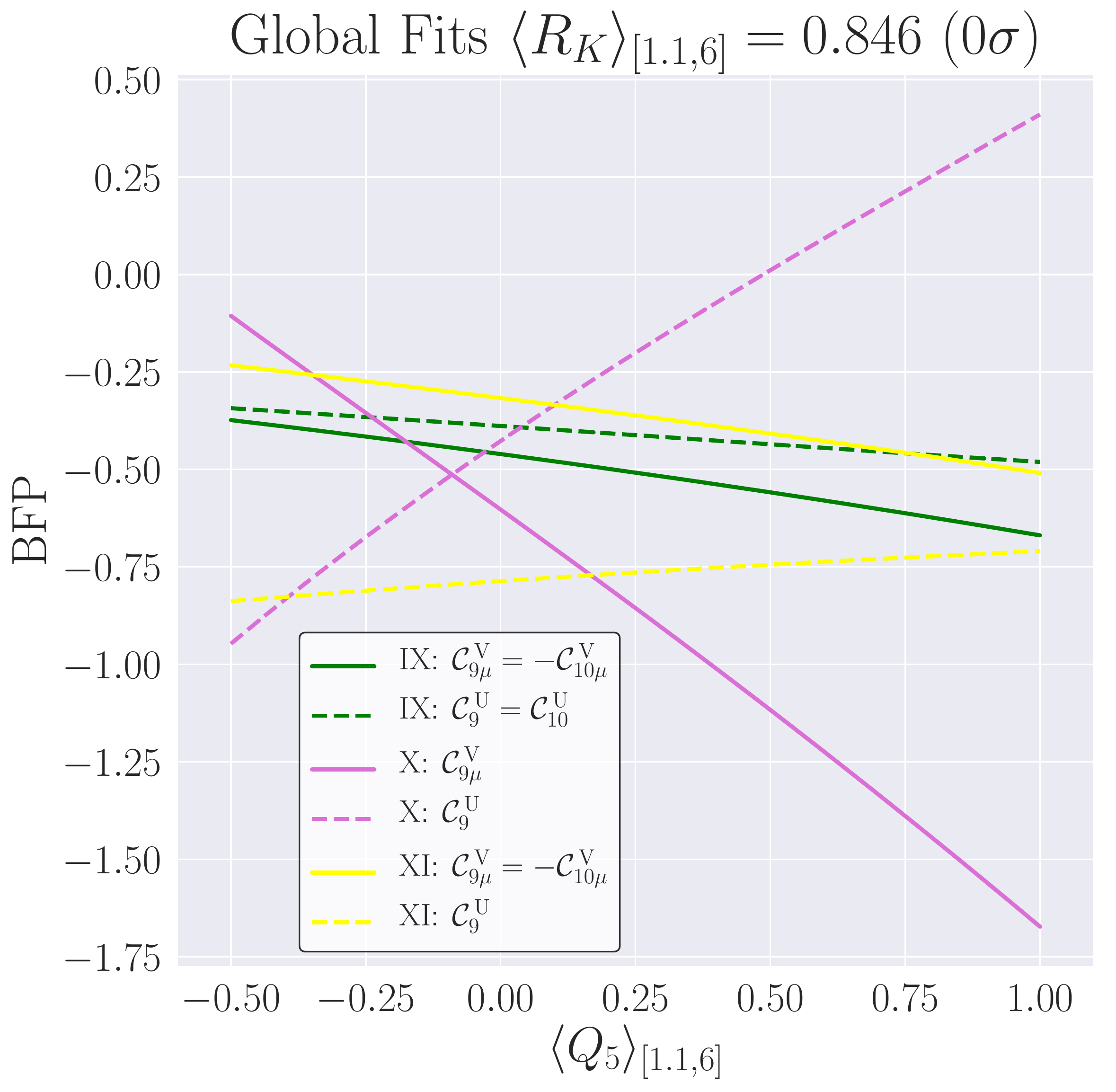}
\end{subfigure}
\begin{subfigure}{.5\textwidth}
  \centering
  \includegraphics[width=0.92\linewidth]{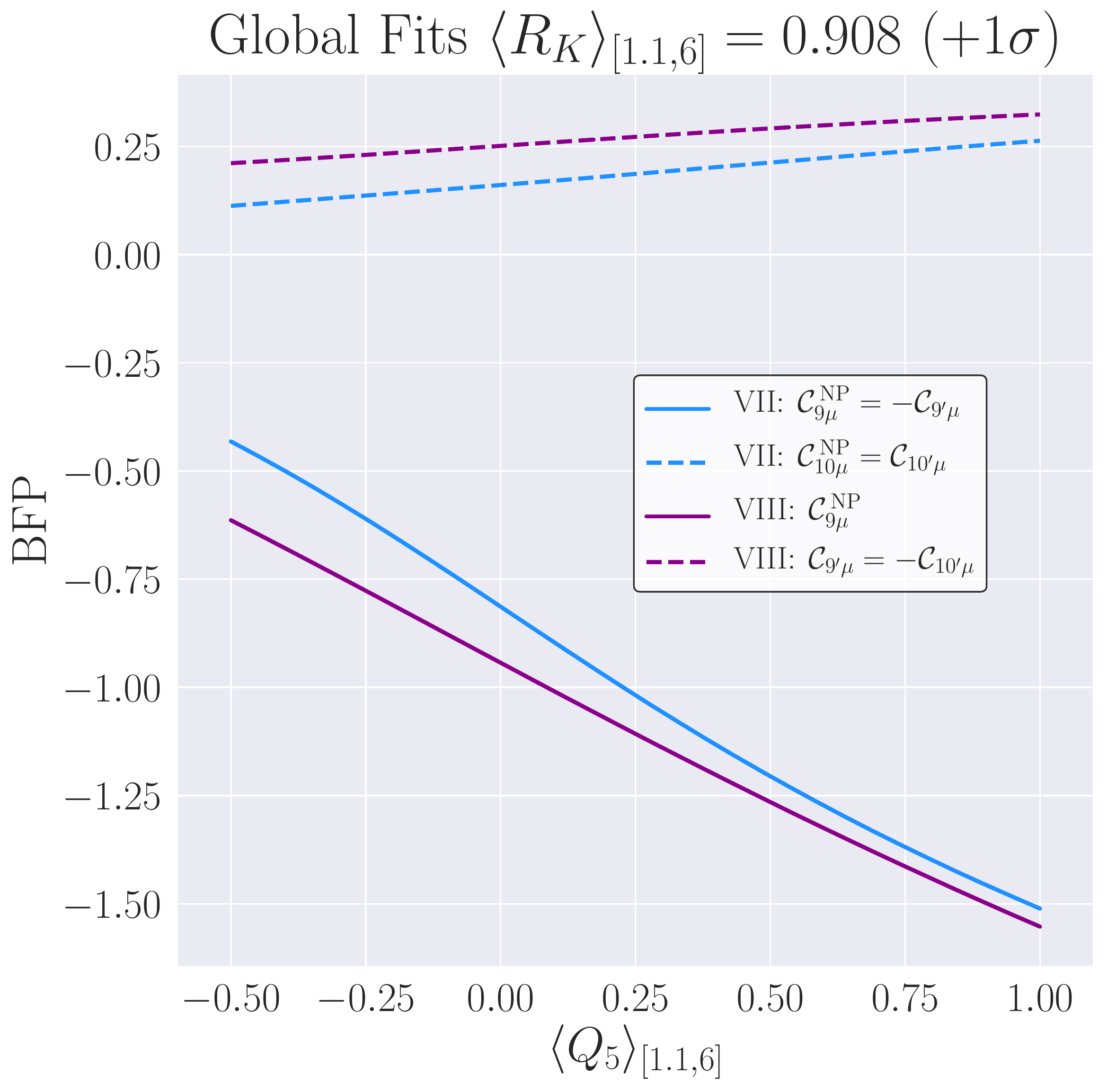}
\end{subfigure}%
\begin{subfigure}{.5\textwidth}
  \centering
  \includegraphics[width=0.92\linewidth]{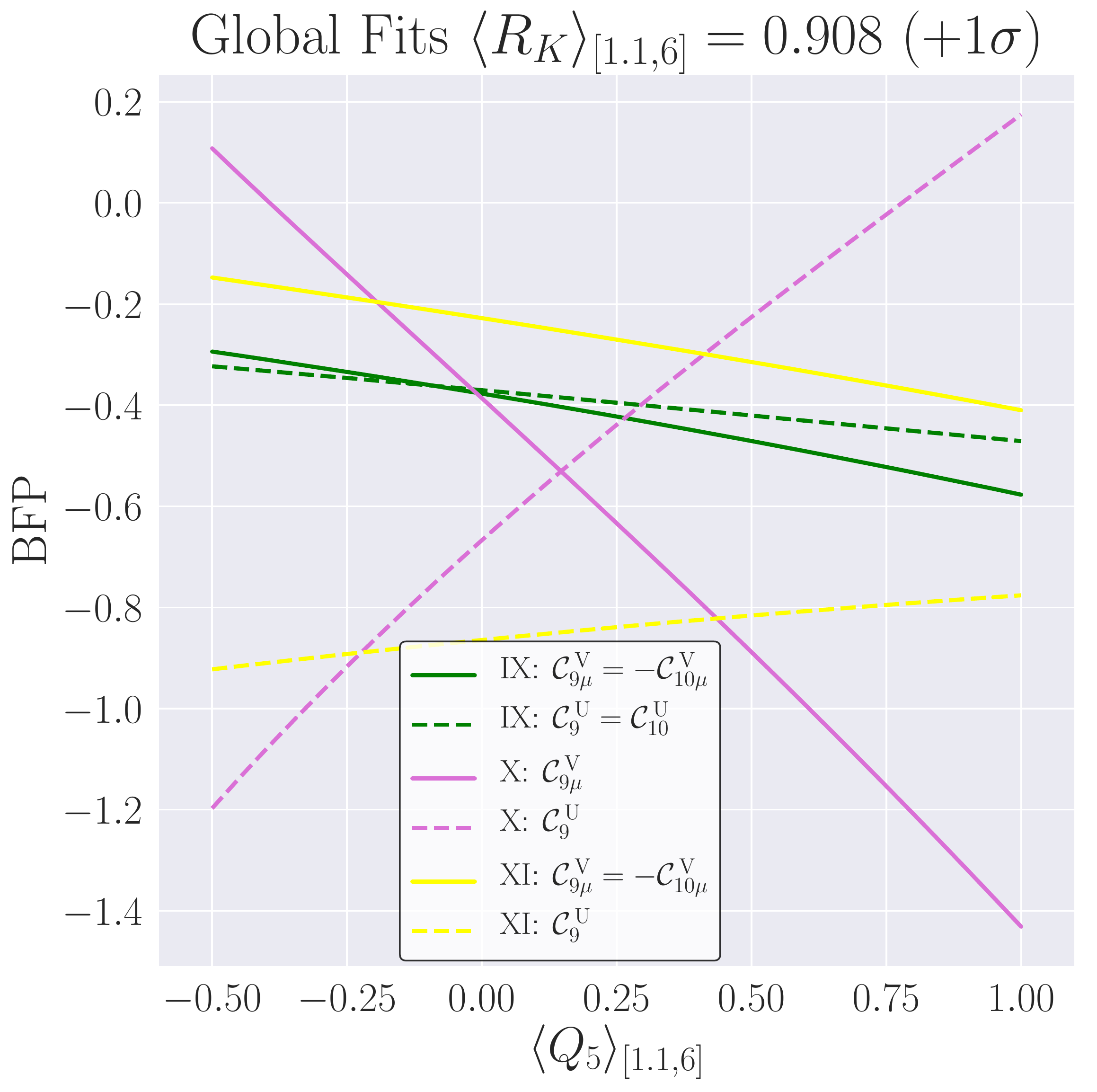}
\end{subfigure}
\caption{Global fit: Impact of $\langle Q_5\rangle_{[1.1,6]}$ on the b.f.p.s of the NP hypotheses under consideration for different values of $\langle R_K\rangle_{[1.1,6]}$.}         
\label{fig:GlobalFitPlotsQ5_2b}
\end{figure}

\begin{figure}[h]
\centering
\begin{subfigure}{.5\textwidth}
  \centering
  \includegraphics[width=0.92\linewidth]{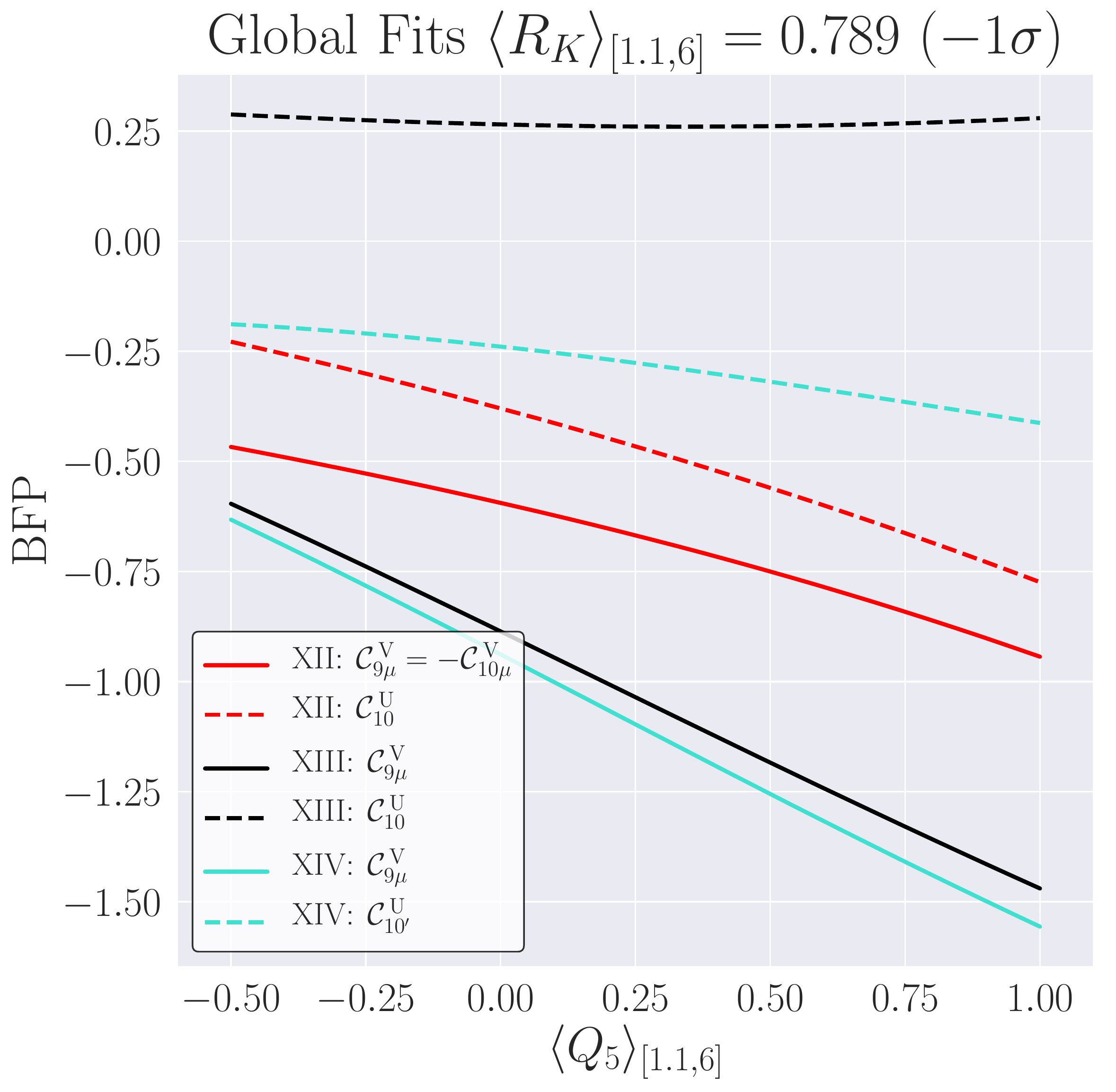}
\end{subfigure}

\begin{subfigure}{.5\textwidth}
  \centering
  \includegraphics[width=0.92\linewidth]{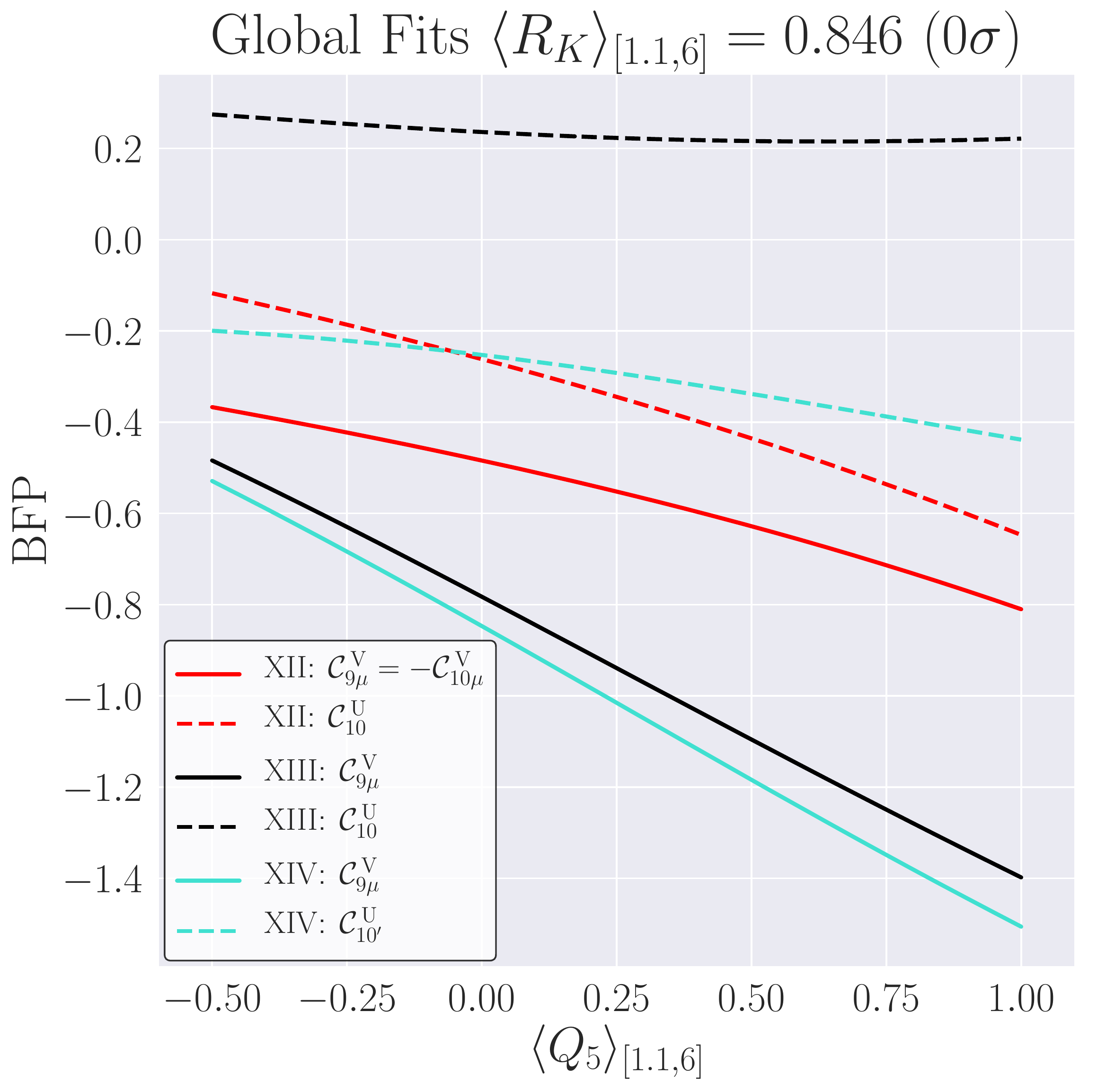}
\end{subfigure}

\begin{subfigure}{.5\textwidth}
\centering
\includegraphics[width=0.92\textwidth]{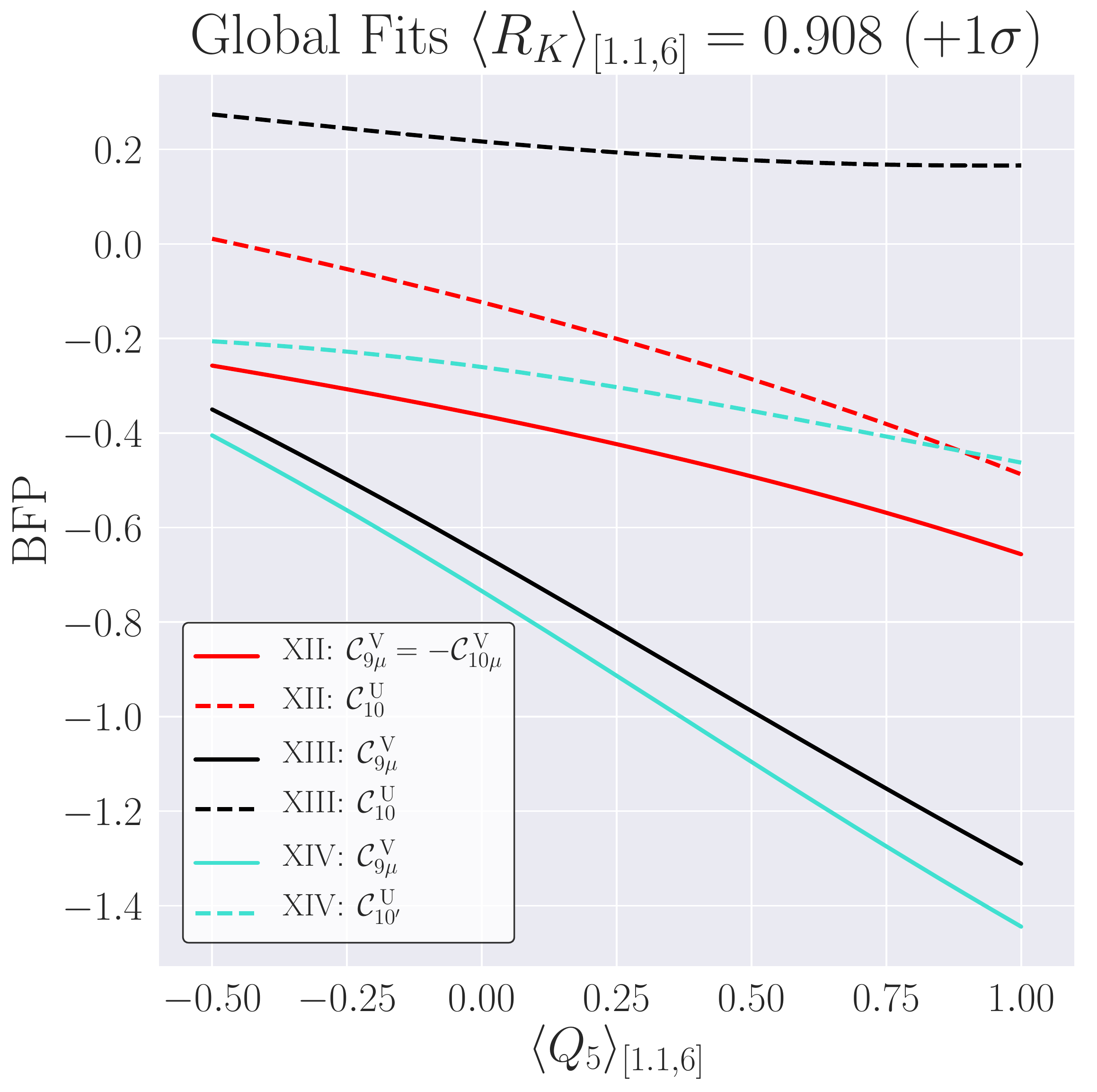}
\end{subfigure}
\caption{Global fit: Impact of $\langle Q_5\rangle_{[1.1,6]}$ on the b.f.p.s of the NP hypotheses under consideration for different values of $\langle R_K\rangle_{[1.1,6]}$.}         
\label{fig:GlobalFitPlotsQ5_2c}
\end{figure}


\clearpage

\begin{thebibliography}{99}


\bibitem{Aaij:2019wad}
  R.~Aaij {\it et al.} [LHCb Collaboration],
  Phys.\ Rev.\ Lett.\  {\bf 122} (2019) no.19,  191801
  [arXiv:1903.09252 [hep-ex]].
  
\bibitem{Abdesselam:2019wac}
  A.~Abdesselam {\it et al.} [Belle Collaboration],
  arXiv:1904.02440 [hep-ex].

\bibitem{Alguero:2019ptt}
  M.~Alguer\'{o}, B.~Capdevila, A.~Crivellin, S.~Descotes-Genon, P.~Masjuan, J.~Matias and J.~Virto,
  arXiv:1903.09578 [hep-ph].


\bibitem{Capdevila:2017bsm}
  B.~Capdevila, A.~Crivellin, S.~Descotes-Genon, J.~Matias and J.~Virto,
  JHEP {\bf 1801} (2018) 093
  [arXiv:1704.05340 [hep-ph]].
  
  

\bibitem{Descotes-Genon:2013wba}
S.~Descotes-Genon, J.~Matias and J.~Virto,
Phys.\ Rev.\ D {\bf 88} (2013) 074002
[arXiv:1307.5683 [hep-ph]].
  

\bibitem{Descotes-Genon:2015uva}
S.~Descotes-Genon, L.~Hofer, J.~Matias and J.~Virto,
JHEP {\bf 1606} (2016) 092
[arXiv:1510.04239 [hep-ph]].

    
\bibitem{Kumar:2019qbv}
  J.~Kumar and D.~London,
  arXiv:1901.04516 [hep-ph].

\bibitem{Ciuchini:2019usw}
  M.~Ciuchini, A.~M.~Coutinho, M.~Fedele, E.~Franco, A.~Paul, L.~Silvestrini and M.~Valli,
  arXiv:1903.09632 [hep-ph].
  
\bibitem{Aebischer:2019mlg}
  J.~Aebischer, W.~Altmannshofer, D.~Guadagnoli, M.~Reboud, P.~Stangl and D.~M.~Straub,
  arXiv:1903.10434 [hep-ph].

\bibitem{Alok:2019ufo}
  A.~K.~Alok, A.~Dighe, S.~Gangal and D.~Kumar,
  JHEP {\bf 1906} (2019) 089
  [arXiv:1903.09617 [hep-ph]].

 \bibitem{Arbey:2019duh}
  A.~Arbey, T.~Hurth, F.~Mahmoudi, D.~M.~Santos and S.~Neshatpour,
  arXiv:1904.08399 [hep-ph].


\bibitem{Ciuchini:2017mik}
  M.~Ciuchini, A.~M.~Coutinho, M.~Fedele, E.~Franco, A.~Paul, L.~Silvestrini and M.~Valli,
  Eur.\ Phys.\ J.\ C {\bf 77} (2017) no.10,  688
  [arXiv:1704.05447 [hep-ph]].


\bibitem{Hurth:2017hxg}
  T.~Hurth, F.~Mahmoudi, D.~Martinez Santos and S.~Neshatpour,
  Phys.\ Rev.\ D {\bf 96} (2017) no.9,  095034
  [arXiv:1705.06274 [hep-ph]].
  

\bibitem{Alguero:2018nvb}
  M.~Alguer\'{o}, B.~Capdevila, S.~Descotes-Genon, P.~Masjuan and J.~Matias,
  Phys.\ Rev.\ D {\bf 99} (2019) no.7,  075017
  [arXiv:1809.08447 [hep-ph]].
  
\bibitem{Capdevila:2017ert}
B.~Capdevila, S.~Descotes-Genon, L.~Hofer and J.~Matias,
JHEP {\bf 1704} (2017) 016
[arXiv:1701.08672 [hep-ph]].
  
\bibitem{Wehle:2016yoi}
S.~Wehle {\it et al.} [Belle Collaboration],
Phys.\ Rev.\ Lett.\  {\bf 118} (2017) no.11,  111801
[arXiv:1612.05014 [hep-ex]].


\bibitem{Datta:2017ezo}
  A.~Datta, J.~Kumar, J.~Liao and D.~Marfatia,
  Phys.\ Rev.\ D {\bf 97} (2018) no.11,  115038
  [arXiv:1705.08423 [hep-ph]].


\bibitem{Altmannshofer:2017bsz}
  W.~Altmannshofer, M.~J.~Baker, S.~Gori, R.~Harnik, M.~Pospelov, E.~Stamou and A.~Thamm,
  JHEP {\bf 1803} (2018) 188
  [arXiv:1711.07494 [hep-ph]].

\bibitem{Misiak:2006zs}
  M.~Misiak {\it et al.},
  Phys.\ Rev.\ Lett.\  {\bf 98} (2007) 022002
  [hep-ph/0609232].
  
\bibitem{Misiak:2006ab}
  M.~Misiak and M.~Steinhauser,
  Nucl.\ Phys.\ B {\bf 764} (2007) 62
  [hep-ph/0609241].
  
\bibitem{DescotesGenon:2011yn}
  S.~Descotes-Genon, D.~Ghosh, J.~Matias and M.~Ramon,
  JHEP {\bf 1106} (2011) 099
  [arXiv:1104.3342 [hep-ph]].
  
\bibitem{Asner:2010qj}
  D.~Asner {\it et al.} [Heavy Flavor Averaging Group],
  arXiv:1010.1589 [hep-ex].
  
    \bibitem{Aaij:2017vbb}
  R.~Aaij {\it et al.} [LHCb Collaboration],
  JHEP {\bf 1708} (2017) 055
  [arXiv:1705.05802 [hep-ex]].
  
\bibitem{Aaij:2016flj}
  R.~Aaij {\it et al.} [LHCb Collaboration],
  JHEP {\bf 1611} (2016) 047
   Erratum: [JHEP {\bf 1704} (2017) 142]
  [arXiv:1606.04731 [hep-ex]].
  
\bibitem{Straub:2015ica}
  A.~Bharucha, D.~M.~Straub and R.~Zwicky,
  JHEP {\bf 1608} (2016) 098
  [arXiv:1503.05534 [hep-ph]].
  
  
\bibitem{Khodjamirian:2010vf}
A.~Khodjamirian, T.~Mannel, A.~A.~Pivovarov and Y.-M.~Wang,
JHEP {\bf 1009} (2010) 089
[arXiv:1006.4945 [hep-ph]].

  

\bibitem{Altmannshofer:2017yso}
  W.~Altmannshofer, P.~Stangl and D.~M.~Straub,
  Phys.\ Rev.\ D {\bf 96} (2017) no.5,  055008
  [arXiv:1704.05435 [hep-ph]].
  
\bibitem{Altmannshofer:2017fio}
  W.~Altmannshofer, C.~Niehoff, P.~Stangl and D.~M.~Straub,
  Eur.\ Phys.\ J.\ C {\bf 77} (2017) no.6,  377
  [arXiv:1703.09189 [hep-ph]].
  
  
\bibitem{Gubernari:2018wyi}
  N.~Gubernari, A.~Kokulu and D.~van Dyk,
  JHEP {\bf 1901} (2019) 150
  [arXiv:1811.00983 [hep-ph]].
  
    
\bibitem{Gonzalez-Solis:2018ooo}
  S.~Gonz\`{a}lez-Sol\'is and P.~Masjuan,
  Phys.\ Rev.\ D {\bf 98} (2018) no.3,  034027
  [arXiv:1805.11262 [hep-ph]].
  
  
  \bibitem{DescotesGenon:2011pb}
  S.~Descotes-Genon, J.~Matias and J.~Virto,
  Phys.\ Rev.\ D {\bf 85} (2012) 034010
  [arXiv:1111.4882 [hep-ph]].

\bibitem{DeBruyn:2012wj}
  K.~De Bruyn, R.~Fleischer, R.~Knegjens, P.~Koppenburg, M.~Merk and N.~Tuning,
  Phys.\ Rev.\ D {\bf 86} (2012) 014027
  [arXiv:1204.1735 [hep-ph]].

\bibitem{DeBruyn:2012jp}
  K.~De Bruyn, R.~Fleischer, R.~Knegjens, M.~Merk, M.~Schiller and N.~Tuning,
  Nucl.\ Phys.\ B {\bf 868} (2013) 351
  [arXiv:1208.6463 [hep-ph]].

\bibitem{Descotes-Genon:2015hea}
  S.~Descotes-Genon and J.~Virto,
  JHEP {\bf 1504} (2015) 045
   Erratum: [JHEP {\bf 1507} (2015) 049]
  [arXiv:1502.05509 [hep-ph]].
  

\bibitem{Dettori:2018bwt}
  F.~Dettori and D.~Guadagnoli,
  Phys.\ Lett.\ B {\bf 784} (2018) 96
  [arXiv:1804.03591 [hep-ph]].
  
  
\bibitem{Aaij:2015esa}
  R.~Aaij {\it et al.} [LHCb Collaboration],
  JHEP {\bf 1509} (2015) 179
  [arXiv:1506.08777 [hep-ex]].
  
  \bibitem{Aaij:2015xza}
  R.~Aaij {\it et al.} [LHCb Collaboration],
  JHEP {\bf 1506} (2015) 115
   Erratum: [JHEP {\bf 1809} (2018) 145]
  [arXiv:1503.07138 [hep-ex]].


\bibitem{Aaij:2018gwm}
  R.~Aaij {\it et al.} [LHCb Collaboration],
  JHEP {\bf 1809} (2018) 146
  [arXiv:1808.00264 [hep-ex]].

\bibitem{Iwasaki:2005sy}
  M.~Iwasaki {\it et al.} [Belle Collaboration],
  Phys.\ Rev.\ D {\bf 72} (2005) 092005
  [hep-ex/0503044].
  
\bibitem{Lees:2013nxa}
 J.~P.~Lees {\it et al.} [BaBar Collaboration],
 Phys.\ Rev.\ Lett.\  {\bf 112} (2014) 211802
  [arXiv:1312.5364 [hep-ex]].

\bibitem{Aaij:2017vad}
  R.~Aaij {\it et al.} [LHCb Collaboration],
  Phys.\ Rev.\ Lett.\  {\bf 118} (2017) no.19,  191801
  [arXiv:1703.05747 [hep-ex]].

\bibitem{Lenz:2010gu}
  A.~Lenz {\it et al.},
  Phys.\ Rev.\ D {\bf 83} (2011) 036004
  [arXiv:1008.1593 [hep-ph]].
  
\bibitem{Charles:2011va}
  J.~Charles {\it et al.},
  Phys.\ Rev.\ D {\bf 84} (2011) 033005
  [arXiv:1106.4041 [hep-ph]].
  
\bibitem{Lenz:2012az}
  A.~Lenz, U.~Nierste, J.~Charles, S.~Descotes-Genon, H.~Lacker, S.~Monteil, V.~Niess and S.~T'Jampens,
  Phys.\ Rev.\ D {\bf 86} (2012) 033008
  [arXiv:1203.0238 [hep-ph]].
  
\bibitem{Charles:2016qtt}
  J.~Charles, S.~Descotes-Genon, V.~Niess and L.~Vale Silva,
  Eur.\ Phys.\ J.\ C {\bf 77} (2017) no.4,  214
  [arXiv:1611.04768 [hep-ph]].


\bibitem{:2005ema}
  [ALEPH and DELPHI and L3 and OPAL and SLD and LEP Electroweak Working Group and SLD Electroweak Group and SLD Heavy Flavour Group Collaborations],
  Phys.\ Rept.\  {\bf 427} (2006) 257
  [hep-ex/0509008].
  
 
\bibitem{demortier}
L. Demortier and L. Lyons, \emph{Everything you always wanted to know about pulls}, CDF Note 5776.


\bibitem{Capdevila:2018jhy}
  B.~Capdevila, U.~Laa and G.~Valencia,
  Eur.\ Phys.\ J.\ C {\bf 79} (2019) no.6,  462
  [arXiv:1811.10793 [hep-ph]].
  
\bibitem{Matias:2012xw}
  J.~Matias, F.~Mescia, M.~Ramon and J.~Virto,
  JHEP {\bf 1204} (2012) 104
  [arXiv:1202.4266 [hep-ph]].
  
\bibitem{Altmannshofer:2008dz}
  W.~Altmannshofer, P.~Ball, A.~Bharucha, A.~J.~Buras, D.~M.~Straub and M.~Wick,
  JHEP {\bf 0901} (2009) 019
  [arXiv:0811.1214 [hep-ph]].
  
   
\bibitem{Capdevila:2016ivx}
B.~Capdevila, S.~Descotes-Genon, J.~Matias and J.~Virto,
JHEP {\bf 1610} (2016) 075
[arXiv:1605.03156 [hep-ph]].

\bibitem{DAmico:2017mtc}
  G.~D'Amico, M.~Nardecchia, P.~Panci, F.~Sannino, A.~Strumia, R.~Torre and A.~Urbano,
  JHEP {\bf 1709} (2017) 010
  [arXiv:1704.05438 [hep-ph]].
  
\bibitem{Geng:2017svp}
  L.~S.~Geng, B.~Grinstein, S.~J\"ager, J.~Martin Camalich, X.~L.~Ren and R.~X.~Shi,
  Phys.\ Rev.\ D {\bf 96} (2017) no.9,  093006
  [arXiv:1704.05446 [hep-ph]].

\bibitem{Arbey:2018ics}
  A.~Arbey, T.~Hurth, F.~Mahmoudi and S.~Neshatpour,
  Phys.\ Rev.\ D {\bf 98} (2018) no.9,  095027
  [arXiv:1806.02791 [hep-ph]].
  
  
\bibitem{Hiller:2017bzc}
  G.~Hiller and I.~Nisandzic,
  Phys.\ Rev.\ D {\bf 96} (2017) no.3,  035003
  [arXiv:1704.05444 [hep-ph]].
  
\bibitem{Alok:2017sui}
  A.~K.~Alok, B.~Bhattacharya, A.~Datta, D.~Kumar, J.~Kumar and D.~London,
  Phys.\ Rev.\ D {\bf 96} (2017) no.9,  095009
  [arXiv:1704.07397 [hep-ph]].
  
  
\bibitem{Bobeth:2017vxj}
  C.~Bobeth, M.~Chrzaszcz, D.~van Dyk and J.~Virto,
  Eur.\ Phys.\ J.\ C {\bf 78} (2018) no.6,  451
  [arXiv:1707.07305 [hep-ph]].
  
  
  
  
  
  












  







\end{thebibliography}
\end{document}